\documentclass[
aps,
prx,
twocolumn,
superscriptaddress,
nofootinbib, longbibliography,
10pt]{revtex4-2}

\usepackage{soul}
\usepackage[labelfont=bf,font=small,singlelinecheck=false,justification=justified,format=plain]{caption}
\usepackage{physics}
\pdfoutput=1
\usepackage{amsmath,amsthm,mathtools, amssymb}
\usepackage{MnSymbol}
\usepackage{dsfont}
\usepackage{graphicx}
\usepackage{bm}
\usepackage{multirow}
\usepackage{cancel}
\usepackage[table,xcdraw]{xcolor}
\usepackage{enumitem}
\usepackage{comment}

\usepackage[colorlinks=true,allcolors=teal]{hyperref}

\newcommand{\Ish}{\!\!\rotatebox[origin=c]{135}{$\curvearrowleft$}}
\newcommand{\Dsh}{\!\!\rotatebox[origin=c]{45}{$\curvearrowright$}}

\newcommand{\declor}[1]{L^{\Dsh}_{#1}(s)}
\newcommand{\inclor}[1]{L^{\Ish}_{#1}(s)}
\newcommand{\declorarg}[1]{L^{\Dsh}_{#1}}
\newcommand{\inclorarg}[1]{L^{\Ish}_{#1}}
\newcommand{\qdeclor}[1]{L^{\Dsh}_{#1|q}(s)}
\newcommand{\qinclor}[1]{L^{\Ish}_{#1|q}(s)}

\renewcommand{\subsubsection}[1]{\textbf{#1}.}

\newtheorem{theorem}{Theorem}

\newtheorem{definition}{Definition}

\theoremstyle{definition}

\begin{document}

\title{Majorization theory for quasiprobabilities}

\author{Twesh Upadhyaya}
\email{tweshu@umd.edu}
\affiliation{Joint Center for Quantum Information and Computer Science, University of Maryland and NIST, College Park, MD 20742, USA}
\affiliation{Department of Physics, University of Maryland, College Park, MD 20742, USA}

\author{Zacharie Van Herstraeten}
\affiliation{QAT team, DIENS, \'Ecole Normale Sup\'erieure, PSL University, CNRS, INRIA, 45 rue d'Ulm, Paris 75005, France}

\author{Jack Davis}
\affiliation{QAT team, DIENS, \'Ecole Normale Sup\'erieure, PSL University, CNRS, INRIA, 45 rue d'Ulm, Paris 75005, France}

\author{Oliver Hahn}
\affiliation{Department of Basic Science, The University of Tokyo, 3-8-1 Komaba, Meguro-ku, Tokyo 153-8902, Japan}

\author{Nikolaos Koukoulekidis}
\affiliation{Duke Quantum Center and Departments of Physics and Electrical and Computer Engineering, Duke University, Durham, NC 27708, USA}

\author{Ulysse Chabaud}
\affiliation{QAT team, DIENS, \'Ecole Normale Sup\'erieure, PSL University, CNRS, INRIA, 45 rue d'Ulm, Paris 75005, France}

\date{August 26, 2026}

\begin{abstract}
    The question of how to characterize disorder arises in a wide variety of settings, including information theory, economics, and quantum thermodynamics.
    The theory of majorization provides an elegant answer, playing a central role in these fields.
    However, the existing majorization framework is inapplicable to functions that both take negative values and are defined on infinite spaces. Yet such functions, in the form of quasiprobability distributions, are ubiquitous in fields such as quantum optics, signal analysis, and bosonic quantum computation.
    Here we develop a notion of majorization that is applicable to such functions, proving that it admits four equivalent characterizations that naturally reduce to the finite case, thereby generalizing a seminal theorem by Hardy, Littlewood, and P\'olya. Moreover, we extend this equivalence to the setting where majorization is considered relative to an arbitrary positive distribution.
    We give several applications of our results in the context of quantum resource theories. These include deriving families of resource monotones and constraining quantum state conversions. We analytically and numerically study examples using the Wigner and Husimi functions, which feature prominently in quantum optics.
    Our results provide a comprehensive majorization framework for assessing the disorder of integrable functions over infinite measure spaces.
\end{abstract}

\maketitle

\section*{Introduction}

\textit{Majorization} quantifies the intuition that one probability distribution may be more ordered than another, capturing the notion of disorder, or information spread, in a more refined manner than other coarse-grained measures such as variance or entropies~\cite{marshall2011inequalities, Moein2019}. 
Initially defined for finite discrete probability distributions over a century ago~\cite{Muirhead1903,Lorenz1905,Dalton1920,Schur1923uber}, majorization has since been generalized to countable discrete distributions~\cite{Pereira2015} and to continuous probability distributions over both finite~\cite{hardy1934inequalities,ryff1963representation} and infinite measure spaces~\cite{Chong1974,Joe1987}. 
Following a breakthrough result by Hardy, Littlewood, and P\'olya~\cite{hardy1934inequalities}, several equivalent characterizations of majorization have been obtained using the concepts of Lorenz curves, stochastic operators, convex functions, and piecewise linear functions~\cite{Schur1923uber, Ostrowski1952sur, Day1973, Chong1974, Ruch1980}.
The definition of majorization has also been extended to \textit{relative majorization}, where probability distributions are analyzed with respect to a reference distribution~\cite{Blackwell1953,Veinott1971least,ruch1976principle,Joe1990}. These frameworks
have wide-ranging applications~\cite{marshall2011inequalities, wang2017majorization}, including in physics, information processing, and economics. Following the seminal work of Nielsen~\cite{nielsen1999conditions}, majorization has been used in quantum information to provide necessary, and sometimes sufficient, conditions for conversions between quantum states. This includes conversions using local operations and classical communication in the context of entanglement~\cite{Owari_CV_Nielsen_2008,Torun_Majorization_QRT_Review_2023,Horodecki2003reversible,horodecki2013fundamental} and using thermal operations in the context of quantum thermodynamics~\cite{Torun_Majorization_QRT_Review_2023,cwiklinski2015limitations,lostaglio2015description,gour2015the,gour2018quantum}. 

Akin to probability distributions are \textit{quasiprobability distributions}, which may violate some probability axioms, most commonly by being normalized but taking negative values.  Such distributions are used to represent quantum states and their transformations, over both finite and infinite measure spaces.  The former is commonly found in the quantum information processing of quantum systems with finite-dimensional Hilbert spaces~\cite{rundle2021overview}, while the latter is found across a variety of fields such as quantum optics~\cite{Leonhardt_1997}, continuous-variable quantum information processing~\cite{rundle2021overview, Braunstein_CVQI_2005}, and signal processing~\cite{dragoman2005applications}.

It was recognized early on in the mathematics literature that majorization theory over \emph{finite} spaces permits distributions with negative values~\cite{hardy1934inequalities, Chong1974, Joe1992}, a fact that has recently been 
exploited in quantum information: majorization has been applied to discrete Wigner quasiprobability distributions with resource-theoretic implications and practical applications for fault-tolerant quantum computing~\cite{koukoulekidis2022constraints, alexander2023general}, while entropies of Kirkwood--Dirac quasiprobabilities have been used to study the thermodynamics of quantum transport~\cite{Upadhyaya2024}.

An analogous extension of majorization to quasiprobabilities on \emph{infinite} measure spaces would be highly desirable, given the many uses of such distributions noted above. Yet, the infinite-measure case is significantly more challenging than the finite-measure one, an obstruction identified early on in the mathematics literature~\cite{Chong1974,Joe1992} and later revisited~\cite{Manjegani2023}. 
Some definitions of majorization in this setting have recently been put forward~\cite{Pereira2015,van2021majorization,Manjegani2023,de2024continuous}.
However, it remains unknown whether there are relationships between the different proposals and whether they inherit desirable properties of majorization from other settings.

In this work, we resolve this open question, formulating a comprehensive theory of majorization for quasiprobabilities on infinite measure spaces.
We prove a generalized Hardy--Littlewood--P\'olya theorem by establishing the 
equivalence of four characterizations based on Lorenz curves, stochastic 
operators, convex functions, and piecewise linear functions.
We further extend these definitions to the setting of relative majorization.
For nonnegative distributions and/or finite measure spaces, 
our definition naturally reduces to existing ones.

Our framework has broad applications, which we illustrate by applying it to several quantum resource theories where the underlying Hilbert spaces are infinite-dimensional. Focusing on two common quasiprobabilities, the Wigner function~\cite{Wigner1932} and the Husimi $Q$ function~\cite{husimi1940some}, we derive constraints on state conversions both analytically and numerically. Our framework lets us consider different sets of experimentally relevant free operations, including passive linear optical networks, Gaussian channels, and Wigner-positive channels, and rules out state conversions that existing methods do not~\cite{Takagi2018, Albarelli2018, Genoni_2008, Marian_2013, PhysRevLett.124.063605, hahn2024classical, hahn2025assessing}.
Moreover, the quasiprobabilistic entropies and divergences stemming from our framework provide a wide range of resource monotones for these resource theories which, unlike many existing monotones, are easily computed for mixed and multipartite states.

\section*{Results and Discussion}

We first provide an informal overview of our main results, including the equivalence of four conditions for majorization of general quasiprobability distributions, which forms the basis of our majorization framework.

We consider quasiprobabilities that are normalized distributions $f: X \rightarrow \mathbb{R}$ over a measure space $X$ with measure $\mu$, where $\mu(X)$ can be infinite.
Intuitively, to compare two distributions via majorization, one first rearranges their values in decreasing order then compares the cumulative sums or integrals, i.e. the \emph{Lorenz curves}. The key difficulty in infinite measure spaces arises due to an `infinite plateau effect'. Roughly speaking, since an integrable function over an infinite space must be close to zero over most of the space, rearranging its values in decreasing order creates an infinite plateau of vanishing values that `pushes to infinity' all the negative values and thus loses all information about them.
Our definition of majorization circumvents this plateau effect.
The key insight we leverage is to utilize \emph{both the increasing and decreasing rearrangements of a quasiprobability}~\cite{Pereira2015,van2021majorization}.
By doing so, we retain information about the positive and negative values of the quasiprobability. 

We prove four equivalent conditions for $f$ to \emph{majorize} $g$, denoted by $f\succ g$, which we now sketch (see Theorem~\ref{major} for the formal statement):
\medskip

\noindent (i) Our first characterization of majorization involves a comparison of Lorenz curves,
\begin{equation*}
    \declor{f} \geq \declor{g} \quad\text{ and }\quad \inclor{f}\leq \inclor{g} \,,
\end{equation*}
for all $s \in [0, \mu(X)]$. This captures the idea that the values of $f$ are more concentrated than those of $g$ and plotting the Lorenz curves offers a visual comparison of the disorder of distributions $f,g$.

\medskip

\noindent (ii) The second characterization of majorization is via a sequence of semidoubly stochastic operators $S_n$ mapping $f$ to $g$,
\begin{equation*}
    \quad S_n f \rightarrow g .
\end{equation*}
Roughly speaking, such operators $S_n$ are mixtures of different permutations so $g$ is formed by scrambling $f$.
These operators can be used to represent physical quantum operations that are \emph{free} in relevant quantum resource theories, such as Gaussian channels.

\medskip

\noindent (iii) A third characterization is in terms of all convex functions $\phi$ that vanish at the origin,
\begin{equation*}
    \int_X \phi(f) \ \mathrm d\mu \ge \int_X \phi(g) \ \mathrm d\mu \,.
\end{equation*}
This family of conditions captures another formulation of disorder: that the entropy of $f$ is smaller than that of $g$.
This rigorous extension of entropic measures to quasiprobabilities generates a large class of monotones for resource theories.

\medskip

\noindent (iv) Finally, the previous condition can be reduced to piecewise linear functions without loss of generality as they approximate any convex function.

\medskip

Majorization imposes a preorder on the set of distributions based on how well they resemble the uniform distribution. We extend this preorder to majorization relative to an arbitrary positive distribution $q$, denoted by $f\succ_q g$, and show that a similar equivalence between four definitions holds (see Theorem~\ref{relmajor} for a formal statement).
As we shall see, this permits the study of quantum channels with specific fixed points represented by $q$ (e.g.\ lossy linear optical networks preserve the vacuum state), thereby extending the reach of our framework to all Gaussian channels.

\section*{Part A: Majorization for quasiprobability distributions}
\label{sec:maths}

In this section, we provide the formal definitions and mathematical results underlying our majorization framework. The reader interested in the application of our framework may skip directly to Part B.

\subsection*{Preliminary material}
\label{sec:prelim}

Let $(X,\mathcal{A},\mu)$ be a $\sigma$-finite measure space, where $X$ is a set, $\mathcal{A}$ is a $\sigma$-algebra of $X$, and $\mu$ is a measure on them (in this article, $X$ is assumed to be $\sigma$-finite and integration is Lebesgue). If $\mu(X)$ is finite, $X$ is referred to as a finite measure space and as an infinite measure space otherwise. 
We consider the space of real-valued functions $f$ in $L^1(X,\mu)$ satisfying $\int_X f \mathrm d\mu=1$, and we refer to such functions as \emph{quasiprobability distributions}. 
Notable examples of quasiprobability distributions in quantum mechanics include the Wigner function~\cite{Wigner1932} and the Husimi $Q$ function~\cite{husimi1940some}, both of which play central roles in phase-space formulations of quantum theory. Note that our formalism excludes some distributions of mathematical interest, such as Wigner functions with infinite negative volume~\cite{De_Gosson_De_Gosson_2021_Feichtinger}. 
The negative volume of $f$ is defined as
\begin{equation}\label{eq:nv}
    NV(f) \coloneqq \frac{1}{2}\left(\int\abs{f} \mathrm{d} \mu-1\right)=\frac{1}{2}\qty(\norm{f}_1-1),
\end{equation}
where $\norm{\cdot}_1$ is the $L^1$-norm.
In order to analyze and compare $L^1$ functions in the context of majorization, we need to define the distribution function and the corresponding decreasing rearrangement.

\begin{definition}[Distribution function and decreasing rearrangement~\cite{Chong1974}]
    The distribution function of $f$ is $D_f(t)\coloneqq\mu\{x: f(x)> t \}$.
    The decreasing rearrangement of $f$ is
    \begin{equation}
        f^\downarrow(u)\coloneqq\inf \{ t: D_f(t)\leq u \}=\sup \{ t: D_f(t)> u \} \,.
    \end{equation}
\end{definition}

Analogously, we introduce the complementary concept of a \emph{co}distribution function.
\begin{definition}[Codistribution function and increasing rearrangement]
    The codistribution function of $f$ is $C_f(t)\coloneqq\mu\{x: f(x)< t \}$. 
    The increasing rearrangement of $f$ is
    \begin{equation}
        f^\uparrow(u)\coloneqq\sup \{ t: C_f(t)\leq u \}=\inf \{ t: C_f(t)> u \} \,.
    \end{equation}
\end{definition}
\noindent $D_f$ is right-continuous and decreasing; $C_f$ is left-continuous and increasing.

The positive and negative parts of $f$ are denoted by $f^+\coloneqq\max\{f,0\}$ and $f^-\coloneqq\min\{f,0\}$.
The positive and negative Lorenz curves of $f$ are
\begin{equation}
    L^{\Dsh}_f(s)\coloneqq \int_0^s (f^+)^\downarrow \ \mathrm{d} u \, \quad \mathrm{and}\quad
    L^{\Ish}_f(s)\coloneqq  \int_0^s (f^-)^\uparrow \ \mathrm{d}  u \,,
\end{equation}
for all $s\in [0,\mu(X)]$, where $\mathrm{d} u$ indicates integration with respect to Lebesgue measure on $\mathbb{R}$. 
As a matter of convention, our positive and negative Lorenz curves only depend on $f^+$ and $f^-$, respectively, in contrast to the literature~\cite{marshall2011inequalities}. As we treat the finite- and infinite-measure cases in a unified manner, it is to be understood that when $\mu(X)$ is infinite, $[0,\mu(X)]$ is replaced by $[0,\mu(X))$.
We denote the indicator function of a set $A$ by $\chi_A$. For clarity, when we introduce a second copy of the measure space, we denote it by $Y=X$.

For our purposes, it is sufficient to focus on integral operators. 
A linear operator $S: L^1(Y,\mu) \rightarrow L^1(X,\mu)$ is an integral operator if there exists a real-valued measurable function $S(x,y)$ on $X \cross Y$, called the (integral) \textit{kernel of} $S$, such that $Sf=g$ can be expressed as
\begin{equation}\label{eq:intgral_operator_kernel_def}
    g(x) = \int_Y S(x,y) f(y) \ \mathrm d\mu(y) \,.
\end{equation}  
\begin{definition}[Positive integral operator]\label{def:positive}
An integral operator $S$ is positive
if $S(x,y) \geq 0$ for all $(x,y) \in X \cross Y$.
\end{definition}
\begin{definition}[Stochastic integral operator]\label{def:stochastic}
A positive integral operator $S$ is stochastic if $\int_X S(x,y) \ \mathrm d\mu(x)=1$ for almost all $y\in Y$.
\end{definition}
\begin{definition}[Semidoubly stochastic integral operator~\cite{Bahrami2020}]\label{def:sds}
A stochastic integral operator $S$ is semidoubly stochastic (SDS) if $\int_Y S(x,y) \ \mathrm d\mu(y) \leq 1$ for almost all $x\in X$ and doubly stochastic if the same holds with equality.
\end{definition}

The need for \emph{semi}double stochasticity in going from finite to infinite measure spaces has been noted in the discrete~\cite{Markus1964} and continuous~\cite{Bahrami2020,Manjegani2023} cases.
As per Theorem 4.11 of~\cite{Manjegani2023}, given any semidoubly stochastic operator $S:L^1(Y)\rightarrow L^1(X)$, there exists a sequence of integral semidoubly stochastic operators converging to $S$ on a finite-dimensional subspace of $L^1(Y)$. When $\mu(X)<\infty$, all semidoubly stochastic operators are doubly stochastic (see Proposition 2.6 of~\cite{Bahrami2020}). 
Intuitively, this can be seen by summing all the elements of a finite matrix either by row or by column and noting that the result must be the same.

At this point, we should emphasize a striking discrepancy between the finite-measure and infinite-measure scenarios. Consider $f$ such that $\int_X f\ \mathrm{d}\mu=1$.
When $X$ has finite measure, the (co)distribution functions $C_f$ and $D_f$ are \textit{almost-everywhere} equivalent, in the sense of the following relation:
\begin{align}
    D_f(t)
    +
    C_f(t)
    =
    \mu(X)
    -
    \underbrace{\mu\Big(\big\lbrace x:f(x)=t\big\rbrace\Big)}_{\text{=0 almost everywhere}}
    .
\end{align}
Thus, both $D_f$ and $C_f$ contain the same information about the level sets of $f$. 
The decreasing and increasing Lorenz curves $L_f^{\uparrow}\coloneqq \int f^\downarrow \mathrm{d} u$ and $L_f^{\downarrow}\coloneqq \int f^\uparrow  \mathrm{d} u$, defined without taking the positive and negative parts of $f$, are then related to each other:
\begin{align}
    L_f^{\uparrow}(s)
    +
    L^{\downarrow}_f\big(\mu(X)-s\big)
    =
    1.
\end{align}
So, for $\mu(X)<\infty$, it suffices to consider just one of these two Lorenz curves to define a majorization relation. 
Further, to connect to our convention where the positive and negative parts of $f$ are considered separately, note that $L_f^{\uparrow}(s) = \min\{\declor{f},1-\inclorarg{f}(\mu(X)-s)\}$ and, similarly, $L_f^{\downarrow}(s) = \max\{\inclor{f},1-\declorarg{f}(\mu(X)-s)\}$. Hence in finite measure spaces,
\begin{equation}\label{eq:samecurve}
\begin{aligned}
&L_f^{\uparrow}(s)\geq L_g^{\uparrow}(s) \iff L_f^{\downarrow}(s)\leq L_g^{\downarrow}(s) \iff     
\declor{f}\geq\declor{g}\\ 
&\quad\mathrm{and} \quad\inclor{f}\leq\inclor{g} \,.
\end{aligned}
\end{equation}

The situation is very different when $X$ has infinite measure.
In such a case, observe that the condition $f\in L^{1}(X,\mu)$ informally forces $f$ to vanish at infinity.
In turn, its (co)distribution function generically diverges at $t=0$.
Hence $D_f$ and $C_f$ are no longer equivalent; both Lorenz curves need to be taken into account when defining a majorization relation. Prior works have proposed extensions of majorization to possibly negative functions; see for example Corollary 1.3 in~\cite{Chong1974} or the comment on page 146 of~\cite{Joe1992};
however they did so by considering $\mu(X)-D_f(t)$ and were hence restricted to the case of finite measure. Our introduction of the codistribution function $C_f$ is a key step enabling us to expand the theory of majorization to quasiprobabilities on infinite measure spaces. As a convention to avoid redundancy, we only consider the positive (negative) parts of $f$ in the decreasing (increasing) Lorenz curves.

\subsection*{Equivalent characterizations of majorization}

With notation in place, we are able to provide four equivalent characterizations of majorization for integrable quasiprobability distributions over infinite measure spaces.
Our theorem can be viewed as a generalization of the theorem by Hardy, Littlewood and P\'olya~\cite{hardy1934inequalities} which concerns discrete probability distributions.

The first of our equivalent definitions of majorization involves a comparison between the Lorenz curves of the distributions, which reflects one of the original definitions of majorization for discrete probability distributions, and provides a visualization of majorization order which we exploit in our numerical plots.
We then employ stochastic operators, which represent simple physical operations that transform between the system states represented by the distributions.
Finally, we stipulate majorization in terms of convex functions, which directly leads to entropic inequalities that respect majorization order.
The convex functions $(\cdot)^+$ and $-(\cdot)^-$ act as building blocks that can linearly approximate any convex function (Theorem 2.7 of~\cite{Joe1990}), and they are highlighted as our fourth definition.

\begin{theorem}\label{major}
Given two functions $f,g\in L^1(X,\mu) $, we say that $f$ majorizes $g$, and write $f \succ g$, if any of the following equivalent statements holds.
\begin{enumerate}
\item $L^{\Dsh}_f(s) \geq L^{\Dsh}_g(s)$ and $L^{\Ish}_f(s) \leq L^{\Ish}_g(s)$ for all $s\in [0,\mu(X)]$, and $\int_X f \ \mathrm d\mu = \int_X g \ \mathrm d\mu$; 
\item There exists a sequence of semidoubly stochastic (integral) operators $(S_n)_{n \in \mathbb{N}}$ such that $S_n f$ converges to $g$ in $L^1(X,\mu)$;
\item $\int_X \phi(f) \ \mathrm d\mu \ge \int_X \phi(g) \ \mathrm d\mu$ for all nonnegative convex functions $\phi: \mathbb{R} \rightarrow \mathbb{R}_{\geq 0}$ satisfying $\phi(0)=0$ such that the integrals converge, and $\int_X f \ \mathrm d\mu = \int_X g \ \mathrm d\mu$;
\item $ \int_X (f-z)^+ \ \mathrm d\mu \ge \int_X (g-z)^+ \ \mathrm d\mu$ and $ \int_X (f+z)^- \ \mathrm d\mu \le \int_X (g+z)^- \ \mathrm d\mu $ for all real numbers $z \ge 0$, and $\int_X f \ \mathrm d\mu = \int_X g \ \mathrm d\mu$.
\end{enumerate}
\end{theorem}
We first make some remarks. In statement 2, we can interchange SDS operators and SDS integral operators, which follows from the fact that the latter can approximate the former arbitrarily well on $L^1$ functions (see Preliminary material).
Finally, the inequality in statement 3 also holds when we add any linear shift to such a convex function, $\phi(x) + ax$ for any constant $a$. This enables the application of statement 3 to convex functions that take negative values.

Our characterizations of majorization reduce to previously known characterizations restricted to probability distributions~\cite{marshall2011inequalities,Chong1974,Joe1987,VanHerstraeten2023continuous,Manjegani2023} or to quasiprobability distributions when $\mu(X)$ is the counting measure
~\cite{hardy1934inequalities, koukoulekidis2023quasi} (see Supplementary Note (SN) 1D). Note that for probabilities, $\phi$ is only required to be convex over $\mathbb{R}_{\geq 0}$, and, since this is not an open interval, $\phi$ is additionally required to be continuous.

\begin{proof} We sketch here the steps of the proof; the complete proof can be found in SN 1B.

\noindent$\bm{1 \implies 2: }$
We first note that by assumption necessarily $NV(f) \geq NV(g)$. When $NV(f) = NV(g)$, our strategy is to construct quasiprobability distributions $f_\epsilon$ and $g_\epsilon$ supported on finite-measure subsets that approximate $f$ and $g$ arbitrarily well as $\epsilon \rightarrow 0$. We then leverage existing results for $L^1$ functions in finite measure spaces to connect the Lorenz curve condition to doubly stochastic operators.

When $NV(f) > NV(g)$, we construct an intermediate function $f_{\textrm{red}}$ which has a positive (negative) Lorenz curve that lies above (below) the positive (negative) Lorenz curve of $g$ and $NV(f_{\textrm{red}}) = NV(g)$, as depicted in Figure~\ref{fig:fred} for the positive Lorenz curve.
\begin{figure}[h]
    \centering
    \includegraphics[width=\linewidth]{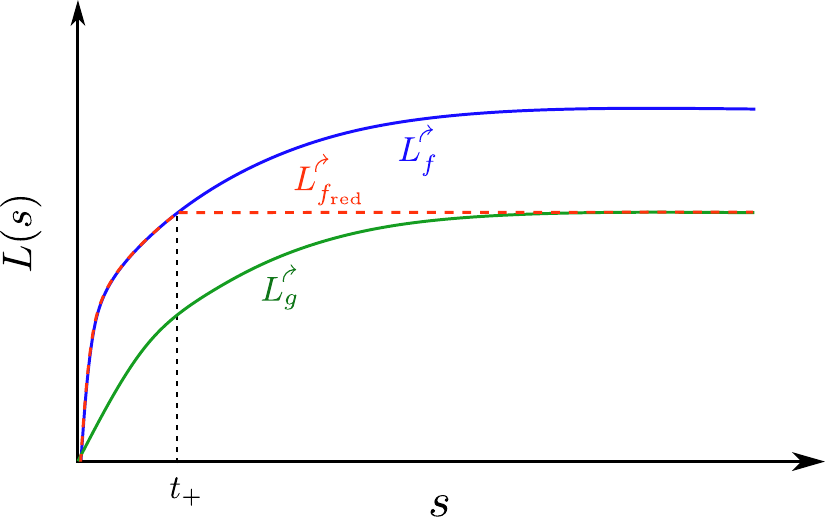}
    \captionof{figure}{Sketch of proof strategy for Theorem~\ref{major}. The reduced Lorenz curve (red) is constructed to lie below the decreasing Lorenz curve of $f$ and reach the same height as that of $g$.}
    \label{fig:fred}
\end{figure}
We show that there exists a semidoubly stochastic operator that maps $f$ to $f_{\textrm{red}}$, and we know already that, because $NV(f_{\textrm{red}}) = NV(g)$,  $f_{\textrm{red}}$ can be mapped to $g$ via a doubly stochastic operator, concluding the proof.

\noindent$\bm{2 \implies 3: }$
Semidouble stochasticity lets us apply Jensen's inequality to the convex function $\phi$.

\noindent$\bm{3 \implies 4: }$
This follows simply from the fact that the functions $(\cdot)^+$ and $-(\cdot)^-$ are convex.

\noindent$\bm{4 \implies 1: }$ We establish and exploit tight bounds that connect the functions $(\cdot)^+$ and $-(\cdot)^-$ to the (co)distribution functions, which lead to the desired implication.
\end{proof}

\subsection*{Relative majorization}
Majorization can be viewed as a preorder of distributions based on how well they resemble the uniform distribution. 
In many applications, we may instead be interested in comparisons relative to an arbitrary distribution, not necessarily the uniform one. 
This motivates the definition of \emph{majorization relative to q}, or just relative majorization, which is a preorder on distributions based on how well they resemble $q$.
This generalization of majorization is also known in the literature of discrete probability distributions as $d$--majorization~\cite{Veinott1971least}, or in the context of quantum thermodynamics as thermo-majorization~\cite{horodecki2013fundamental}.
Here, we extend our characterization of majorization for quasiprobability distributions over infinite measure spaces to relative majorization.

Let $f,g\in L^1(X,\mu) $ and let $q$ be any strictly positive, measurable function. We define a rescaled integration measure $\nu$ to formulate relative majorization,
\begin{equation}\label{eq:nu}
     \nu(A) \coloneqq \int_X \chi_A \, q(x) \, \mathrm d\mu(x) \,,\quad \text{for all measurable } A \subseteq X \,.
\end{equation}

Other quantities are defined analogously to regular majorization.
\begin{definition}[Relative distribution function and decreasing rearrangement]\label{def:rel_dist}
    The distribution function of $f$ relative to $q$ is $D^q_f(t)\coloneqq\nu\{x: f(x)> tq(x) \} $.
The decreasing rearrangement of $f$ relative to $q$ is 
\begin{align}
    f^{\downarrow,q}(u)&\coloneqq\inf \{ t: D^q_f(t)\leq u \}=\sup \{ t: D^q_f(t)> u \}\,.
\end{align}
\end{definition}
\begin{definition}[Relative codistribution function and increasing rearrangement]\label{def:rel_codist}
   The codistribution function of $f$ relative to $q$ is $C^q_f(t)\coloneqq\nu\{x: f(x)< tq(x) \}$.
The increasing rearrangement of $f$ relative to $q$ is
\begin{align}
    f^{\uparrow,q}(u)&\coloneqq\sup \{ t: C^q_f(t)\leq u \}=\inf \{ t: C^q_f(t)> u \} \,.
\end{align}
\end{definition}
The positive and negative Lorenz curves of $f$ relative to $q$ are
\begin{align}
    \qdeclor{f}\coloneqq \int_0^s (f^+)^{\downarrow,q} \mathrm du \,\quad \mathrm{and}\quad \qinclor{f}\coloneqq \int_0^s (f^-)^{\uparrow,q} \ \mathrm du \,,
\end{align}
for all $ s\in [0,\nu(X)]$, where $\mathrm d u$ is the Lebesgue measure on $\mathbb{R}$.

We also generalize the notion of SDS operators to accommodate relative majorization.
\begin{definition}[Semi-$q$-stochastic operator]\label{def:SqS}
A stochastic integral operator $S$ is semi-$q$-stochastic (S$q$S) if $(Sq)(x) \leq q(x)$ for almost all $x\in X$ and $q$-stochastic if the same holds with equality.
\end{definition}

We extend all four equivalent characterizations of Theorem~\ref{major} to relative majorization.
\begin{theorem}\label{relmajor}
    Given two functions $f,g\in L^1(X,\mu) $ and a strictly positive, measurable function $q$, we say that $f$ majorizes $g$ relative to $q$, and write $f \succ_q g$, if any of the following equivalent statements holds.
    \begin{enumerate}
    \item $\qdeclor{f} \geq \qdeclor{g}$ and $\qinclor{f} \leq \qinclor{g}$ for all $s\in [0,\nu(X)]$,
    and $\int_X f \ \mathrm d\mu = \int_X g \ \mathrm d\mu$;
    \item There exists a sequence of stochastic integral operators $(S_n)_{n \in \mathbb{N}}$ such that $S_n f$ converges to $g$ in $L^1(X,\mu)$ and $S_n q \le q$ for all $n \in \mathbb{N}$;
    \item  $\int_X q \phi\left(\frac{f}{q}\right) \ \mathrm d\mu \ge \int_X q \phi\left(\frac{g}{q}\right) \ \mathrm d\mu$ for all nonnegative convex functions $\phi: \mathbb{R} \rightarrow \mathbb{R}_{\geq 0}$ satisfying $\phi(0)=0$ such that the integrals converge, and $\int_X f \ \mathrm d\mu = \int_X g \ \mathrm d\mu$;
    \item $ \int_X (f-zq)^+ \ \mathrm d\mu \ge \int_X (g-zq)^+ \ \mathrm d\mu $ and $ \int_X (f+zq)^- \ \mathrm d\mu \le \int_X (g+zq)^- \ \mathrm d\mu  $ for all real numbers $z \ge 0$, and $\int_X f \ \mathrm d\mu = \int_X g \ \mathrm d\mu$.
    \end{enumerate}
\end{theorem}

The proof of this result builds upon Theorem~\ref{major} and is detailed in SN 1C.
Some remarks are in order. Multiplying $q$ by a positive constant has no effect on the relative majorization preorder.
As in Theorem~\ref{major}, the inequality in statement 3 immediately extends to linear shifts of such convex functions.
Relative majorization reduces to regular majorization for $q(x)=1$. Note that when $q\in L^1(X)$, $\nu(X)$ is finite. 

In SN 1D, we explicitly write out Theorem~\ref{relmajor} for the special case of the counting measure, which allows us to treat discrete distributions.
In particular, our approach recovers the embedding map construction commonly used in quantum thermodynamics~\cite{horodecki2013fundamental,lostaglio2019introductory} for finite, discrete distributions with thermal distributions as reference.
In fact, one could consider an even more general notion of relative majorization with two reference distributions $q,q'$, such that $f$ majorizes $g$ if and only if $S_nf \rightarrow g$ and $S_nq \le q'$ for a sequence of stochastic operators $(S_n)_{n\in\mathbb{N}}$, which we leave to future work.

\subsection*{Properties of majorization}
\label{sec:prop}

In this section we discuss important mathematical properties of (relative) majorization as defined above (see SN 1E for additional properties).
Majorization is a preorder: it is reflexive, $f \succ_q f$, and transitive, $f\succ_q g$ and $g\succ_q h$ $\Rightarrow$ $f\succ_q h$, for all $f,g,h \in L^1(X,\mu)$ and strictly-positive measurable functions $q$.
If $f \not\succ_q g$ and $g \not\succ_q f$, we say that $f$ and $g$ are \emph{incomparable relative to $q$}. 
If $f \succ_q g$ and $g \succ_q f$, we say that $f$ and $g$ are \emph{equivalent relative to $q$}.
The positive (negative) Lorenz curve is concave (convex) and continuous.

\textbf{Schur-convex and Schur-concave functionals}
\label{schurfunc}
Schur-convex functionals are coarse--grained measures that partially capture the majorization preorder, and can constitute resource monotones within the context of a resource theory (see "Results and Discussion Part B").
A \emph{Schur-convex} functional $F:L^1(X)\rightarrow \mathbb{R}$ is a functional that respects the majorization preorder: $f \succ g \implies F(f) \geq F(g)$, and is \textit{Schur-concave} if the inequality is flipped. 
The Shannon entropy is a well-known example of a Schur-concave functional, measuring the expected amount of information needed to describe the state of a random variable. Schur-convex functionals are often easier than Lorenz curves to compute analytically.

By the third equivalent definition of majorization (see Theorem~\ref{major}), a convex function $\phi$ satisfying some additional requirements can be used to construct a Schur-convex functional. The composition of a Schur-convex functional with an increasing function is also Schur-convex. With these two rules, we can build a wide array of Schur-convex/concave functionals on quasiprobabilities. 
For example, the function $\phi_\alpha(t) = \abs{t}^\alpha$ is convex over $\mathbb{R}$ for $\alpha\geq1$, nonnegative, and vanishes at the origin. Thus the $L^p$-norm with $p=\alpha$,
\begin{equation}
    \Vert f\Vert_{\alpha} \coloneqq \left( \int_X \abs{f(x)}^\alpha \mathrm{d} \mu(x) \right)^{\frac{1}{\alpha}},    
\end{equation}
is a Schur-convex functional for $\alpha\geq1$. Similarly, the $\alpha$-R\'{e}nyi entropy, 
\begin{align}
    H_\alpha(f)\coloneqq\frac{\alpha}{1-\alpha}\log\Vert f\Vert_{\alpha},
\end{align}
is Schur-concave since $\frac{1}{1-\alpha} \log(\cdot)$ is a decreasing function. 
A different entropy called the $\alpha$-Tsallis entropy, $\frac{1}{\alpha-1} \left( 1- \int_X \abs{f(x)}^\alpha \mathrm{d} \mu(x) \right)$, is similarly Schur-concave for $\alpha>1$.
Interestingly, $\phi_\alpha(t)$ is neither concave nor convex over $\mathbb{R}$ for $\alpha<1$. Hence the $\alpha$-R\'{e}nyi and $\alpha$-Tsallis entropies are not Schur-concave for quasiprobability distributions when $\alpha < 1$, in contrast to the case of probability distributions where they remain Schur-concave for all $\alpha > 0$. For quasiprobabilities with nonzero negative volume, both entropies diverge as $\alpha\rightarrow 1$. 

Similarly to how majorization gives rise to entropic functions, relative majorization gives rise to relative entropies, also known as divergences. 
For example, the $\alpha$-R\'{e}nyi divergence,
\begin{equation}
    D_\alpha(f||q) \coloneqq \frac{1}{\alpha-1} \log \int_X \abs{f(x)}^\alpha q(x)^{1-\alpha} \mathrm d\mu(x),
\end{equation}
satisfies $D_\alpha(f||q) \ge D_\alpha(g||q)$ whenever $f \succ_q g$, for $\alpha>1$ (see statement 3 of Theorem~\ref{relmajor} and note $\abs{f(x)}^\alpha q(x)^{1-\alpha} = \abs{f(x)/q(x)}^{\alpha} q(x)$).

The previous examples showcase how some entropic functions for probability distributions can be converted into entropic functions for quasiprobability distributions by taking the absolute value of the argument of the underlying convex function $\phi$, extending it to a convex function over the entire real line. 
It is worth noting that this is not the only way to build such extensions. For example, the power divergence~\cite{Joe1990}, defined by the convex function $\phi(t) = 2(\alpha(\alpha+1))^{-1} (t^{1+\alpha}-t)$ can be extended to quasiprobability distributions for $\alpha > 0$ via $\phi(t) = 2(\alpha(\alpha+1))^{-1} (\abs{t}^{1+\alpha}-t)$, where we have only taken the absolute value of one occurrence of $t$.

Monotones can also be straightforwardly derived from geometrical features of the Lorenz curve.
For instance, the asymptotic value of the positive Lorenz curve equals $1+NV(f)$ (for a normalized function $f$), where $NV$ denotes the total negative volume (see Eq.~\eqref{eq:nv}), which is Schur-convex.
Hence, our majorization framework directly implies the monotonicity of negative volume for comparable distributions.
The negative volume of the Wigner function of quantum states is a commonly used monotone in the resource theory of non-Gaussianity~\cite{kenfack2004negativity, Takagi2018,Albarelli2018} (see SN 4A for more details).
Other Schur-convex functionals defined by features of the Lorenz curves include the extreme values $\max f^+$ and $-\min f^-$ corresponding to the slopes of the positive and negative Lorenz curves at the origin, as well as $G(f) \coloneqq \frac{1}{s}$, where $s$ is the smallest value at which the Lorenz curve reaches 1 and $G=0$ if there is no such $s$.
In the context of quantum resource theory, these monotones provide ways to quantify the nonclassicality of quantum states.

\textbf{Incomparability of functions with equivalent \texorpdfstring{$L^2$}{}-norms.}\label{Sec:L2_incomparability}

As a preorder, majorization allows for incomparable distributions, i.e., cases where no majorization relation holds in either direction. Intuitively, this happens when it is not possible to objectively label one distribution as more disordered than the other: their relative ordering depends on the specific Schur-concave function used to quantify disorder.
In this spirit, we prove a general incomparability theorem (see SN 2 for the proof).
\begin{theorem}
\label{th:incomparable}
Let $f,g\in L^{1}(X,\mu)\cap L^{\infty}(X,\mu)$ be normalized functions with equal $L^2$-norms, and assume that both their decreasing and increasing rearrangements are absolutely continuous.
Then, $f$ and $g$ are either incomparable ($f\not\succ g$ and $f\not\prec g$) or equivalent ($f\succ g$ and $f\prec g$) under majorization. 
\end{theorem}

Although Theorem 3 allows for both incomparability and equivalence, the latter is a much stronger condition: it requires the Lorenz curves of the two functions to match exactly, while incomparability occurs as soon as they cross. This suggests that equivalence is a rather special case while incomparability is generic. In this sense, the strength of Theorem 3 lies primarily in its ability to certify incomparability. 

Importantly, Theorem 3 implies that the Wigner functions of two quantum states with equal purity are generically incomparable, since the squared $L^2$-norm of a Wigner function is equal to the state's purity (see SN 2).

\section*{Part B: Application to state conversion}
\label{sec:app}
Our equivalent definitions of majorization apply generally to integrable functions on infinite measure spaces, and enable in particular the study of disorder in quasiprobabilistic (i.e.\ functional) representations of quantum states. Hereafter, we use majorization theory for these representations to study quantum state conversions, adopting an approach inspired by \textit{quantum resource theories}~\cite{ChitambarGourResourceReview2019}, which have been an active research area in quantum information theory. These provide a unified framework for quantifying useful quantum features under physically constrained operations. More precisely, a quantum resource theory specifies a subset of operations as \emph{free}, which usually corresponds to operations that can be easily performed experimentally. For example, local operations and classical communication are free in resource theories of entanglement because they do not require using entanglement or sending quantum states, while Gaussian channels are free in resource theories of non-Gaussianity because they can be implemented with simple optical components. Free operations can be identified in other contexts such as quantum computation~\cite{lloyd1999quantum}, quantum error correction~\cite{PhysRevA.97.022341}, and quantum metrology~\cite{grochowski2025optimal}.
One of the key questions in any quantum resource theory is: When can one quantum state $\rho$ be converted into another state $\sigma$ using only \textit{free operations}?

Given a resource theory, we choose a quasiprobability representation $\Upsilon$ that maps quantum states $\rho$ to distributions $f_{ \rho}$ over an appropriately selected space $X$. This induces a corresponding kernel representation of quantum channels, or completely positive trace-preserving maps (see SN 3). 
When the free operations of the resource theory are represented by stochastic kernels, which is the case in many quantum resource theories of interest, majorization theory allows us to draw conclusions about possible transformations using these operations.

We provide a general methodology to apply our majorization framework:
\begin{enumerate}
    \item Pick a quasiprobability representation $\Upsilon$ and restrict attention to quantum states that are mapped to integrable functions.
    \item Identify quantum channels that have SDS kernels. Analyze state conversions with these channels using majorization.
    \item Identify quantum channels that have S$q$S kernels, for instance by identifying the fixed points of quantum channels with stochastic kernels. Analyze state conversions with majorization relative to $q$.
\end{enumerate}

From a resource-theoretic point of view, once a representation $\Upsilon$ is picked, this methodology allows us to assess quantum state conversion under several classes of free operations. For majorization, the set of free operations includes all quantum channels with SDS kernels with respect to $\Upsilon$. For relative majorization with respect to a reference distribution $q$, the set of free operations includes all the quantum channels with S$q$S kernels with respect to $\Upsilon$, where $q$ is the quasiprobability distribution of a fixed point of the channel.  
Then, if $f_{ \rho}$ does not majorize $f_{ \sigma}$, no free operation can convert $ \rho$ into $ \sigma$. 
If, in addition, $f_{ \sigma}$ and $f_{ \rho}$ are incomparable (i.e.\ their Lorenz curves cross), then neither $ \rho$ nor $ \sigma$ can be converted into the other with free operations. Moreover, Schur-convex functionals (see Schur-convex and Schur-concave functionals) directly provide monotones for the corresponding resource theory.
It is generally difficult to characterize exactly the set of all channels with a stochastic kernel. For example, it is an open question even to identify all the Wigner-positive states; identifying all channels with Wigner-positive kernels is at least as hard. However, one does not need to exactly characterize this set to apply our majorization framework. Instead, one just needs to show that the set contains the operations designated as free. One strategy to achieve this is to pick a representation $\Upsilon$ that is covariant under unitary free operations and consider dilated channels similar to the approach in SN 5.

\subsection*{Wigner majorization}

Hereafter, we show how to apply our framework by focusing on continuous-variable (CV) quantum mechanics associated with a finite number of canonical degrees of freedom, known physically as a bosonic quantum system. 
For the purposes of this section, we set aside most technical considerations related to convergence and boundedness, which, while mathematically significant, do not influence physical applications in which all representations are typically `well-behaved'. 

Each choice of representation defines a different subset of functions in $L^1(\mathbb R^{2n})$; in general, the number of such choices is vast~\cite{Cohen_2013}.
We illustrate our majorization framework by focusing primarily on the Wigner quasiprobability representation~\cite{Wigner1932} (see SN 4A for a detailed background), which is arguably the most prominent quasiprobability representation in quantum mechanics. In that case, we show that majorization provides insightful monotones and bounds for state conversions within existing quantum resource theories. However, our framework is general and accommodates other quasiprobability distributions as well. In SN 6, we extend our regular and relative majorization analysis to the Husimi $Q$ function~\cite{husimi1940some}, another important quasiprobability distribution.

An important step in our approach is to identify classes of channels whose Wigner kernels are either SDS or S$q$S. 
Among them, we distinguish the well-known family of Gaussian channels, characterized by a rescaling matrix $\bm{X}$, a displacement in phase space, and a noise matrix, and their generalization to positive Gaussian-dilatable channels, which share the same rescaling matrix $\bm X$ but involve an arbitrary Wigner-positive (i.e.\ possibly non-Gaussian) environment state.
The latter reduce to Gaussian channels when the environment is Gaussian. We prove that such channels are SDS if and only if $\vert\det\bm{X}\vert \geq 1$.
A detailed analysis is provided in SN 4B and SN 5, where we also discuss other examples of channels with stochastic Wigner kernels, such as certain measure-and-prepare channels.

Hereafter, we consider both regular majorization and relative majorization of Wigner functions for different reference distributions $q$ (see SN 7 for a list of all functions used). In each case, we summarize examples of the corresponding free operations, i.e. operations whose kernel in the Wigner representation is either SDS or S$q$S. We plot numerically comparable and incomparable Lorenz curves of states of interest. 
Particular attention is given to Fock states, which in certain settings exhibit a (relative) majorization hierarchy.
We also present carefully selected examples that highlight the necessity of considering both Lorenz curves and demonstrate that our framework imposes stronger constraints than previously known monotones. 
Finally, we explore a multimode example and examine majorization relative to thermal and unbounded reference operators.

\textbf{Regular Wigner majorization.}
For regular majorization, the set of free operations includes amplifying Gaussian channels
and more generally (mixtures of) amplifying Gaussian-dilatable channels with Wigner-positive environments. 

\begin{figure*}[t]
\centering
    \includegraphics[width=0.8\linewidth]{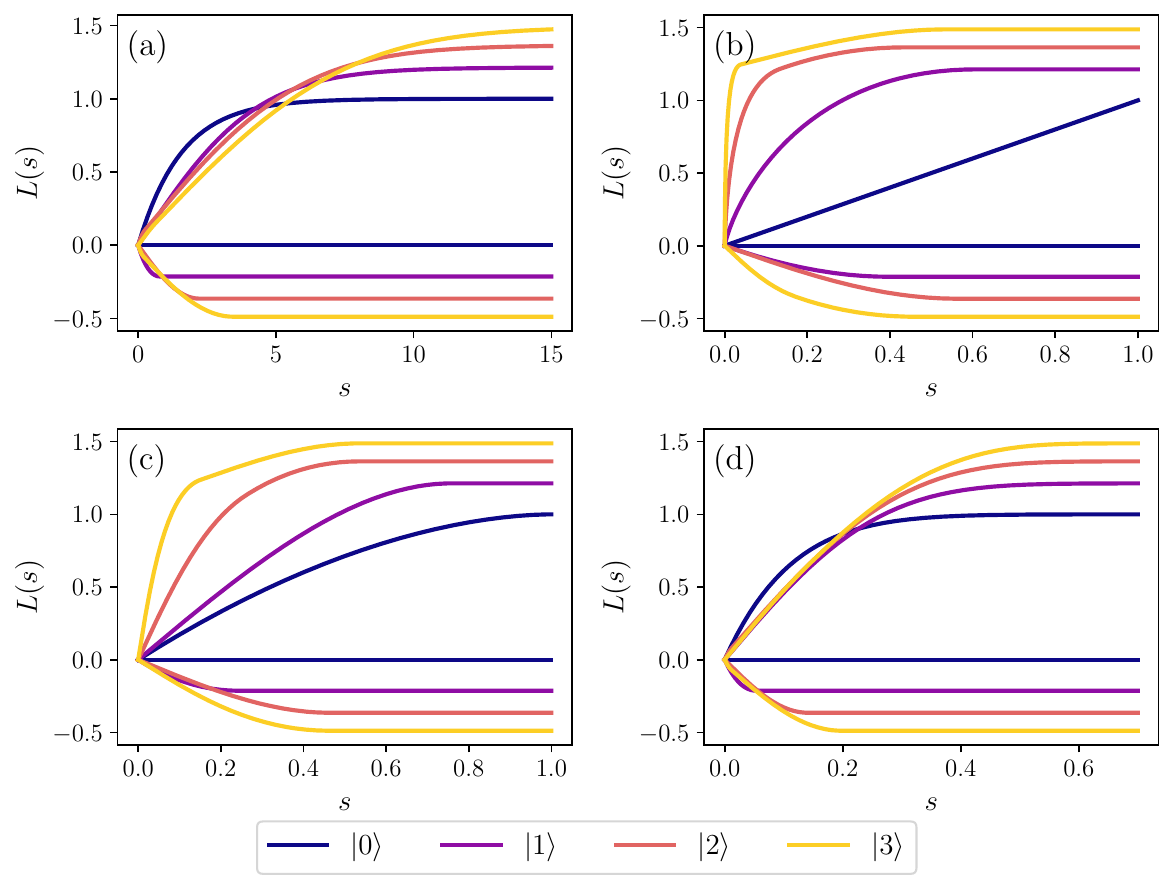}
    \captionof{figure}{Positive and negative Wigner Lorenz curves of first few Fock states. (a) Regular majorization: all the curves cross, showing the incomparability of Fock states. (b) Majorization relative to vacuum: a hierarchy forms with higher Fock states majorizing lower ones. (c) Majorization relative to thermal state at $\bar{n}=0.4$: some of the curves cross, indicating a partial breakdown of the Fock hierarchy. (d) Majorization relative to thermal state at $\bar{n}=4$: the Fock hierarchy collapses further as the curves start to resemble their regular majorization counterparts.
    }
    \label{fig:Fockhierarchy}
\end{figure*}

\paragraph*{Incomparability of Fock states.}
The Wigner functions of Fock states are all incomparable via regular majorization, as illustrated in Figure~\ref{fig:Fockhierarchy}(a). This is in direct connection with Theorem~\ref{th:incomparable}, which tells us that all pure states are either incomparable or equivalent.

\begin{figure}[h!]
    \centering
    \includegraphics[width=0.9\linewidth]{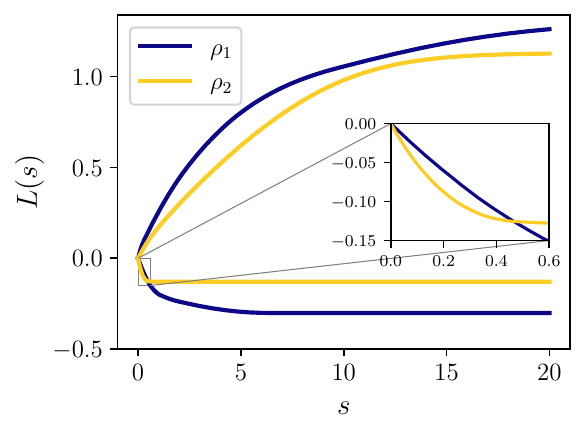}
\captionof{figure}{Positive and negative Wigner Lorenz curves of two quantum states. The states are $\rho_1=\frac{4}{5} \dyad{\alpha=2}_\mathrm{cat}+ \frac{1}{5}\dyad{7}$ and $\rho_2=\frac{1}{2}\dyad{a=5,n=3}_\mathrm{ON}+\frac{1}{2}\dyad{1}$. While the positive curves do not intersect, the negative ones do, as highlighted in the inset.}
\label{fig:negativeLorenz}
\end{figure}

\paragraph*{Importance of both Lorenz curves.} Our definition of majorization crucially relies upon considering both increasing and decreasing rearrangements. As Figure~\ref{fig:negativeLorenz} illustrates, the information about negative Lorenz curves cannot be omitted in general. In this example, the positive Lorenz curve of $\rho_1=\frac{4}{5} \dyad{\alpha=2}_\mathrm{cat}+ \frac{1}{5}\dyad{7}$ lies above that of $\rho_2=\frac{1}{2}\dyad{a=5,n=3}_\mathrm{ON}+\frac{1}{2}\dyad{1}$, but the negative Lorenz curves intersect. Transforming the former state to the latter using a free operation is hence prohibited, a fact that is made evident by considering the negative Lorenz curves.

\begin{figure}[h!]
\centering
\includegraphics[width=0.9\linewidth]{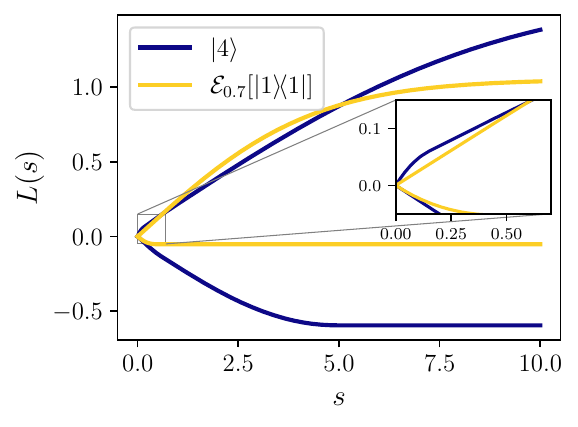}
\captionof{figure}{Positive and negative Wigner Lorenz curves of the Fock state $\ket{4}\bra{4}$ and a lossy Fock state $\mathcal{E}_{0.7}[\ket{1}\bra{1}]$. The crossing of the curves, highlighted in the inset, indicates a transformation from the first to the second is not possible with any amplifying Gaussian channel.}
\label{fig:2way}
\end{figure}

\paragraph*{Strong constraints from majorization.}
Majorization provides strong restrictions on the interconvertibility of quantum states. Consider the task of transforming $\dyad{4}$ to a lossy Fock state $\dyad{1}$ with an amplifying Gaussian channel (i.e.~$\abs{\det \bm X}\geq1$). Any such channel is a free operation in resource theories of non-Gaussianity. Hence, several existing monotones can be used to study the possibility of this conversion. In Table~\ref{tab:wigner_vanilla}, we compute several state-of-the-art monotones: Wigner negativity~\cite{kenfack2004negativity,Takagi2018,Albarelli2018}, stellar rank~\cite{PhysRevLett.124.063605}, and relative entropy of non-Gaussianity~\cite{Genoni_2008,Marian_2013} (we remark that the different monotones we consider are not entirely interchangeable, as they each apply to different sets of free operations, e.g.~decreasing stellar rank rules out conversions under any Gaussian channel, not just the amplifying ones).
Each of the monotones is decreasing and thus fails to rule out the transformation. Three of the Schur-convex functionals we defined, the squared $L^2$-norm, the maximum, and the negative minimum, also fail to rule out the conversion. 
Nonetheless, our majorization theory lets us deduce the transformation is in fact impossible; in Figure~\ref{fig:2way}, we plot Lorenz curves for the two states and find that they intersect. This illustrates the novel insights our theory provides, as well as the importance of considering the entirety of the Lorenz curves. 

\begin{table}
\centering
\resizebox{\columnwidth}{!}{%
\renewcommand{\arraystretch}{1.1}
\begin{tabular}{lcc|}
\cline{2-3}
\multicolumn{1}{l|}{}                                                                  & \multicolumn{1}{c|}{\cellcolor[HTML]{EFEFEF}$\ket{4}\bra{4}$} & \cellcolor[HTML]{EFEFEF}$\mathcal{E}_{0.7}[\ket{1}\bra{1}]$ \\ \hline
\rowcolor[HTML]{FFFFFF} 
\multicolumn{1}{|l}{\cellcolor[HTML]{FFFFFF}\textbf{Existing monotones}}               & \multicolumn{1}{l}{\cellcolor[HTML]{FFFFFF}}                  & \multicolumn{1}{l|}{\cellcolor[HTML]{FFFFFF}}               \\ \hline
\multicolumn{1}{|l|}{\cellcolor[HTML]{EFEFEF}Wigner negative volume}                   & \multicolumn{1}{c|}{0.596}                                    & 0.052                                                       \\ \hline
\multicolumn{1}{|l|}{\cellcolor[HTML]{EFEFEF}Relative entropy of non-Gaussianity}      & \multicolumn{1}{c|}{2.502}                                    & 0.541                                                       \\ \hline
\multicolumn{1}{|l|}{\cellcolor[HTML]{EFEFEF}Stellar rank}                             & \multicolumn{1}{c|}{4}                                        & 1                                                           \\ \hline
\rowcolor[HTML]{FFFFFF} 
\multicolumn{1}{|l}{\cellcolor[HTML]{FFFFFF}\textbf{Schur-convex functionals}} & \multicolumn{1}{l}{\cellcolor[HTML]{FFFFFF}}                  & \multicolumn{1}{l|}{\cellcolor[HTML]{FFFFFF}}               \\ \hline
\multicolumn{1}{|l|}{\cellcolor[HTML]{EFEFEF}Squared $L^2$-norm (Purity)}                   & \multicolumn{1}{c|}{1}                                        & 0.580                                                       \\ \hline
\multicolumn{1}{|l|}{\cellcolor[HTML]{EFEFEF}Wigner maximum}                           & \multicolumn{1}{c|}{0.318}                                    & 0.123                                                       \\ \hline
\multicolumn{1}{|l|}{\cellcolor[HTML]{EFEFEF}Negative Wigner minimum}                           & \multicolumn{1}{c|}{0.129}                                   & 0.127                                                      \\ \hline
\end{tabular}%
}
\captionof{table}{
Existing and majorization-derived monotones for amplifying Gaussian channels. Left column is calculated for the Fock state $\ketbra{4}{4}$ and right for the lossy Fock state $\mathcal{E}_{0.7}(\dyad{1})$, where $\mathcal{E}_{\eta}$ is the pure loss channel with transmittance $\eta$. 
Each measure is higher for the former state, thus not ruling out conversion into the latter using an amplifying Gaussian channel. However this conversion is in fact prohibited by our majorization theory because of the intersection of Lorenz curves.
}

\label{tab:wigner_vanilla}
\end{table}

\textbf{Wigner majorization relative to the vacuum state.}

Next, we consider majorization relative to the Wigner function of the vacuum. 
The set of free operations includes Gaussian channels that have vacuum as a fixed point. Lossy linear optical networks (LONs) are an important such class, see SN 4B for more details.

\paragraph*{Fock-state hierarchy.}
While the Wigner functions of Fock states are all pairwise incomparable via regular majorization, they are comparable relative to vacuum. An interesting phenomenon arises, namely a \emph{Fock-state-majorization hierarchy}, illustrated for the first few Fock states in Figure~\ref{fig:Fockhierarchy}(b).
We see that $W_{\ket{n+1}}$ majorizes $W_{\ket{n}}$ relative to $W_{\ket{0}}$. This implies that a conversion from higher to lower Fock states is not ruled out using LONs, even though it is ruled out when we restrict to amplifying Gaussian channels, i.e.\ Gaussian channels with $\abs{\det \bm X} \geq 1$. 
In SN 6, we prove that for the Husimi $Q$ function, a Fock-state hierarchy also emerges for both majorization relative to vacuum and regular majorization.

\begin{figure*}[t]
    \centering
    \includegraphics[width=0.9\linewidth]{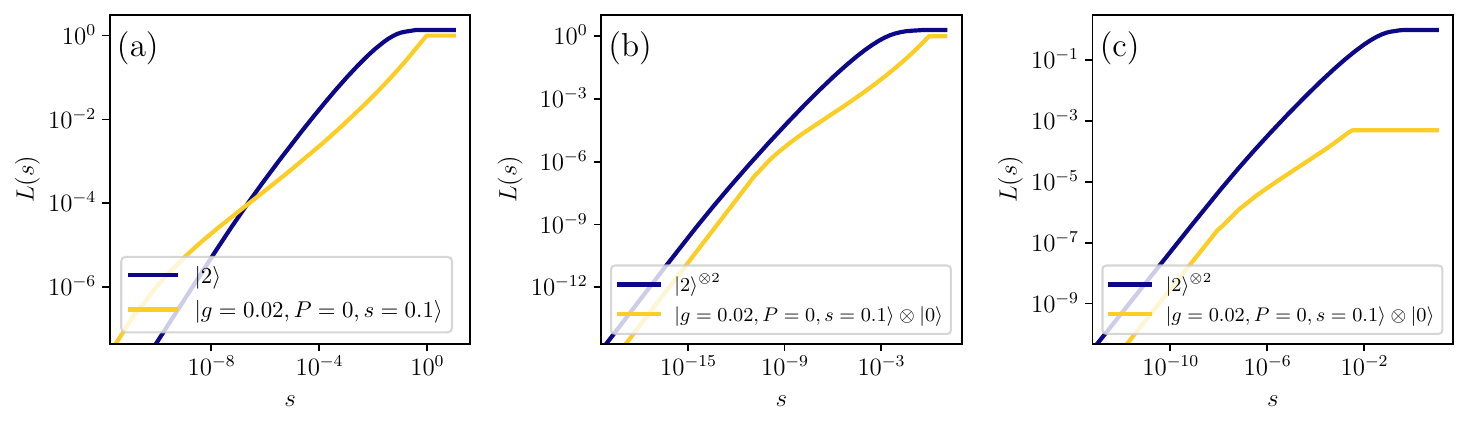}
    \captionof{figure}{Log-log plots of Wigner Lorenz curves for Fock and squeezed cubic phase states. (a) Log-log plot of positive Wigner Lorenz curves for Fock state $\ket{2}$ and squeezed cubic phase state $\ket{g=0.02,P=0,s=0.1}$, relative to vacuum. The curves are incomparable. Log-log plots of the positive (b) and the absolute value of the negative (c) Wigner Lorenz curves for $\ket{2}^{\otimes 2}$ and $\ket{g=0.02,P=0,s=0.1} \otimes \ket{0}$, relative to vacuum. The curves are comparable.} 
    \label{fig:2wayrelative}
\end{figure*}

\paragraph*{Cubic phase state generation.}
The framework we have introduced is not limited to single-mode Wigner functions.
Consider the task of using free operations to convert Fock states into squeezed cubic phase states, the latter of which are important for universal quantum computation with Gaussian operations~\cite{Takagi2018}. It is impossible to generate $\ket{g=0.02,P=0,s=0.1}$ from one copy of $\ket{2}$ due to the intersection of Lorenz curves in Figure~\ref{fig:2wayrelative}(a). However, the state conversion is not ruled out with free operations given two copies of $\ket{2}$ (Figure~\ref{fig:2wayrelative}), as evidenced by the comparability of the positive (b) and negative (c) Lorenz curves. As this example highlights, there are interesting phenomena in multimode state conversion which majorization allows us to probe.

\textbf{Wigner majorization relative to thermal states.}

Generalizing from vacuum, we now consider majorization relative to thermal states at positive temperatures (i.e.~positive mean photon number $\bar{n}$).
The set of free operations includes several attenuating Gaussian channels, such as the thermal loss channel, and the dephasing channel.
Consider an imperfect, lossy LON that has some thermal noise. Such a channel has a thermal state rather than vacuum as a fixed point, and is hence a free operation for this section. 

\paragraph*{Collapse of Fock-state hierarchy.} The Fock-state hierarchy relative to vacuum is not robust. Even at small $\bar{n}$, some Fock states become incomparable (Figure~\ref{fig:Fockhierarchy}(c)). 
Fock states tautologically majorize the vacuum $\ket{0}$ relative to itself, yet they do not relative to sufficiently hot thermal states. In Table~\ref{tab:0N_crossing_thermal}, we determine numerically the smallest photon numbers at which the first few Fock states become incomparable with the vacuum state.
As the temperature is increased past this threshold, curves that are incomparable seem to remain that way.
The collapse of the Fock hierarchy with increasing temperature is not surprising, as in the limit $\bar{n} \rightarrow \infty$, relative majorization becomes equivalent to regular majorization. This equivalence follows because the reference thermal Wigner function approaches the uniform distribution, and majorization relative to the uniform distribution is simply regular majorization. We illustrate this effect for the first few Fock states in Figure~\ref{fig:Fockhierarchy}(d).

\begin{table}[h!]
    \centering
    \begin{tabular}{c|c c c c c}
       $\ket{n}$  & 1&2&3&4&5 \\ \hline
         $\bar{n}$&   0.64& 1.23&1.80&2.36&2.90  
    \end{tabular}
    \captionof{table}{Mean photon number $\bar{n}$ of thermal state relative to whose Wigner function the Wigner Lorenz curve of Fock state $\ket{n}$ starts to cross the Wigner Lorenz curve of the vacuum $\ket{0}$.}
    \label{tab:0N_crossing_thermal}
\end{table}

\textbf{Wigner majorization relative to unbounded operators.}
\label{sec:unboundedmaj}
To showcase the generality of our majorization framework, we now consider reference thermal distributions at \emph{negative} temperature.
Fixed points of the kernel of a channel, which are distributions in phase space, need not correspond to density operators. Thus, some amplifying Gaussian channels, despite having no quantum states as fixed points, are free operations in this section. 
Many Gaussian-dilatable channels also have negative-temperature thermal fixed points.
An important example of such a channel is the quantum-limited phase conjugation channel, characterized by the matrices $\bm X=-\kappa \sigma_3$ and $\bm Y=(1+\kappa^2) \mathds 1$ for a parameter $\kappa\geq0$~\cite{Ivan2011-pm}. The channel satisfies $\abs{\det \bm X}=\kappa^2$ and has $W(x,p)= e^{\frac{\kappa^2-1}{\kappa^2+1}(x^2+p^2)}$ 
as a fixed point. 

\begin{figure*}[t]
\centering
\includegraphics[width=0.9\linewidth]{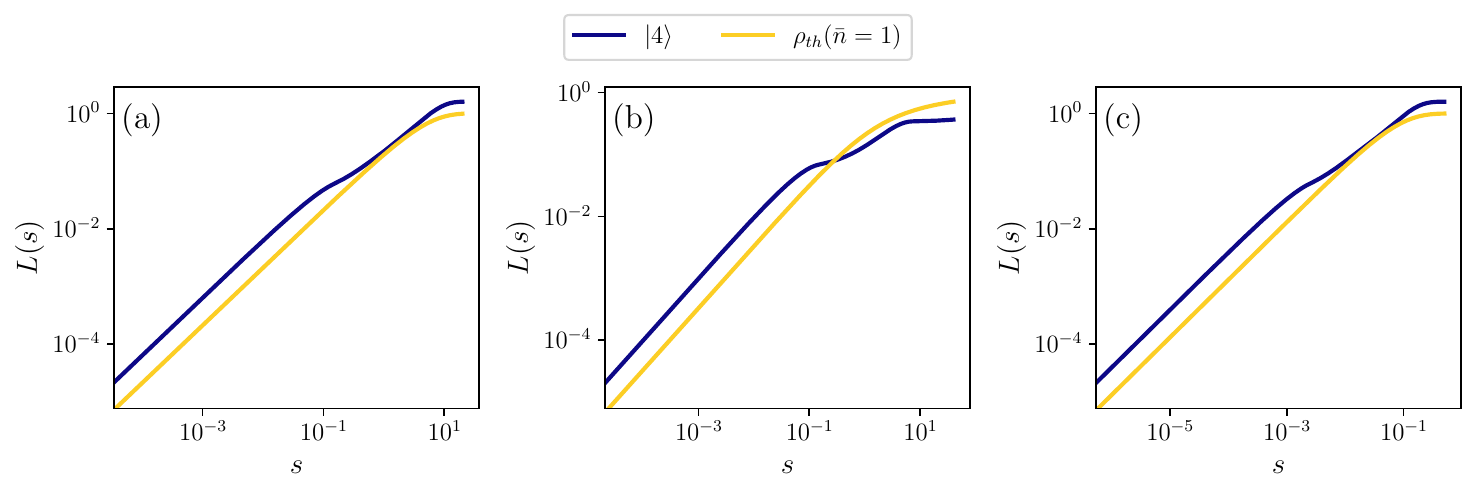}
\captionof{figure}{Log-log plots of positive Wigner Lorenz curves of Fock state $\ket{4}$ and thermal state $\rho_{th}( \bar{n}=1)$. (a) Regular majorization: the curves are comparable at large positive temperature. (b) Majorization relative to thermal distribution at $\bar{n}=-1$: the curves are incomparable relative to a thermal distribution at negative temperature. (c) Majorization relative to thermal distribution at $\bar{n}=-20$: the curves are again comparable.}
\label{fig:unbounded}
\end{figure*}

\paragraph*{Majorization relative to thermal operators.} 
As we shall illustrate, negative temperature reference distributions are not just a theoretical curiosity. Consider converting $\ket{4}$ into $\rho_{th}(\bar{n}=1)$ with a free operation. As the latter is Wigner-positive, it suffices to consider only the positive Lorenz curves to determine comparability. The Lorenz curves for regular majorization (which can be thought of as majorization relative to an infinite positive temperature distribution) indicate this transformation could be possible using any of the corresponding free operations (Figure~\ref{fig:unbounded}(a)). However, the Lorenz curves relative to a thermal distribution at negative temperature (i.e. hotter than any positive temperature) intersect (Figure~\ref{fig:unbounded}(b)). Numerically we observe that the curves remain incomparable unless $\bar{n}$ becomes sufficiently negative (Figure~\ref{fig:unbounded}(c)). Thus, we can rule out the conversion using free operations with thermal fixed points in the temperature range where the curves are incomparable.

\section*{Conclusion}
\label{sec:concl}

In this work we have introduced majorization and relative majorization for integrable quasiprobability distributions over infinite measure spaces. 
We have shown that both flavors of majorization admit four equivalent definitions, based on Lorenz curves, stochastic operators, convex functions, and piecewise linear functions. We then analyzed properties of (relative) majorization, establishing entropic functionals and other monotones as features that directly arise from the majorization preorder, and deriving a strong constraint on the comparability of functions in terms of their $L^2$-norms.

Applied to quasiprobability representations in quantum mechanics, our framework provides insight into state conversions within quantum resource theories. We study several examples with the Wigner and Husimi functions numerically and analytically. In particular, we show our framework rules out state conversions which are not ruled out by existing techniques.

Our majorization results can be used to assess the preparation of bosonic states with restricted operations. We have demonstrated this for cubic phase states, used in universal quantum computation. Similarly, our results apply to quantum error correction, where Gaussian conversions~\cite{PhysRevA.97.022341} are often used to prepare bosonic states. These include GKP and cat states, which are useful for correcting errors~\cite{gottesman2001encoding,aghaee2025scaling} and reducing volume overheads~\cite{putterman2025hardware}. Our framework can also address, in the context of quantum metrology, preparation of probe states such as the N00N state~\cite{dowling2008quantum} using phase-insensitive operations~\cite{grochowski2025optimal,slussarenko2017unconditional,you2021scalable,nielsen2023deterministic}. 

More broadly, potential applications of our majorization preorder include: assessing thermalization for solutions to differential equations over unbounded measure spaces; evaluating truncated series approximations of positive functions where the truncation induces negativity, for example Edgeworth series; and comparing distributions of electric charge or time-frequency signal representations and characterizing physical processes by their action on such distributions.

A promising avenue is to generalize our results to other variants of majorization. These include: relative majorization with different reference states, which allows the study of quantum operations without specifying fixed points; weak majorization, which allows comparing normalized and subnormalized states; catalytic majorization~\cite{gour2021entropy}, which accounts for using an additional state as a catalyst to enable an otherwise impossible transformation; and group majorization~\cite{giovagnoli_1985,giovagnoli_cyclic_1996}, which refines the group of stochastic operators to a subgroup, such as the symplectic group which is central in the study of Gaussian unitaries. For example, with weak and relative majorization, one could bound the rates of probabilistic distillation protocols.

Some families of quasiprobability distributions of interest, beyond the examples we consider here, are not in $L^1$, and thus not captured by our framework, so relaxing some of the technical assumptions in our main theorems may be desirable.
Allowing for $L^2$ instead of just $L^1$ functions would cover all Wigner functions directly, rather than the dense subset currently covered by our framework. Extending to more general distributions would enable applicability to more singular $s$-ordered quasiprobabilities. Furthermore, characterizations for complex-valued distributions would bring the Kirkwood--Dirac distribution into the fold.

Our results rule out the existence of a Wigner function that majorizes all others. Thus, the Wigner majorization preorder is rather different from the Husimi one, where the Lieb-Solovej-Wehrl theorem establishes that the vacuum majorizes all other states~\cite{Wehrl1979,Lieb1978,Lieb2014}. Nevertheless, we have shown that interesting majorization relations arise when comparing subsets of Wigner functions or when using relative majorization. Numerically, we observed that if $W_\rho \succ_{e^{-\beta \hat{n}}} W_\sigma$ then $W_\rho \succ_{e^{-\beta' \hat{n}}} W_\sigma$ for all $\beta' > \beta$. It would be interesting to prove this fact analytically.
Our framework offers a possible route to generalize an open conjecture formulated in Ref.~\cite{VanHerstraeten2023continuous}, namely that the Wigner function of the vacuum $W_0$ majorizes that of any Wigner-positive state. More generally, our results suggest that Wigner majorization, like Husimi majorization, may provide a way to refine uncertainty relations.

On the more theoretical side, we have found a large class of Wigner-positive channels that have SDS or S$q$S kernels. This raises the question of characterizing all the quantum channels with Wigner SDS or S$q$S kernels. Progress on this question would allow majorization to be used as not just a necessary but a sufficient condition for state conversions.

Expanding on the significant role of majorization on probabilities in quantum entanglement, quantum thermodynamics and quantum computing, our results provide a rigorous framework to characterize disorder in unbounded systems.

\section*{Acknowledgments}

T.U.\ thanks Artemy Kolchinsky, Matteo Lostaglio, and Nicole Yunger Halpern for helpful discussions and the organizers of FADEx 2024 for the travel opportunity facilitating this collaboration.
Z.V.H.\ gratefully acknowledges Nicolas Cerf, who supervised his PhD work, for numerous discussions and for his contributions to the early development of continuous majorization for quasiprobability distributions. Several conceptual ideas exploited in the present manuscript were shaped during that work, where the extension of continuous majorization accounting for both Lorenz curves was proposed. 
T.U.\ acknowledges the support of the Natural Sciences and Engineering Research Council of Canada (NSERC) through the Doctoral Postgraduate Scholarship.
U.C., Z.V.H.\ and J.D.\ acknowledge funding from the European Union's Horizon Europe Framework Programme (EIC Pathfinder Challenge project Veriqub) under Grant Agreement No.~101114899.
N.K.\ acknowledges support from the NSF QLCI grant OMA-2120757.
O.H.\ acknowledges support from CREST Grant Number JPMJCR23I3, Japan.

\section*{Author contributions}
T.U., Z.V.H., N.K.\ developed the proofs. T.U., Z.V.H., O.H. performed the numerical simulations. T.U., Z.V.H., J.D., O.H., U.C. conducted the Wigner function analysis. All authors contributed to the initial conceptualization of the project and the writing of the manuscript.

\section*{Competing interests}
The authors declare no competing interests. 

\section*{Data availability statement}
Data for all figures is available from the corresponding author on request.

\section*{Code availability statement}
The code used to generate plots of Lorenz curves has been made available as an open-source software package at \url{https://github.com/tweshu/major-qurve}.

\bibliography{mainbib.bib}

@article{PhysRevA.97.022341,
  title = {Generating grid states from {Schr\"odinger}-cat states without postselection},
  author = {Weigand, Daniel J. and Terhal, Barbara M.},
  journal = {Phys. Rev. A},
  volume = {97},
  issue = {2},
  pages = {022341},
  numpages = {13},
  year = {2018},
  month = {Feb},
  publisher = {American Physical Society},
  doi = {10.1103/PhysRevA.97.022341},
  url = {https://link.aps.org/doi/10.1103/PhysRevA.97.022341}
}

@book{Cohen_2013, 
    address={Basel}, 
    title={{The Weyl Operator and its Generalization}}, 
    rights={https://www.springernature.com/gp/researchers/text-and-data-mining}, 
    ISBN={978-3-0348-0293-2}, 
    url={https://link.springer.com/10.1007/978-3-0348-0294-9}, 
    DOI={10.1007/978-3-0348-0294-9}, 
    publisher={Springer Basel}, 
    author={Cohen, Leon}, 
    year={2013}
}

@book{Leonhardt_1997, address={Cambridge}, series={Cambridge studies in modern optics}, title={Measuring the quantum state of light}, ISBN={978-0-521-49730-5}, publisher={Cambridge Univ. Press}, author={Leonhardt, Ulf}, year={1997}, collection={Cambridge studies in modern optics}}

@article{Ferrie_Emerson_2009, 
    title={{Framed Hilbert space: hanging the quasi-probability pictures of quantum theory}}, 
    volume={11}, 
    ISSN={1367-2630}, 
    DOI={10.1088/1367-2630/11/6/063040}, 
    number={6}, 
    journal={New Journal of Physics}, 
    author={Ferrie, Christopher and Emerson, Joseph}, 
    year={2009}, 
    month=jun, 
    pages={063040} 
}

@article{husimi1940some,
  title={Some formal properties of the density matrix},
  author={Husimi, K{\^o}di},
  journal={Proceedings of the Physico-Mathematical Society of Japan. 3rd Series},
  volume={22},
  number={4},
  pages={264--314},
  year={1940},
  publisher={The Physical Society of Japan, The Mathematical Society of Japan},
  doi={https://doi.org/10.11429/ppmsj1919.22.4_264}
}

@article{cahill1969density,
	Author = {Cahill, Kevin E and Glauber, Roy J},
	Journal = {Physical Review},
	Number = {5},
	Pages = {1882},
	Publisher = {APS},
	Title = {Density operators and quasiprobability distributions},
	Volume = {177},
	Year = {1969},
	Doi = {10.1103/PhysRev.177.1882}
}

@article{Sudarshan_1963,
  title = {Equivalence of Semiclassical and Quantum Mechanical Descriptions of Statistical Light Beams},
  author = {Sudarshan, E. C. G.},
  journal = {Physical Review Letters},
  volume = {10},
  issue = {7},
  pages = {277--279},
  numpages = {0},
  year = {1963},
  month = {Apr},
  publisher = {American Physical Society},
  doi = {10.1103/PhysRevLett.10.277},
  url = {https://link.aps.org/doi/10.1103/PhysRevLett.10.277}
}

@article{Glauber_1963,
  title = {Coherent and Incoherent States of the Radiation Field},
  author = {Glauber, Roy J.},
  journal = {Phys. Rev.},
  volume = {131},
  issue = {6},
  pages = {2766--2788},
  numpages = {0},
  year = {1963},
  month = {Sep},
  publisher = {American Physical Society},
  doi = {10.1103/PhysRev.131.2766},
  url = {https://link.aps.org/doi/10.1103/PhysRev.131.2766}
}

@article{royer1977Wigner,
	Author = {Royer, Antoine},
	Journal = {Physical Review A},
	Number = {2},
	Pages = {449},
	Publisher = {APS},
	Title = {{W}igner function as the expectation value of a parity operator},
	Volume = {15},
	Year = {1977},
	Doi ={10.1103/PhysRevA.15.449}
}

@article{Grossmann_1976, 
    title={Parity operator and quantization of $\delta$-functions}, 
    volume={48}, 
    ISSN={0010-3616, 1432-0916}, 
    DOI={10.1007/BF01617867}, 
    number={3}, 
    journal={Communications in Mathematical Physics}, 
    author={Grossmann, A.}, 
    year={1976}, 
    month=oct, 
    pages={191–194}
}

@article{De_Gosson_De_Gosson_2021_Feichtinger, 
    title={On the Non-Uniqueness of Statistical Ensembles Defining a Density Operator and a Class of Mixed Quantum States with Integrable {Wigner} Distribution}, 
    volume={3}, 
    rights={https://creativecommons.org/licenses/by/4.0/}, 
    ISSN={2624-960X}, 
    DOI={10.3390/quantum3030031}, 
    number={3}, 
    journal={Quantum Reports}, 
    publisher={MDPI AG}, 
    author={De Gosson, Charlyne and De Gosson, Maurice}, 
    year={2021}, 
    month=aug, 
    pages={473–481}
}

@article{Dias_Prata_2019_Wigner_coordinate, 
    title={Quantum mappings acting by coordinate transformations on {Wigner} distributions}, 
    volume={35}, 
    ISSN={0213-2230, 2235-0616}, 
    DOI={10.4171/rmi/1056}, 
    number={2}, 
    journal={Revista Matemática Iberoamericana}, 
    publisher={European Mathematical Society - EMS - Publishing House GmbH}, 
    author={Dias, Nuno Costa and Prata, Joao Nuno},
    year={2019}, 
    month=feb, 
    pages={317–337} 
}

@article{ChitambarGourResourceReview2019,
  title = {Quantum resource theories},
  author = {Chitambar, Eric and Gour, Gilad},
  journal = {Reviews of Modern Physics},
  volume = {91},
  issue = {2},
  pages = {025001},
  numpages = {48},
  year = {2019},
  month = {Apr},
  publisher = {American Physical Society},
  doi = {10.1103/RevModPhys.91.025001},
  url = {https://link.aps.org/doi/10.1103/RevModPhys.91.025001}
}

@article{rundle2021overview,
  title={Overview of the phase space formulation of quantum mechanics with application to quantum technologies},
  author={Rundle, Russell P and Everitt, Mark J},
  journal={Advanced Quantum Technologies},
  volume={4},
  number={6},
  pages={2100016},
  year={2021},
  publisher={Wiley Online Library},
  doi={https://doi.org/10.1002/qute.202100016}
}

@article{Sabapathy_Guassian_dilatable_2017,
  title = {{Non-Gaussian} operations on bosonic modes of light: Photon-added {Gaussian} channels},
  author = {Sabapathy, Krishna Kumar and Winter, Andreas},
  journal = {Physical Review A},
  volume = {95},
  issue = {6},
  pages = {062309},
  numpages = {17},
  year = {2017},
  month = {Jun},
  publisher = {American Physical Society},
  doi = {10.1103/PhysRevA.95.062309},
  url = {https://link.aps.org/doi/10.1103/PhysRevA.95.062309}
}

@article{Lami_Sabapathy_Winter_2018, 
    title={All phase-space linear bosonic channels are approximately {Gaussian} dilatable}, 
    volume={20}, 
    DOI={10.1088/1367-2630/aae738}, 
    number={11}, 
    journal={New Journal of Physics}, 
    publisher={IOP Publishing}, 
    author={Lami, Ludovico and Sabapathy, Krishna Kumar and Winter, Andreas}, 
    year={2018}, 
    month=nov, 
    pages={113012}
}

@article{nielsen1999conditions,
  title = {Conditions for a Class of Entanglement Transformations},
  author = {Nielsen, M. A.},
  journal = {Physical Review Letters},
  volume = {83},
  issue = {2},
  pages = {436--439},
  numpages = {0},
  year = {1999},
  month = {Jul},
  publisher = {American Physical Society},
  doi = {10.1103/PhysRevLett.83.436},
  url = {https://link.aps.org/doi/10.1103/PhysRevLett.83.436}
}

@article{brandao2015second,
  title={The second laws of quantum thermodynamics},
  author={Brand\~ao, Fernando and Horodecki, Micha{\l} and Ng, Nelly and Oppenheim, Jonathan and Wehner, Stephanie},
  journal={Proceedings of the National Academy of Sciences},
  volume={112},
  number={11},
  pages={3275--3279},
  year={2015},
  publisher={National Academy of Sciences}
}

@article{veitch2014resource,
  title={The resource theory of stabilizer quantum computation},
  author={Veitch, Victor and Mousavian, SA Hamed and Gottesman, Daniel and Emerson, Joseph},
  journal={New Journal of Physics},
  volume={16},
  number={1},
  pages={013009},
  year={2014},
  publisher={IOP Publishing}
}

@article{gottesman2001encoding,
  title = {Encoding a qubit in an oscillator},
  author = {Gottesman, Daniel and Kitaev, Alexei and Preskill, John},
  journal = {Phys. Rev. A},
  volume = {64},
  issue = {1},
  pages = {012310},
  numpages = {21},
  year = {2001},
  month = {Jun},
  publisher = {American Physical Society},
  doi = {10.1103/PhysRevA.64.012310},
  url = {https://link.aps.org/doi/10.1103/PhysRevA.64.012310}
}

@article{aghaee2025scaling,
  title={Scaling and networking a modular photonic quantum computer},
  author={Aghaee Rad, H and Ainsworth, T and Alexander, RN and Altieri, B and Askarani, MF and Baby, R and Banchi, L and Baragiola, BQ and Bourassa, JE and Chadwick, RS and others},
  journal={Nature},
  volume={638},
  number={8052},
  pages={912--919},
  year={2025},
  publisher={Nature Publishing Group UK London},
  url={https://www.nature.com/articles/s41586-024-08406-9}
}

@article{grochowski2025optimal,
  title = {Optimal Phase-Insensitive Force Sensing with Non-Gaussian States},
  author = {Grochowski, Piotr T. and Filip, Radim},
  journal = {Phys. Rev. Lett.},
  volume = {135},
  issue = {23},
  pages = {230802},
  numpages = {11},
  year = {2025},
  month = {Dec},
  publisher = {American Physical Society},
  doi = {10.1103/7pyw-tgjd},
  url = {https://link.aps.org/doi/10.1103/7pyw-tgjd}
}

@article{dowling2008quantum,
  title={Quantum optical metrology--the lowdown on high-{N00N} states},
  author={Dowling, Jonathan P},
  journal={Contemporary physics},
  volume={49},
  number={2},
  pages={125--143},
  year={2008},
  publisher={Taylor \& Francis},
  url={https://www.tandfonline.com/doi/abs/10.1080/00107510802091298}
}

@article{nielsen2023deterministic,
  title = {Deterministic Quantum Phase Estimation beyond {N00N} States},
  author = {Nielsen, Jens A. H. and Neergaard-Nielsen, Jonas S. and Gehring, Tobias and Andersen, Ulrik L.},
  journal = {Phys. Rev. Lett.},
  volume = {130},
  issue = {12},
  pages = {123603},
  numpages = {6},
  year = {2023},
  month = {Mar},
  publisher = {American Physical Society},
  doi = {10.1103/PhysRevLett.130.123603},
  url = {https://link.aps.org/doi/10.1103/PhysRevLett.130.123603}
}

@article{slussarenko2017unconditional,
  title={Unconditional violation of the shot-noise limit in photonic quantum metrology},
  author={Slussarenko, Sergei and Weston, Morgan M and Chrzanowski, Helen M and Shalm, Lynden K and Verma, Varun B and Nam, Sae Woo and Pryde, Geoff J},
  journal={Nature Photonics},
  volume={11},
  number={11},
  pages={700--703},
  year={2017},
  publisher={Nature Publishing Group UK London},
  url={https://www.nature.com/articles/s41566-017-0011-5}
}

@article{you2021scalable,
    author = {You, Chenglong and Hong, Mingyuan and Bierhorst, Peter and Lita, Adriana E. and Glancy, Scott and Kolthammer, Steve and Knill, Emanuel and Nam, Sae Woo and Mirin, Richard P. and Magaña-Loaiza, Omar S. and Gerrits, Thomas},
    title = {Scalable multiphoton quantum metrology with neither pre- nor post-selected measurements},
    journal = {Applied Physics Reviews},
    volume = {8},
    number = {4},
    pages = {041406},
    year = {2021},
    month = {10},
    issn = {1931-9401},
    doi = {10.1063/5.0063294},
    url = {https://doi.org/10.1063/5.0063294},
}

@article{Braunstein_CVQI_2005,
  title = {Quantum information with continuous variables},
  author = {Braunstein, Samuel L. and van Loock, Peter},
  journal = {Reviews of Modern Physics},
  volume = {77},
  issue = {2},
  pages = {513--577},
  numpages = {0},
  year = {2005},
  month = {Jun},
  publisher = {American Physical Society},
  doi = {10.1103/RevModPhys.77.513},
  url = {https://link.aps.org/doi/10.1103/RevModPhys.77.513}
}

@article{Owari_CV_Nielsen_2008,
    author = {Owari, Masaki and Braunstein, Samuel L. and Nemoto, Kae and Murao, Mio},
    title = {$\varepsilon$-convertibility of entangled states and extension of {Schmidt} rank in infinite-dimensional systems},
    year = {2008},
    issue_date = {January 2008},
    publisher = {Rinton Press, Incorporated},
    volume = {8},
    number = {1},
    issn = {1533-7146},
    journal = {Quantum Information and Computation},
    month = jan,
    pages = {30–52},
    numpages = {23},
    url={https://dl.acm.org/doi/abs/10.5555/2011752.2011755}
}

@article{Torun_Majorization_QRT_Review_2023, 
    title={A compendious review of majorization-based resource theories: quantum information and quantum thermodynamics}, 
    volume={47}, 
    ISSN={1300-0101}, 
    DOI={10.55730/1300-0101.2744}, 
    number={4}, 
    journal={Turkish Journal of Physics}, 
    publisher={The Scientific and Technological Research Council of Turkey (TUBITAK-ULAKBIM) - DIGITAL COMMONS JOURNALS}, 
    author={Torun, G\"{o}khan and Pusuluk, Onur and Müstecaplio\v{g}lu, \"{O}zg\"{u}r Esat}, 
    year={2023}, 
    month=aug, 
    pages={141–182}
}

@phdthesis{koukoulekidis2023quasi,
  title={Quasi-probability representations of quantum computing},
  author={Koukoulekidis, Nikolaos},
  doi={10.25560/106874},
  url = {http://hdl.handle.net/10044/1/106874},
  school = {Imperial College London},
  month={Jan},
  year={2023}
}

@article{lostaglio2019introductory,
  title={An introductory review of the resource theory approach to thermodynamics},
  author={Lostaglio, Matteo},
  journal={Reports on Progress in Physics},
  volume={82},
  number={11},
  pages={114001},
  year={2019},
  publisher={IOP Publishing},
  doi={10.1088/1361-6633/ab46e5}
}

@article{giovagnoli_1985,
	title = {G-majorization with applications to matrix orderings},
	volume = {67},
	issn = {0024-3795},
	url = {https://www.sciencedirect.com/science/article/pii/0024379585901909},
	doi = {https://doi.org/10.1016/0024-3795(85)90190-9},
	pages = {111--135},
	journal = {Linear Algebra and its Applications},
	author = {Giovagnoli, A. and Wynn, H. P.},
	year = {1985}
}

@article{giovagnoli_cyclic_1996,
	title = {Cyclic majorization and smoothing operators},
	volume = {239},
	issn = {0024-3795},
	url = {https://www.sciencedirect.com/science/article/pii/S0024379596900130},
	doi = {10.1016/S0024-3795(96)90013-0},
	pages = {215--225},
	journal={Linear Algebra and its Applications},
	author = {Giovagnoli, Alessandra and Wynn, Henry P.},
	year = {1996}
}

@article{Blackwell1953,
    author = {David Blackwell},
    title = {{Equivalent Comparisons of Experiments}},
    volume = {24},
    journal = {The Annals of Mathematical Statistics},
    number = {2},
    publisher = {Institute of Mathematical Statistics},
    pages = {265--272},
    year = {1953},
    doi = {10.1214/aoms/1177729032},
    url = {https://doi.org/10.1214/aoms/1177729032}
}

@incollection{wang2017majorization,
  title={Majorization theory and applications},
  Editor = {Serpedin, E and Chen, T and Rajan, D},  
  author={Wang, Jiaheng and Palomar, Daniel},
  booktitle={Mathematical Foundations for Signal Processing, Communications, and Networking},
  pages={561--598},
  year={2017},
  publisher={CRC Press}
}

@article{lostaglio2015description,
  title = {Description of quantum coherence in thermodynamic processes requires constraints beyond free energy},
  author = {Lostaglio, Matteo and Jennings, David and Rudolph, Terry},
  journal = {Nature Communications},
  volume = {6},
  year = {2015},
  month = {Mar},
  pages = {6383},
  url = {https://doi.org/10.1038/ncomms7383}
}

@article{gour2018quantum,
	title = {Quantum majorization and a complete set of entropic conditions for quantum thermodynamics},
	volume = {9},
	issn = {2041-1723},
	url = {https://doi.org/10.1038/s41467-018-06261-7},
	doi = {10.1038/s41467-018-06261-7},
	pages = {5352},
	number = {1},
	journal = {Nature Communications},
	author = {Gour, Gilad and Jennings, David and Buscemi, Francesco and Duan, Runyao and Marvian, Iman},
	year = {2018}
}

@article{Schur1923uber,
  title = {{\"U}ber eine Klasse von Mittelbildungen mit Anwendungen auf die Determinantentheorie},
  author = {Schur, Issai},
  journal = {Berliner Mathematische Gesellschaft},
  volume = {22},
  pages = {9-20},
  year = {1923},
  publisher = {Berliner Mathematische Gesellschaft}
}

@article{Ostrowski1952sur,
  title = {Sur quelques applications des fonctions convexes et concaves au sens de I. Schur},
  author = {Ostrowski, Alexander M.},
  journal = {Journal de Mathématiques Pures et Appliquées},
  volume = {31},
  issue = {9},
  pages = {253-292},
  year = {1952},
  publisher = {Gauthier-Villars},
  url={https://www.numdam.org/item/?id=JMPA_1952_9_31__253_0}
}

@article{Veinott1971least,
author = {Veinott, Arthur F.},
title = {Least d-Majorized Network Flows with Inventory and Statistical Applications},
journal = {Management Science},
volume = {17},
number = {9},
pages = {547-567},
year = {1971},
doi = {10.1287/mnsc.17.9.547},
url = {https://doi.org/10.1287/mnsc.17.9.547}
}

@article{Horodecki2003reversible,
  title = {Reversible transformations from pure to mixed states and the unique measure of information},
  author = {Horodecki, Micha\l{} and Horodecki, Pawe\l{} and Oppenheim, Jonathan},
  journal = {Physical Review A},
  volume = {67},
  issue = {6},
  pages = {062104},
  numpages = {9},
  year = {2003},
  month = {Jun},
  publisher = {American Physical Society},
  doi = {10.1103/PhysRevA.67.062104},
  url = {https://link.aps.org/doi/10.1103/PhysRevA.67.062104}
}

@article{gour2015the,
  title = {The resource theory of informational nonequilibrium in thermodynamics},
  author = {Gour, Gilad and M{\"u}ller, Markus P. and Narasimhachar, Varun and Spekkens, Robert W. and Yunger Halpern, Nicole},
  journal = {Physics Reports},
  volume = {583},
  issn = {0370-1573},
  pages = {1--58},
  year = {2015},
  month = {Apr},
  url = {https://doi.org/10.1016/j.physrep.2015.04.003}
}

@article{horodecki2013fundamental,
	title = {Fundamental limitations for quantum and nanoscale thermodynamics},
	volume = {4},
	issn = {2041-1723},
	url = {https://doi.org/10.1038/ncomms3059},
	doi = {10.1038/ncomms3059},
	pages = {2059},
	number = {1},
	journal = {Nature Communications},
	shortjournal = {Nature Communications},
	author = {Horodecki, Michał and Oppenheim, Jonathan},
	year = {2013}
}

@article{alexander2023general,
  title = {General Entropic Constraints on {Calderbank-Shor-Steane} Codes within Magic Distillation Protocols},
  author = {Alexander, Rhea and Gvirtz-Chen, Si and Koukoulekidis, Nikolaos and Jennings, David},
  journal = {PRX Quantum},
  volume = {4},
  issue = {2},
  pages = {020359},
  numpages = {36},
  year = {2023},
  month = {Jun},
  publisher = {American Physical Society},
  doi = {10.1103/PRXQuantum.4.020359},
  url = {https://link.aps.org/doi/10.1103/PRXQuantum.4.020359}
}

@article{cwiklinski2015limitations,
  title = {Limitations on the Evolution of Quantum Coherences: Towards Fully Quantum Second Laws of Thermodynamics},
  author = {\ifmmode \acute{C}\else \'{C}\fi{}wikli\ifmmode \acute{n}\else \'{n}\fi{}ski, Piotr and Studzi\'{n}ski, Micha\l{} and Horodecki, Micha\l{} and Oppenheim, Jonathan},
  journal = {Physical Review Letters},
  volume = {115},
  issue = {21},
  pages = {210403},
  numpages = {5},
  year = {2015},
  month = {Nov},
  publisher = {American Physical Society},
  doi = {10.1103/PhysRevLett.115.210403},
  url = {https://link.aps.org/doi/10.1103/PhysRevLett.115.210403}
}

@article{Albarelli2018,
  title = {Resource theory of quantum non-Gaussianity and Wigner negativity},
  author = {Albarelli, Francesco and Genoni, Marco G. and Paris, Matteo G. A. and Ferraro, Alessandro},
  journal = {Physical Review A},
  volume = {98},
  issue = {5},
  pages = {052350},
  numpages = {17},
  year = {2018},
  month = {Nov},
  publisher = {American Physical Society},
  doi = {10.1103/PhysRevA.98.052350},
  url = {https://link.aps.org/doi/10.1103/PhysRevA.98.052350}
}

@article{Takagi2018,
  title = {Convex Resource Theory of Non-{{Gaussianity}}},
  author = {Takagi, Ryuji and Zhuang, Quntao},
  year = {2018},
  month = jun,
  journal = {Physical Review A},
  volume = {97},
  number = {6},
  pages = {062337},
  issn = {2469-9926, 2469-9934},
  doi = {10.1103/PhysRevA.97.062337}
}

@article{Upadhyaya2024,
  title = {Non-{{Abelian Transport Distinguishes Three Usually Equivalent Notions}} of {{Entropy Production}}},
  author = {Upadhyaya, Twesh and Braasch, William F. and Landi, Gabriel T. and Yunger Halpern, Nicole},
  year = {2024},
  month = sep,
  journal = {PRX Quantum},
  volume = {5},
  number = {3},
  publisher = {American Physical Society (APS)},
  issn = {2691-3399},
  url = {https://link.aps.org/doi/10.1103/PRXQuantum.5.030355},
  copyright = {https://creativecommons.org/licenses/by/4.0/}
}

@article{Markus1964,
  title = {{{The Eigen- and Singular Values of the Sum and Product of Linear Operators}}},
  author = {Markus, A S},
  year = {1964},
  month = aug,
  journal = {Russian Mathematical Surveys},
  volume = {19},
  number = {4},
  pages = {91--120},
  issn = {0036-0279, 1468-4829},
  doi = {10.1070/RM1964v019n04ABEH001154}
}

@article{Manjegani2023,
	Author = {Manjegani, Seyed Mahmoud and Moein, Shirin},
	Da = {2023/02/21},
	Doi = {10.1186/s13660-023-02935-z},
	Id = {Manjegani2023},
	Isbn = {1029-242X},
	Journal = {Journal of Inequalities and Applications},
	Number = {1},
	Pages = {27},
	Title = {Majorization and semidoubly stochastic operators on ${L}^1({X})$},
	Ty = {JOUR},
	Url = {https://doi.org/10.1186/s13660-023-02935-z},
	Volume = {2023},
	Year = {2023}}

@article{Chong1974, title={{Some Extensions of a Theorem of Hardy, Littlewood and Pólya and Their Applications}}, volume={26}, DOI={10.4153/CJM-1974-126-1}, number={6}, journal={Canadian Journal of Mathematics}, author={Chong, Kong-Ming}, year={1974}, pages={1321–1340}}

@article{ryff1963representation,
  title={On the representation of doubly stochastic operators},
  author={Ryff, John V},
  year={1963},
  journal={Pacific Journal of Mathematics},
  volume={13},
  number={4},
  pages={1379–1386},
  doi={http://dx.doi.org/10.2140/pjm.1963.13.1379}
}

@article{kenfack2004negativity,
	Author = {Kenfack, Anatole and {\.Z}yczkowski, Karol},
	Journal = {Journal of Optics B: Quantum and Semiclassical Optics},
	Number = {10},
	Pages = {396},
	Publisher = {IOP Publishing},
	Title = {Negativity of the {W}igner function as an indicator of non-classicality},
	Volume = {6},
	Year = {2004},
	Doi = {10.1088/1464-4266/6/10/003}
}

@article{Benedict1995,
  title = {Generalized Parity and Quasi-Probability Density Functions},
  author = {Benedict, M G and Czirjak, A},
  year = {1995},
  month = aug,
  journal = {Journal of Physics A: Mathematical and General},
  volume = {28},
  number = {16},
  pages = {4599--4608},
  publisher = {IOP Publishing},
  issn = {0305-4470, 1361-6447},
  doi = {10.1088/0305-4470/28/16/017}
}

@article{ruch1976principle,
  title={The principle of increasing mixing character and some of its consequences},
  author={Ruch, Ernst and Mead, Alden},
  journal={Theoretica chimica acta},
  volume={41},
  pages={95--117},
  year={1976},
  publisher={Springer},
  doi={https://doi.org/10.1007/BF01178071}
}

@article{fubini1907sugli,
 author = {Fubini, G.},
 title = {Sugli integrali multipli},
 fjournal = {Accademia dei Lincei, Rendiconti, V. Serie},
 journal = {Rom. Acc. L. Rend. (5)},
 issn = {0001-4435},
 volume = {16},
 number = {1},
 pages = {608--614},
 year = {1907},
 zbMATH = {2643959},
 JFM = {38.0343.02}
}

@article{Ruch1980,
title = {{Generalization of a theorem by Hardy, Littlewood, and Pólya}},
journal = {Journal of Mathematical Analysis and Applications},
volume = {76},
number = {1},
pages = {222-229},
year = {1980},
issn = {0022-247X},
doi = {https://doi.org/10.1016/0022-247X(80)90075-X},
url = {https://www.sciencedirect.com/science/article/pii/0022247X8090075X},
author = {Ernst Ruch and Rudolf Schranner and Thomas H Seligman}
}

@article{Joe1990,
title = {Majorization and divergence},
journal = {Journal of Mathematical Analysis and Applications},
volume = {148},
number = {2},
pages = {287-305},
year = {1990},
issn = {0022-247X},
doi = {https://doi.org/10.1016/0022-247X(90)90002-W},
url = {https://www.sciencedirect.com/science/article/pii/0022247X9090002W},
author = {Harry Joe},
}

@article{Joe1987,
  title = {Majorization, Randomness and Dependence for Multivariate Distributions},
  author = {Joe, Harry},
  year = {1987},
  journal = {The Annals of Probability},
  volume = {15},
  number = {3},
  number = {3},
 pages = {1217--1225},
 publisher = {Institute of Mathematical Statistics},
  url={https://www.jstor.org/stable/2244051}
}

@article{Bahrami2020,
	Author = {Bahrami, Farid and Manjegani, Seyed Mahmoud and Moein, Shirin},
	Da = {2021/03/01},
	Doi = {10.1007/s40840-020-00971-2},
	Id = {Bahrami2021},
	Isbn = {2180-4206},
	Journal = {Bulletin of the Malaysian Mathematical Sciences Society},
	Number = {2},
	Pages = {693--703},
	Title = {Semi-doubly Stochastic Operators and Majorization of Integrable Functions},
	Ty = {JOUR},
	Url = {https://doi.org/10.1007/s40840-020-00971-2},
	Volume = {44},
	Year = {2021}}

@article{VanHerstraeten2023continuous,
  doi = {10.22331/q-2023-05-24-1021},
  url = {https://doi.org/10.22331/q-2023-05-24-1021},
  title = {Continuous majorization in quantum phase space},
  author = {Van Herstraeten, Zacharie and Jabbour, Michael G. and Cerf, Nicolas J.},
  journal = {{Quantum}},
  issn = {2521-327X},
  publisher = {{Verein zur F{\"{o}}rderung des Open Access Publizierens in den Quantenwissenschaften}},
  volume = {7},
  pages = {1021},
  month = may,
  year = {2023}
}

@phdthesis{van2021majorization,
  title={Majorization theoretical approach to quantum uncertainty},
  author={Van Herstraeten, Zacharie},
  year={2021},
  school={Universit{\'e} libre de Bruxelles},
  url={http://quic.ulb.ac.be/_media/publications/2021_phd_thesis_zacharie_van_herstraeten.pdf}
}

@book{de_gosson_wigner_2017,
  title = {The {{Wigner Transform}}},
  author = {{de Gosson}, Maurice},
  year = {2017},
  month = may,
  publisher = {{World Scientific (Europe)}},
  doi = {10.1142/q0089},
  isbn = {978-1-78634-308-6 978-1-78634-310-9},
  langid = {english}
}

@article{koukoulekidis2022constraints,
    title = {Constraints on magic state protocols from the statistical mechanics of {Wigner} negativity},
    volume = {8},
    issn = {2056-6387},
    url = {https://doi.org/10.1038/s41534-022-00551-1},
    doi = {10.1038/s41534-022-00551-1},
    pages = {42},
    number = {1},
    journal = {npj Quantum Information},
    author = {Koukoulekidis, Nikolaos and Jennings, David},
    year = {2022}
}

@article{brif1999,
  title={Phase-space formulation of quantum mechanics and quantum-state reconstruction for physical systems with {Lie-group} symmetries},
  author={Brif, C and Mann, A},
  journal={Physical Review A},
  volume={59},
  number={2},
  pages={971},
  year={1999},
  publisher={APS}
}

@BOOK{marshall2011inequalities,
  title = {Inequalities: Theory of Majorization and Its Applications},
  author = {Marshall, Albert W. and Olkin, Ingram and Arnold, Barry C.},
  year = {2011},
  publisher = {Springer}
}

@BOOK{hardy1934inequalities,
  title = {Inequalities},
  author = {Hardy, G. H. and Littlewood, J. E. and P{\'o}lya, G.},
  year = {1934},
  publisher = {Cambridge University Press}
}

@article{de2024continuous,
  title = {Continuous majorization in quantum phase space for {Wigner-positive} states and proposals for {Wigner-negative} states},
  author = {de Boer, Jan and Di Giulio, Giuseppe and Keski-Vakkuri, Esko and Tonni, Erik},
  journal = {Phys. Rev. A},
  volume = {112},
  issue = {3},
  pages = {032405},
  numpages = {30},
  year = {2025},
  month = {Sep},
  publisher = {American Physical Society},
  doi = {10.1103/w561-h3z5},
  url = {https://link.aps.org/doi/10.1103/w561-h3z5}
}

@article{Wigner1932,
    title = {On the Quantum Correction For Thermodynamic Equilibrium},
    author = {{W}igner, E.},
    journal = {Physical Review},
    volume = {40},
    issue = {5},
    pages = {749--759},
    numpages = {0},
    year = {1932},
    month = {Jun},
    publisher = {American Physical Society},
    doi = {10.1103/PhysRev.40.749},
    url = {https://link.aps.org/doi/10.1103/PhysRev.40.749}
}

@ARTICLE{gour2021entropy,
  author={Gour, Gilad and Tomamichel, Marco},
  journal={IEEE Transactions on Information Theory}, 
  title={Entropy and Relative Entropy From Information-Theoretic Principles}, 
  year={2021},
  volume={67},
  number={10},
  pages={6313-6327},
  doi={10.1109/TIT.2021.3078337}}

@article{Pereira2015,
  title = {Extending a Characterization of Majorization to Infinite Dimensions},
  author = {Pereira, Rajesh and Plosker, Sarah},
  year = {2015},
  month = mar,
  journal = {Linear Algebra and its Applications},
  volume = {468},
  pages = {80--86},
  issn = {00243795},
  doi = {10.1016/j.laa.2014.01.026}
}

@article{dragoman2005applications,
  title={Applications of the Wigner distribution function in signal processing},
  author={Dragoman, Daniela},
  journal={EURASIP Journal on Advances in Signal Processing},
  volume={2005},
  pages={1--15},
  year={2005},
  publisher={Springer},
  doi={https://doi.org/10.1155/ASP.2005.1520}
}

@article{Day1973,
  title={Decreasing rearrangements and doubly stochastic operators},
  author={Day, Peter W},
  journal={Transactions of the American Mathematical Society},
  volume={178},
  pages={383--392},
  year={1973},
  url={https://www.jstor.org/stable/1996707}
}

@article{Rahimi-Keshari2016,
  title = {Sufficient {{Conditions}} for {{Efficient Classical Simulation}} of {{Quantum Optics}}},
  author = {{Rahimi-Keshari}, Saleh and Ralph, Timothy C. and Caves, Carlton M.},
  year = {2016},
  month = jun,
  journal = {Physical Review X},
  volume = {6},
  number = {2},
  pages = {021039},
  issn = {2160-3308},
  doi = {10.1103/PhysRevX.6.021039},
  copyright = {http://creativecommons.org/licenses/by/3.0/}
}

@book{Rudin1987,
  title = {Real and Complex Analysis},
  author = {Rudin, Walter},
  year = {1987},
  edition = {3rd ed},
  publisher = {McGraw-Hill},
  address = {New York},
  isbn = {978-0-07-054234-1},
  lccn = {QA300 .R82 1987}
}

@article{Lorenz1905, 
    title={Methods of Measuring the Concentration of Wealth}, 
    volume={9}, 
    ISSN={1522-5437}, 
    DOI={10.1080/15225437.1905.10503443}, 
    number={70}, 
    journal={Publications of the American Statistical Association}, 
    author={Lorenz, M. O.}, 
    year={1905}, 
    month={jun}, 
    pages={209–219}
}

@article{Dalton1920, 
    title={The Measurement of the Inequality of Incomes}, 
    volume={30}, 
    ISSN={00130133}, 
    DOI={10.2307/2223525}, 
    number={119}, 
    journal={The Economic Journal}, 
    author={Dalton, Hugh}, 
    year={1920}, 
    month={Sep},
    pages={348} 
}

@article{Muirhead1903,
  title = {Some {{Methods}} Applicable to {{Identities}} and {{Inequalities}} of {{Symmetric Algebraic Functions}} of {\emph{n}} {{Letters}}},
  author = {Muirhead, R. F.},
  year = {1903},
  journal = {Proceedings of the Edinburgh Mathematical Society},
  volume = {21},
  pages = {144--157},
  publisher = {Cambridge University Press (CUP)},
  issn = {0013-0915, 1464-3839},
  doi = {10.1017/s001309150003460x},
  copyright = {https://www.cambridge.org/core/terms}
}

@article{Joe1992,
  title = {Generalized {{Majorization Orderings}} and {{Applications}}},
  author = {Joe, Harry},
  year = {1992},
  journal = {Lecture Notes-Monograph Series},
  volume = {22},
  eprint = {4355739},
  eprinttype = {jstor},
  pages = {145--158},
  publisher = {Institute of Mathematical Statistics},
  issn = {07492170}
}

@article{Marian_2013,
  title = {Relative entropy is an exact measure of non-Gaussianity},
  author = {Marian, Paulina and Marian, Tudor A.},
  journal = {Physical Review A},
  volume = {88},
  issue = {1},
  pages = {012322},
  numpages = {6},
  year = {2013},
  month = {Jul},
  publisher = {American Physical Society},
  doi = {10.1103/PhysRevA.88.012322},
  url = {https://link.aps.org/doi/10.1103/PhysRevA.88.012322}
}

@article{Moein2019,
  title = {A Simplified and Unified Generalization of Some Majorization Results},
  author = {Moein, Shirin and Pereira, Rajesh and Plosker, Sarah},
  year = {2019},
  month = oct,
  journal = {Journal of Mathematical Analysis and Applications},
  volume = {478},
  number = {2},
  pages = {1049--1058},
  issn = {0022247X},
  doi = {10.1016/j.jmaa.2019.05.065}
}

@article{hahn2024classical,
  doi = {10.22331/q-2025-10-13-1881},
  url = {https://doi.org/10.22331/q-2025-10-13-1881},
  title = {Classical simulation and quantum resource theory of non-{G}aussian optics},
  author = {Hahn, Oliver and Takagi, Ryuji and Ferrini, Giulia and Yamasaki, Hayata},
  journal = {{Quantum}},
  issn = {2521-327X},
  publisher = {{Verein zur F{\"{o}}rderung des Open Access Publizierens in den Quantenwissenschaften}},
  volume = {9},
  pages = {1881},
  month = oct,
  year = {2025}
}

@BOOK{Hassani2013-ue,
  title     = "Mathematical physics: A modern introduction to its foundations",
  author    = "Hassani, Sadri",
  publisher = "Springer International Publishing",
  address   = "Cham, Switzerland",
  edition   =  2,
  month     =  jul,
  year      =  2013,
  doi       = "10.1007/978-3-319-01195-0",
  isbn      = "9783319011943,9783319011950",
}

@BOOK{Stromberg2015-au,
  title     = "An introduction to classical real analysis",
  author    = "Stromberg, Karl R",
  publisher = "American Mathematical Society",
  address   = "Providence, RI",
  series    = "AMS Chelsea Publishing",
  month     =  dec,
  year      =  2015,
  doi       = "10.1090/chel/376",
  isbn      = "9781470425449,9781470427252"
}

@misc{viaclovsky:mit-18.125-lec17,
  author       = {Jeff Viaclovsky},
  title        = {{Measure and Integration, Lecture 17}},
  howpublished = {MIT OpenCourseWare},
  year         = {2003},
  url          = {https://ocw.mit.edu/courses/18-125-measure-and-integration-fall-2003/6f21af6c40de1eccd70349bd3a3b0095_18125_lec17.pdf},
}

@article{PhysRevLett.124.063605,
  title = {Stellar Representation of Non-Gaussian Quantum States},
  author = {Chabaud, Ulysse and Markham, Damian and Grosshans, Fr\'ed\'eric},
  journal = {Physical Review Letters},
  volume = {124},
  issue = {6},
  pages = {063605},
  numpages = {6},
  year = {2020},
  month = {Feb},
  publisher = {American Physical Society},
  doi = {10.1103/PhysRevLett.124.063605},
  url = {https://link.aps.org/doi/10.1103/PhysRevLett.124.063605}
}

@article{hahn2025assessing,
  doi = {10.22331/q-2026-05-05-2095},
  url = {https://doi.org/10.22331/q-2026-05-05-2095},
  title = {Assessing non-{G}aussian quantum state conversion with the stellar rank},
  author = {Hahn, Oliver and Garnier, Maxime and Ferrini, Giulia and Ferraro, Alessandro and Chabaud, Ulysse},
  journal = {{Quantum}},
  issn = {2521-327X},
  publisher = {{Verein zur F{\"{o}}rderung des Open Access Publizierens in den Quantenwissenschaften}},
  volume = {10},
  pages = {2095},
  month = may,
  year = {2026}
}

@article{Genoni_2008,
  title = {Quantifying the non-Gaussian character of a quantum state by quantum relative entropy},
  author = {Genoni, Marco G. and Paris, Matteo G. A. and Banaszek, Konrad},
  journal = {Physical Review A},
  volume = {78},
  issue = {6},
  pages = {060303},
  numpages = {4},
  year = {2008},
  month = {Dec},
  publisher = {American Physical Society},
  doi = {10.1103/PhysRevA.78.060303},
  url = {https://link.aps.org/doi/10.1103/PhysRevA.78.060303}
}

@article{lloyd1999quantum,
  title = {Quantum Computation over Continuous Variables},
  author = {Lloyd, Seth and Braunstein, Samuel L.},
  journal = {Phys. Rev. Lett.},
  volume = {82},
  issue = {8},
  pages = {1784--1787},
  numpages = {0},
  year = {1999},
  month = {Feb},
  publisher = {American Physical Society},
  doi = {10.1103/PhysRevLett.82.1784},
  url = {https://link.aps.org/doi/10.1103/PhysRevLett.82.1784}
}

@article{putterman2025hardware,
  title={Hardware-efficient quantum error correction via concatenated bosonic qubits},
  author={Putterman, Harald and Noh, Kyungjoo and Hann, Connor T and MacCabe, Gregory S and Aghaeimeibodi, Shahriar and Patel, Rishi N and Lee, Menyoung and Jones, William M and Moradinejad, Hesam and Rodriguez, Roberto and others},
  journal={Nature},
  volume={638},
  number={8052},
  pages={927--934},
  year={2025},
  publisher={Nature Publishing Group UK London},
  url={https://www.nature.com/articles/s41586-025-08642-7}
}

@ARTICLE{Weedbrook2012-qu,
  title     = "Gaussian quantum information",
  author    = "Weedbrook, Christian and Pirandola, Stefano and García-Patrón,
               Raúl and Cerf, Nicolas J and Ralph, Timothy C and Shapiro,
               Jeffrey H and Lloyd, Seth",
  journal   = "Reviews of Modern Physics",
  publisher = "American Physical Society",
  volume    =  84,
  number    =  2,
  pages     = "621--669",
  month     =  may,
  year      =  2012,
  doi       = "10.1103/RevModPhys.84.621",
  issn      = "0034-6861"
}

@book{Walls2008,
  title = {Quantum {{Optics}}},
  author = {Walls, D.F and Milburn, Gerard J},
  year = {2008},
  edition = {2nd ed.},
  publisher = {Springer Berlin Heidelberg},
  address = {Berlin, Heidelberg},
  doi = {10.1007/978-3-540-28574-8},
  isbn = {3-540-28574-1}
}

@incollection{Cerf2007-chap2,
  author    = {Eisert, J. and Wolf, M. M.},
  title     = {Gaussian Quantum Channels},
  booktitle = {Quantum Information With Continuous Variables Of Atoms And Light},
  editor    = {Cerf, Nicolas J and Leuchs, Gerd and Polzik, Eugene S},
  publisher = {World Scientific},
  year      = {2007},
  chapter   = {2},
  pages     = {23--42} 
}

@ARTICLE{Caruso2008-dt,
  title     = "Multi-mode bosonic Gaussian channels",
  author    = "Caruso, F and Eisert, J and Giovannetti, V and Holevo, A S",
  journal   = "New J. Phys.",
  publisher = "IOP Publishing",
  volume    =  10,
  number    =  8,
  pages     =  083030,
  month     =  aug,
  year      =  2008,
  doi       = "10.1088/1367-2630/10/8/083030",
  issn      = "1367-2630",
}

@ARTICLE{Ivan2011-pm,
  title     = "Operator-sum representation for bosonic {Gaussian} channels",
  author    = "Ivan, J Solomon and Sabapathy, Krishna Kumar and Simon, R",
  journal   = "Physical Review A",
  publisher = "American Physical Society",
  volume    =  84,
  number    =  4,
  pages     =  042311,
  month     =  oct,
  year      =  2011,
  doi       = "10.1103/PhysRevA.84.042311",
  issn      = "1050-2947"
}

@article{Lieb1978,
  title = {Proof of an Entropy Conjecture of {{Wehrl}}},
  author = {Lieb, Elliott H},
  year = {1978},
  month = aug,
  journal = {Communications in Mathematical Physics},
  volume = {62},
  number = {1},
  pages = {35--41},
  issn = {0010-3616, 1432-0916},
  doi = {10.1007/BF01940328},
  copyright = {http://www.springer.com/tdm}
}

@article{Wehrl1979,
  title = {On the Relation between Classical and Quantum-Mechanical Entropy},
  author = {Wehrl, Alfred},
  year = {1979},
  month = dec,
  journal = {Reports on Mathematical Physics},
  volume = {16},
  number = {3},
  pages = {353--358},
  issn = {00344877},
  doi = {10.1016/0034-4877(79)90070-3},
  copyright = {https://www.elsevier.com/tdm/userlicense/1.0/}
}

@article{Lieb2014,
  title = {Proof of an Entropy Conjecture for {{Bloch}} Coherent Spin States and Its Generalizations},
  author = {Lieb, Elliott H. and Solovej, Jan Philip},
  year = 2014,
  journal = {Acta Mathematica},
  volume = {212},
  number = {2},
  pages = {379--398},
  issn = {0001-5962},
  doi = {10.1007/s11511-014-0113-6},
}

\end{document}


\title{Supplementary Information --- Majorization theory for quasiprobabilities}

\author{Twesh Upadhyaya}
\email{tweshu@umd.edu}
\affiliation{Joint Center for Quantum Information and Computer Science, University of Maryland and NIST, College Park, MD 20742, USA}
\affiliation{Department of Physics, University of Maryland, College Park, MD 20742, USA}

\author{Zacharie Van Herstraeten}
\affiliation{QAT team, DIENS, \'Ecole Normale Sup\'erieure, PSL University, CNRS, INRIA, 45 rue d'Ulm, Paris 75005, France}

\author{Jack Davis}
\affiliation{QAT team, DIENS, \'Ecole Normale Sup\'erieure, PSL University, CNRS, INRIA, 45 rue d'Ulm, Paris 75005, France}

\author{Oliver Hahn}
\affiliation{Department of Basic Science, The University of Tokyo, 3-8-1 Komaba, Meguro-ku, Tokyo 153-8902, Japan}

\author{Nikolaos Koukoulekidis}
\affiliation{Duke Quantum Center and Departments of Physics and Electrical and Computer Engineering, Duke University, Durham, NC 27708, USA}

\author{Ulysse Chabaud}
\affiliation{QAT team, DIENS, \'Ecole Normale Sup\'erieure, PSL University, CNRS, INRIA, 45 rue d'Ulm, Paris 75005, France}

\begin{abstract}
    This Supplementary Information contains additional background, detailed statements, and proofs for the results stated in the main text.
\end{abstract}

\maketitle


\section*{Supplementary Note 1: Equivalent notions of majorization}
\label{app:proofs}

Here, we prove Theorem~1 and Theorem~2 on equivalent characterizations of (relative) majorization, explicitly restate the results for the counting measure, and discuss properties of the majorization preorder.

\subsection{Preliminary lemmas}

We first provide four lemmas concerning known results for majorization on probability distributions that we utilize in the proof of Theorem~1, as well as an additional technical lemma which allows us to work on nonatomic measure spaces without loss of generality.
The first two majorization lemmas are simple restatements of the results, whereas the last two are generalized slightly to fit the needs of our main proof. For a real number $u$, $u^-$ ($u^+$) indicate approaching $u$ from the left (right) and should not be confused with the notation for positive and negative parts of a function.
Given a subset $A \subseteq X$ of a $\sigma$-finite measure space $X$, we denote by $A^\perp$ the complement of $A$ in $X$.

The first lemma establishes an equivalence between majorizing Lorenz curves and doubly stochastic operators for probability distributions on finite measure spaces. The original statement of the lemma relates to the Lorenz curve $L^{\uparrow}$ which, for finite measure spaces, is equivalent to considering both the increasing and decreasing Lorenz curves $L^{\Dsh}, L^{\Ish}$,
in the sense that $\smash{L^{\uparrow}_f(s)\geq L^{\uparrow}_g(s)\Longleftrightarrow L^{\downarrow}_f(s)\leq L^{\downarrow}_g(s)\Longleftrightarrow L^{\Dsh}_f(s)\geq L^{\Dsh}_g(s)\ \text{and}\ L^{\Ish}_f(s)\leq L^{\Ish}_g(s)}$.
In our restatement of the lemma, we add this equivalent formulation. 
\begin{lemma}[Restatement of Theorem 4.9 in~\cite{Day1973}]\label{daylemma}
    Let $(X,\mu)$ be a finite measure space.
    For $f,g \in L^1(X,\mu)$, $L_f^\uparrow(s) \geq L_g^\uparrow(s) $, or, equivalently, $L^{\Dsh}_f(s) \geq L^{\Dsh}_g(s)$ and $L^{\Ish}_f(s) \leq L^{\Ish}_g(s)$, for all $s \in [0,\mu(X)]$ and $\int_X f \ \mathrm d\mu = \int_X g \ \mathrm d\mu$ if and only if there exists a doubly stochastic operator $S: L^1(X,\mu) \rightarrow L^1(X,\mu)$ such that $Sf = g$.
\end{lemma}

The second lemma concerns partitions of an operator over disjoint subsets that cover the measure space.
\begin{lemma}[Restatement of Fact 5 in~\cite{Bahrami2020}]\label{bahramilemma}
    Let $X$ be a $\sigma$-finite measure space and $A \subseteq X$ a measurable subset of $X$. 
    Given two bounded operators $T_A: L^1(A) \rightarrow L^1(A)$, $T_{A^\perp}: L^1({A^\perp}) \rightarrow L^1({A^\perp})$, the operator $T$ defined by the action
    \begin{equation}
        T(f) = T_A(f \chi_A) \chi_A + T_{A^\perp}(f \chi_{A^\perp}) \chi_{A^\perp} \,,
    \end{equation}
    for all $f \in L^1(X)$, is also a bounded operator on $L^1(X)$.
    Moreover, if $T_A, T_{A^\perp}$ are (semi)doubly stochastic, then $T$ is also (semi)doubly stochastic.
\end{lemma}

The third lemma connects the positive and negative parts of distributions to their distribution and codistribution functions, respectively.
\begin{lemma}[Generalization of Corollaries 1.2 and 1.3 in~\cite{Chong1974}]\label{chonglemma1}
    Given a measurable function $f$ on a $\sigma$-finite measure space $X$ with measure $\mu$, and any real number $z \ge 0$, we have
    \begin{align}
    \int_X (f-z)^+ \ \mathrm d\mu&= \int_z^\infty D_f(t) \ \mathrm dt \,; \\
    \int_X (f+z)^- \ \mathrm d\mu &= - \int_{-\infty}^{-z} C_f(t) \ \mathrm dt \,.
    \end{align}
\end{lemma}
\begin{proof}
We take a different proof approach than~\cite{Chong1974}.
The first equation is derived as follows,
\begin{align}
    \int_X (f-z)^+ \ \mathrm d\mu 
    &= \int_X \max\{ f-z,0 \} \ \mathrm d\mu \\
    &=\int_X \int_0^\infty \chi(x,t)_{t<\max\{f(x)-z,0\}} \ \mathrm dt \ \mathrm d\mu(x)\\
    &=\int_0^\infty \int_X  \chi(x,t)_{t<\max\{ f(x)-z,0 \}} \ \mathrm d\mu(x) \ \mathrm dt \\
    &=\int_0^\infty \mu\{x: f(x)-z>t \} \ \mathrm dt \\
    &=\int_z^\infty \mu\{ f>t \} \ \mathrm dt = \int_z^\infty D_f(t) \ \mathrm dt, 
\end{align}
where in the second line we rewrite the integral with an indicator function using the layer cake representation and in the third line we exchange integrals by Fubini's theorem~\cite{fubini1907sugli,Rudin1987}. 
Similarly,
\begin{align}
    \int_X (f+z)^- \ \mathrm d\mu &= \int_X \min\{ f+z,0 \} \ \mathrm d\mu \\
    &= -\int_X \int_{-\infty}^0 \chi(x,t)_{t>\min\{ f(x)+z,0 \}} \ \mathrm dt \ \mathrm d\mu(x)\\
    &= -\int_{-\infty}^0 \int_X \chi(x,t)_{t>\min\{ f(x)+z,0 \}} \ \mathrm d\mu(x) \ \mathrm dt \\
    &= -\int_{-\infty}^0 \mu\{ f+z<t \} \ \mathrm dt \\
    &= -\int_{-\infty}^{-z} \mu\{ f<t \} \ \mathrm dt = - \int_{-\infty}^{-z} C_f(t) \ \mathrm dt.
\end{align}
\end{proof}

The fourth lemma provides tight bounds on shifted decreasing and increasing rearrangements of a distribution in terms of its distribution and codistribution functions.
\begin{lemma}[Generalization of Lemma 1.5 in~\cite{Chong1974}]\label{chonglemma2}
Given a measurable function $f$ on a $\sigma$-finite measure space $X$ with measure $\mu$,
\begin{align}
    \int_0^s (f^\downarrow-z) \ \mathrm d u &\leq \int_z^\infty D_f(t) \ \mathrm d t \, ;\\
    \int_0^s (f^\uparrow+z) \ \mathrm d u &\geq -\int_{-\infty}^{-z} C_f(t) \ \mathrm d t \, ,
\end{align}
for any $z\geq 0$, $s\in [0,\mu(X))$. 
The first inequality holds as equality if $D_f(z) \leq s \leq D_f(z^-)$ and the second inequality holds as equality if $C_f(-z) \leq s \leq C_f(-z^+)$.
\end{lemma}
\begin{proof}
Our proof closely follows~\cite{Chong1974}.
We have
\begin{align}
    \int_0^s (f^\downarrow -z) \ \mathrm d u &\leq \int_0^s (f^\downarrow-z)^+ \ \mathrm d u\\
    &\leq \int_0^{\mu(X)} {(f-z)^+}^\downarrow \ \mathrm d u\\
    &= \int_X (f-z)^+ \ \mathrm d \mu \\
    &= \int_z^\infty D_f(t) \ \mathrm dt \,,
\end{align}
where the second inequality follows because $s \le \mu(X)$, $f^\downarrow-z = (f-z)^\downarrow$ and $((\cdot)^+)^\downarrow = ((\cdot)^\downarrow)^+$, and the last equality follows from Lemma~\ref{chonglemma1}.

Now assume $ D_f(z) \leq s \leq D_f(z^-)$. Recall that $D_f(z)=\mu\{x: f(x)> z \}$ so $D_f(z^-)=\mu\{x: f(x)\geq z \}$. 
\begin{align}
    \int_0^s (f^\downarrow-z) \ \mathrm d u &= \int_{[0,s)} (f^\downarrow-z) \ \mathrm d u\\
    &=\int_{f^\downarrow  \geq z} (f^\downarrow-z) \ \mathrm d u\\
    &=\int_0^{\mu(X)} (f^\downarrow-z)^+ \ \mathrm d u\\
    &=\int_z^\infty D_{f^\downarrow}(t) \ \mathrm d t\\
    &=\int_z^\infty D_{f}(t) \ \mathrm d t \,,
\end{align}
where the second line follows from the condition on $s$, 
as $f^\downarrow-z$ is nonnegative on that interval, the fourth line follows from Lemma \ref{chonglemma1} and the fifth line follows since $f$ and $f^\downarrow$ have the same distribution function.

Analogously,
\begin{align}
    \int_0^s (f^\uparrow+z) \ \mathrm d u&\geq \int_0^s (f^\uparrow+z)^- \ \mathrm du\\
    &\geq \int_0^{\mu(X)} {(f+z)^-}^\uparrow \ \mathrm du\\
    &= \int_X (f+z)^- \ \mathrm d\mu \\
    &= -\int_{-\infty}^{-z} C_f(t) \ \mathrm dt \,.
\end{align}
Assuming that $C_f(-z) \leq s \leq C_f(-z^+)$, we obtain the analogous bound,
\begin{align}
\int_0^s (f^\uparrow+z) \mathrm du &= \int_{[0,s)} (f^\uparrow+z) \ \mathrm du\\
&=\int_{f^\uparrow \leq -z} (f^\uparrow+z) \ \mathrm du\\
&=\int_0^{\mu(X)} (f^\uparrow+z)^- \ \mathrm du\\
&=-\int_{-\infty}^{-z} C_{f^\uparrow}(t) \ \mathrm dt\\
&=-\int_{-\infty}^{-z} C_{f}(t) \ \mathrm dt \,.
\end{align}

\end{proof}

In the proof of the main theorem, working on a nonatomic measure space will be important for our construction of truncated Lorenz curves. However, we may want to consider quasiprobabilities over measure spaces that are not nonatomic, for example for discrete-variable systems. In the next lemma, we show that we can always extend to a nonatomic space. 
For any $\sigma$-finite $(X,\mu)$, define the product space $\bar{X}=X \cross [0,1]$ with the product measure $\bar{\mu}=\mu \cross u$, where $u$ denotes the Lebesgue measure on the unit interval. This new measure space is guaranteed to be nonatomic for any $(X,\mu)$. 
\begin{lemma}\label{atomiclift}
      For any $f\in L^1(X,\mu)$, define its extension $\bar{f}$ to the product space via $\bar{f}(x_1,x_2)=f(x_1)$. $\bar{f}$ is in $L^1(\bar{X},\bar{\mu})$ and $D_{\bar{f}}=D_f, C_{\bar{f}}=C_f$. Further, for $f,g\in L^1(X,\mu)$, if there is a sequence of semidoubly stochastic operators $\bar{S}_n: L^1(\bar{X},\bar{\mu}) \rightarrow L^1(\bar{X},\bar{\mu})$ such that $\bar{S}_n \bar{f}$ converges to $\bar{g}$ in $L^1(\bar{X},\bar{\mu})$, then there is a sequence of semidoubly stochastic operators $S_n: L^1(X,\mu) \rightarrow L^1(X,\mu)$ such that $S_n f$ converges to $g$ in $L^1(X,\mu)$.
\end{lemma}

\begin{proof}
We have that 
\begin{align}
    \int_{\bar{X}} \bar{f} \mathrm{d} \bar{\mu} = \int_X f  \mathrm{d} \mu
\end{align}
We obtain immediately that 
\begin{align}
    D_{\bar{f}}(t)&=\bar{\mu} \{ (x_1,x_2): \bar{f}(x_1,x_2)>t \}\\
    &=\bar{\mu} \{ (x_1,x_2): f(x_1)>t \}\\
    &=\mu \{ x_1: f(x_1)>t \}\\
    &=D_f(t),
\end{align}
as $u([0,1])=1$. An analogous argument shows $C_{\bar{f}}(t)=C_{f}(t)$. In particular, note this implies the positive and negative Lorenz curves of $\bar{f}$ are identical to those of $f$.

Throughout the rest of the proof we reorder integrals freely by Fubini's theorem~\cite{fubini1907sugli,Rudin1987}.
Suppose there is a sequence of semidoubly stochastic operators $\bar{S}_n$ taking $\bar{f}$ to $\bar{g}$,
\begin{equation}
 \int_{[0,1]} \int_X \bar{S}_n(x_1,x_2,y_1,y_2) \bar{f}(y_1,y_2) \mathrm{d} \mu(y_1) \mathrm d u(y_2) \rightarrow \bar{g}(x_1,x_2),
\end{equation}
where the convergence is in $L^1(\bar{X},\bar{\mu})$.
Define the reduced operators
\begin{equation}
 S_n(x_1,y_1)=\int_{[0,1]} \int_{[0,1]} \bar{S}_n(x_1,x_2,y_1,y_2) \mathrm{d} u(x_2) \mathrm{d} u(y_2).
\end{equation}
For any function $h\in L^1(X,\mu)$, 
\begin{align}
    S_n h &= \int_X S_n(x_1,y_1) h(y_1) \mathrm{d} \mu(y_1)\\
    &= \int_X \int_{[0,1]} \int_{[0,1]} \bar{S}_n(x_1,x_2,y_1,y_2) \mathrm{d} u(x_2) \mathrm{d} u(y_2) h(y_1) \mathrm{d} \mu(y_1)\\
    &= \int_{[0,1]} \int_X  \int_{[0,1]} \bar{S}_n(x_1,x_2,y_1,y_2)   \bar{h}(y_1,y_2) \mathrm{d} u(y_2) \mathrm{d} \mu(y_1) \mathrm{d} u(x_2) \\
    &= \int_{[0,1]} (\bar{S}_n \bar{h})(x_1,x_2) \mathrm{d} u(x_2).
\end{align}
By a similar argument as above, $\bar{h} \in L^1(\bar{X},\bar{\mu})$, so by assumption $\bar{S}_n \bar{h} \in L^1(\bar{X},\bar{\mu})$. That is, the integral\\ $\int_X \int_{[0,1]} \abs{(\bar{S}_n \bar{h})(x_1,x_2)} \mathrm{d} u(x_2) \mathrm{d} \mu(x_1) $ converges. But this implies that  $\int_{[0,1]} (\bar{S}_n \bar{h})(x_1,x_2) \mathrm{d} u(x_2) \in L^1(X,\mu)$. Thus, we have shown that $S_n$ is an operator from $L^1(X,\mu)$ to $L^1(X,\mu)$.

That $S_n$ is positive is immediate from the positivity of $\bar{S}_n$. 
$\bar{S}_n$ is stochastic so 
\begin{equation}
     \int_X \int_{[0,1]} \bar{S}_n(x_1,x_2,y_1,y_2) \mathrm{d} \mu(x_1) \mathrm{d} u(x_2)=1,
\end{equation}
for almost all $(y_1,y_2) \in \bar{X}$. Let $A$ be the measure zero set of points $(y_1,y_2) \in\bar{X}$ where this equation does not hold.
$S_n$ satisfies
\begin{align}
    \int_X S_n(x_1,y_1) \mathrm{d} \mu(x_1) &= \int_X \int_{[0,1]} \int_{[0,1]} \bar{S}_n(x_1,x_2,y_1,y_2) \mathrm{d} u(x_2) \mathrm{d} u(y_2) \mathrm{d} \mu(x_1)\\
    &= \int_{[0,1]} 1 \ \mathrm{d} u(y_2), 
\end{align}
where the integrand in the second line is to be understood in the sense of equality almost everywhere (see above).
To show the stochasticity of $S_n$, we need to show that this expression is equal to 1 for almost all $y_1\in X$. For each $y_1$, define $A'(y_1)=\{y_2\in[0,1]: (y_1,y_2)\in A \}$. We have that 
\begin{align}
    0&=\bar{\mu}(A)\\
    &=\int_X u(A'(y_1)) \mathrm{d} \mu(y_1).
\end{align}
Since measures are nonnegative, this implies the set $\{y_1: u(A'(y_1)) >0 \}$ has measure zero. But if $u(A'(y_1)) =0$, then $\int_{[0,1]} 1 \ \mathrm{d} u(y_2)=1$. Therefore, $\int_X S_n(x_1,y_1) \mathrm{d} \mu(x_1) =1$ for almost all $y_1\in X$ so $S_n$ is stochastic.

Similarly, we can check that 
\begin{align}
    \int_X S_n(x_1,y_1) \mathrm{d} \mu(y_1) \leq 1,
\end{align}
for almost all $x_1 \in X$, so $S_n$ is semidoubly stochastic.

Finally, we have that 
\begin{align}
    (S_n f)(x_1) &= \int_X S_n(x_1,y_1) f(y_1) \mathrm{d} \mu(y_1)\\
    &= \int_{[0,1]} \int_X  \int_{[0,1]} \bar{S}_n(x_1,x_2,y_1,y_2)   \bar{f}(y_1,y_2) \mathrm{d} u(y_2) \mathrm{d} \mu(y_1) \mathrm{d} u(x_2) \\
    &= \int_{[0,1]} \bar{g}(x_1,x_2)+h_n(x_1,x_2) \mathrm{d} u(x_2) \\
    &= g(x_1) + \int_{[0,1]} h_n(x_1,x_2) \mathrm{d} u(x_2) 
\end{align}
where $\int_{\bar{X}} \abs{h_n(x_1,x_2)}\rightarrow 0$ by assumption so $S_n f$ converges to $g$ in $L^1(X,\mu)$.
\end{proof}

We now move to the proof of our main theorem.

\subsection{Proof of Theorem~1}\label{app:thm1proof}

\begin{proof}

We prove each implication sequentially.

\noindent$\bm{1 \implies 2:}$ Without loss of generality, take $f$ and $g$ to be normalized to $1$.
Observe that $NV(f) \geq NV(g)$, where $NV(\cdot)= \frac{1}{2} (\int_X \abs{\cdot} \mathrm d\mu -1)$ is the negative volume.
To see this, note that $NV(f) < NV(g)$ leads to $L^{\Dsh}_f(s) < L^{\Dsh}_g(s)$ for sufficiently large $s$, which contradicts the assumption. 

Consider first the case where $NV(f) = NV(g)$.
Our strategy is to build approximations to $f$ and $g$ that are nonzero only on finite-measure domains. The approximations are built by zeroing out the function values of smallest magnitude (the tails of the functions) to retain some of the structure of the Lorenz curves.

The positive Lorenz curve is continuous, so define $s_\epsilon$ such that $\declorarg{f}(s_\epsilon)=1+NV(f)-\epsilon$ for small $\epsilon > 0$. The distribution function $D_{f^+}$ is right-continuous, therefore for small $\epsilon > 0$ exactly one of the following two conditions is met:
\begin{enumerate}
    \item there exists $t_\epsilon$ such that $D_{f^+}(t_\epsilon) = s_\epsilon$.
    Set $A' = \emptyset$~;
    \item there exists $t_\epsilon$ such that $D_{f^+}(t_\epsilon) < s_\epsilon \leq D_{f^+}(t_\epsilon^-)$. By Lemma~\ref{atomiclift}, we can assume the measure space is nonatomic. Thus, we can pick a subset $A' \subseteq \{x:f(x)=t_\epsilon\}$ that satisfies $\mu(A') = s_\epsilon - D_{f^+}(t_\epsilon) > 0$.
\end{enumerate}
Define the domain $A_{+,\epsilon} \coloneqq \{x: f^+(x) > t_\epsilon\} \cup A'$ which by construction has finite measure $s_\epsilon$. 
The Lorenz curve of $f^+$ restricted to this domain is truncated and matches that of $f^\plus$ up to $1+NV(f)-\epsilon$.
A similar argument using the codistribution $C_{f^-}$ gives a domain $A_{-,\epsilon}$ of finite measure, such that $f^-$ restricted to this domain has a negative Lorenz curve matching $f^-$ up to $-NV(f)+\epsilon$.

Define $\Af \coloneqq A_{+,\epsilon} \cup A_{-,\epsilon}$, and the quasiprobability distribution restricted to this domain,
\begin{equation}
    \fee(x) \coloneqq f(x)\chi_{ A_{f,\epsilon}}(x) \,.
    \end{equation}
Repeat the above construction for $g$, to obtain domain $\Ag$ and quasiprobability distribution $\gee(x) \coloneqq g(x)\chi_{\Ag}(x)$, analogously.

Formally, we consider $\fee,\gee$ as functions restricted to the domain $E \coloneqq \Af \cup \Ag $. 
Note that $\mu(E)<\infty$. 
By construction, $\fee,\gee$ are $L^1$-integrable and normalized, as equal contributions of positive and negative parts are zeroed out in going from $f,g$ to $\fee,\gee$.
By assumption, $L^{\Dsh}_{\fee}(s) \geq L^{\Dsh}_{\gee}(s)$ and $L^{\Ish}_{\fee}(s) \leq L^{\Ish}_{\gee}(s)$ for all $s\in [0,\mu(E)]$.
Hence, by Lemma~\ref{daylemma}, there exists a doubly stochastic operator $T: L^1(E) \rightarrow L^1(E)$ satisfying $T\fee = \gee$. Additionally, Lemma~\ref{bahramilemma} implies that the operator $S$ defined by the action
\begin{equation}
    S(f) \coloneqq T(f \chi_E) \chi_E + R(f \chi_{E^\perp}) \chi_{E^\perp} \,,
\end{equation}
is doubly stochastic for any arbitrary choice of doubly stochastic $R:L^1(E^\perp) \rightarrow L^1(E^\perp)$. 

Consider a sequence of such operators $(S_n)_{n \in \mathbb{N}}$ for $\epsilon_n=2^{-n-2}$ and set $f_{\epsilon_n}^* \coloneqq f - f_{\epsilon_n}$ and analogously for $g$. Note that $\norm{f_{\epsilon_n}^*}_1\leq 2\epsilon_n$ and $\norm{g_{\epsilon_n}^*}_1\leq 2\epsilon_n$.
Then, we obtain
\begin{align}
    \norm{S_n f - g}_1 &= \norm{S_n(f_{\epsilon_n}+f_{\epsilon_n}^*) - g}_1\\
    &= \norm{T_n f_{\epsilon_n} + S_n f_{\epsilon_n}^* - (g_{\epsilon_n} + g_{\epsilon_n}^*)}_1\\
    &= \norm{S_n f_{\epsilon_n}^* - g_{\epsilon_n}^*}_1\\
    &\le \norm{S_n f_{\epsilon_n}^*}_1 + \norm{ g_{\epsilon_n}^*}_1\\
    &\leq \norm{f_{\epsilon_n}^*}_1 + \norm{ g_{\epsilon_n}^*}_1\\
    &\le 2^{-n}.
\end{align}
Therefore, as $n\rightarrow +\infty$, $S_n f \rightarrow g$ in $L^1(X,\mu)$.
\vspace{10pt}

Now consider the case when $NV(f)>NV(g)$. 
Our proof strategy is to build arbitrarily good approximations to an intermediate function $f_{\text{red}}$, with positive (negative) Lorenz curve lying above (below) the positive (negative) Lorenz curve of $g$ and having the same negative volume as $g$, as depicted in Figure~1 of the main text for the positive Lorenz curve. 
Then, the previous step of the proof implies that there exists a sequence of doubly stochastic operators $(S_n)_{n \in \mathbb{N}}$ such that $S_n f_{\text{red}} \rightarrow g$ in $L^1$. 
The composition of semidoubly stochastic operators is itself semidoubly stochastic~\cite{Manjegani2023}, so it remains to show that there exists a sequence of semidoubly stochastic operators $(T_m)_{m \in \mathbb{N}}$ such that $T_m f \rightarrow f_{\text{red}}$ in $L^1(X)$.

Let $s_+$ and $s_-$ be the points at which $L^{\Dsh}_f(s_+) = 1+NV[g]$, and $L^{\Ish}_f(s_-) = -NV[g]$, respectively. 
Following the above construction for obtaining the distribution $\fee$, we can define a domain $A$ such that $f_{\text{red}}(x)=f(x) \chi_{A}$ satisfies $L^{\Dsh}_{f_{\text{red}}}(s)=L^{\Dsh}_f(s)$ for all $s \in [0,s_+]$ and $L^{\Ish}_{f_{\text{red}}}(s)=L^{\Ish}_f(s)$ for all $s \in [0,s_-]$. As before, $\mu(A)<\infty$.
By construction, $f_{\text{red}}$ is normalized, as an equal amount of positive and negative contributions are set to 0. 
We now show that $f_{\text{red}}$ can be reached from $f$ via a sequence of semidoubly stochastic operators.
Let $B_n \subset A^\perp$ be a sequence of subsets of increasing but finite measure, such that $B_n \subset B_{n+1}$ and every point in $A^\perp$ is eventually included in some $B_n$.
Using Lemma~\ref{bahramilemma}, we define a sequence of semidoubly stochastic operators acting as identity on $A \cup (A^\perp-B_n)=X-B_n=B_n^\perp$ and via the kernel $T'_n(x,y) = \frac{1}{\mu(B_n)}$ on $B_n$,
\begin{equation}
    T_n f = \id(f \chi_{B_n^\perp }) + T'_n(f \chi_{B_n})\chi_{B_n} \,.
\end{equation}
The operator $T_n$ preserves the large absolute values of $f$ while smearing out the small positive and negative values of $f$ in the limit $n\rightarrow\infty$. 
Because $f$ is integrable, it is possible to choose the sequence $B_n$ such that $T_n f \rightarrow f_{\text{red}}$ in $L^1$. 

\noindent$\bm{2\implies 3:}$ 
By stochasticity, $\int_Y f =\int_X S_n f $. Since $S_n f$ converges to $g$ in $L^1$, by the triangle inequality,
\begin{align}
    \abs{\int_X S_n f - g \ \mathrm d \mu} &\leq \int_X \abs{ S_n f - g }\mathrm d \mu\\
    &\rightarrow 0,
\end{align}
so $f$ and $g$ have the same normalization.

The measure defined by $\nu_{n,x}(A) = \int_A S_n(x,y) \mathrm d \mu(y)$ satisfies $\nu_{n,x}(Y)\leq 1$, for almost all $x$, because $S_n$ is semidoubly stochastic. 
We can assume $\nu_{n,x}(Y)>0$, as the desired inequality follows trivially otherwise. 
We apply Jensen's inequality to the renormalized measure,
\begin{align}
    \phi( S_n f )(x)&= \phi\left( \int_Y S_n(x,y) f(y) \ \mathrm d\mu(y)  \right)\\
    &= \phi \left( \int_Y \nu_{n,x}(Y) f(y) \ \frac{1}{\nu_{n,x}(Y)}\mathrm d\nu_{n,x}(y) \right)\\
    &\leq  \int_Y \phi \left[ \nu_{n,x}(Y) f(y)\right] \ \frac{1}{\nu_{n,x}(Y)}\mathrm d\nu_{n,x}(y). 
\end{align}
By assumption $\phi$ is convex and vanishes at the origin. Thus,
\begin{equation}
    \phi \left[ \nu_{n,x}(Y) f(y)\right] \leq \nu_{n,x}(Y) \phi \left[ f(y)\right] + \left[1-\nu_{n,x}(Y)\right] \phi(0) = \nu_{n,x}(Y) \phi  \left[ f(y)\right] \,,
\end{equation}
leading to the inequality,
\begin{equation}
    0\leq\phi( S_n f )(x) \leq \int_Y \phi \left( f(y)\right) \ \mathrm d\nu_{n,x}(y)\eqqcolon h_n(x), \label{convexub}
\end{equation}
since $\phi$ is nonnegative. 
We compute
\begin{align}
    \int_X h_n(x) \ \mathrm d\mu(x)&= \int_X \int_Y \phi \left( f(y)\right) \ S_n(x,y) \mathrm d\mu(y) \mathrm d\mu(x)\\
    &=  \int_Y \phi \left( f(y)\right) \ \int_X S_n(x,y)  \mathrm d\mu(x) \mathrm d\mu(y)\\
    &=  \int_Y \phi \left( f(y)\right) \ \mathrm d\mu(y), \label{jensenfubini}
\end{align}
where the second line follows from Fubini's theorem~\cite{fubini1907sugli,Rudin1987}, the third line follows from stochasticity of $S_n$, and the final integral exists by assumption.

Since $S_n f$ converges to $g$ in $L^1$, there exists a subsequence such that $S_{n_i} f$ converges pointwise to $g$ almost everywhere.
Since $\phi$ is convex, it is also continuous on $\mathbb{R}$ so $\phi(S_{n_i} f)$ converges pointwise to $\phi(g)$ almost everywhere. Since the limit exists almost everywhere the limit inferior equals it. Then, we have that
\begin{align}
    \int_X \phi(g) \mathrm d \mu &=\int_X \liminf_{i\rightarrow \infty} \phi(S_{n_i} f) \ \mathrm d \mu\\
    &\leq  \liminf_{i\rightarrow \infty} \int_X \phi(S_{n_i} f) \ \mathrm d \mu\\
    &\leq  \int_Y \phi(f) \ \mathrm d \mu,
\end{align}
where the second line follows by Fatou's lemma and the third line by Eqs.~\eqref{convexub} and~\eqref{jensenfubini}.

The same inequality extends to any linear shift of the convex function: $\phi_a(x)\coloneqq \phi(x)+ a x$ for any constant $a$. The proof is immediate:
\begin{align}
    \int_X \phi_a(f) \mathrm d\mu&= \int_X \phi(f) \mathrm d\mu + a\int_X f \mathrm d\mu\\
    &\geq \int_X \phi(g) \mathrm d\mu + a\int_X f \mathrm d\mu\\
    &= \int_X \phi(g) \mathrm d\mu + a\int_X g \mathrm d\mu= \int_X \phi_a(g) \mathrm d\mu,
\end{align}
where the second line follows from the inequality for $\phi$ and the third by normalization.

\noindent$\bm{3\implies 4:}$
The functions $s \mapsto (s-z)^+$ and $s \mapsto -(s+z)^-$ for $z\geq0$ are convex, nonnegative, and vanish at the origin.
Therefore, by assumption
\begin{align}
    \int_X (f(x)-z)^+ \ \mathrm d\mu(x) &\geq \int_X (g(x)-z)^+ \ \mathrm d\mu(x),\\
    \int_X (f(x)+z)^- \ \mathrm d\mu(x) &\leq \int_X (g(x)+z)^- \ \mathrm d\mu(x)
\end{align}
for all $z\geq0$, where all integrals exist because $f,g\in L^1$. Normalization follows by assumption.

\noindent$\bm{4\implies 1:}$
By assumption, $\int_X (g-z)^+ \ \mathrm d\mu \leq \int_X (f-z)^+ \ \mathrm d\mu$. By Lemma~\ref{chonglemma1}, this is equivalent to $\int_z^\infty D_{\gp} \leq \int_z^\infty D_{\fp} $. For any $s\in (0,\mu(X))$, set $z=(f^+)^\downarrow(s)\geq 0$, which guarantees that $D_f(z) \leq s \leq D_f(z^-)$. Then, by Lemma~\ref{chonglemma2},
\begin{align}
    \int_0^s (g^\downarrow -z) \mathrm d u &\leq  \int_z^\infty D_g(t) \ \mathrm d t\\
    &\leq \int_z^\infty D_f(t) \ \mathrm d t\\
    &=\int_0^s (f^\downarrow -z) \mathrm d u \,,
\end{align}
whence $L^{\Dsh}_f(s) \geq L^{\Dsh}_g(s)$ for all $s$ follows.

Similarly, we have $\int_X (g+z)^- \mathrm d\mu \geq \int_X (f+z)^- \mathrm d\mu$ if and only if $ - \int_{-\infty}^{-z} C_g \geq - \int_{-\infty}^{-z} C_f$. Setting $z=-(f^-)^\uparrow(s)\geq 0$, we guarantee that $C_f(-z) \leq s \leq C_f(-z^+)$, leading to
\begin{align}
    \int_0^s (g^\uparrow + z) \mathrm d u &\geq  - \int_{-\infty}^{-z} C_g(t) \  \mathrm dt\\
    &\geq - \int_{-\infty}^{-z} C_f(t) \  \mathrm dt \\
    &=\int_0^s (f^\uparrow +z) \mathrm d u \,,
\end{align}
whence $L^{\Ish}_f(s) \leq L^{\Ish}_g(s)$ for all $s$ follows. 

The normalization condition holds by assumption.
\end{proof}

\subsection{Proof of Theorem~2}\label{app:thm2proof}

\begin{proof}
We express each condition in a way that allows us to use Theorem~1 for the proof. 
This follows the ideas in~\cite{Joe1990} of reducing relative majorization to regular majorization via changing the measure as per Eq.~(9) in the main text.
Note that normalization is preserved in the new measure,
\begin{equation}
    \int_X f \mathrm{d} \mu = \int_X g \mathrm{d} \mu=\int_X \frac{f}{q}\mathrm{d} \nu=\int_X \frac{g}{q}\mathrm{d} \nu \,.
\end{equation}

Statement 1 is equivalent to
\begin{equation}
    \declor{\frac{f}{q}} \geq \declor{\frac{g}{q}} \text{ and } \inclor{\frac{f}{q}}\leq \inclor{\frac{g}{q}} \,,\quad \text{for all } s \in [0,\nu(X)] \,,
\end{equation}
by the definitions of the relative (co)distribution functions (Eqs.~(10) and (11) in the main text) and the rescaled measure in Eq.~(9) in the main text. 

Consider S$q$S integral operators $S_n$ on $L^1(X,\mu)$.
Define a new sequence of integral operators via the kernels $T_n(x,y) = S_n(x,y)/q(x)$. These operators are positive and well-defined by virtue of $q$ being strictly positive. The operator $T_n$ is SDS on $L^1(X,\nu)$ because 
\begin{align}
    \int_X T_n(x,y) \ \mathrm d\nu(x) &= \int_X S_n(x,y) \ \mathrm d\mu(x) = 1 \,,\quad \text{and}\\
    \int_Y T_n(x,y) \ \mathrm d\nu(y) &= \frac{1}{q(x)} \int_Y S_n(x,y) q(y) \ \mathrm d\mu(y) \le 1 \,
\end{align}
for almost all $y$ and $x$ respectively.
Additionally, 
\begin{align}
    \int_Y T_n(x,y) \frac{f(y)}{q(y)} \ \mathrm d\nu(y)
    \qquad\qquad\qquad\qquad
    \\= \frac{1}{q(x)} \int_Y S_n(x,y) f(y) \ \mathrm d\mu(y) \rightarrow \frac{g(x)}{q(x)} \,.
\end{align}
Therefore, statement 2 is equivalent to the existence of semidoubly stochastic $T_n$ satisfying $T_n \frac{f}{q}\rightarrow \frac{g}{q}$ in $L^1(X, \nu)$. 

Statement 3 is equivalent to $\int_X \phi\left(\frac{f}{q}\right) \ \mathrm d\nu \geq \int_X \phi\left(\frac{g}{q}\right) \ \mathrm d\nu$.

Statement 4 is equivalent to $ \int_X \left(\frac{f}{q}-z\right)^+ \ \mathrm d\nu \geq \int_X \left(\frac{g}{q}-z\right)^+ \ \mathrm d\nu $ and $\int_X \left(\frac{g}{q}+z\right)^- \ \mathrm d\nu \geq \int_X \left(\frac{f}{q}+z\right)^- \ \mathrm d\nu$, for all $z \geq 0$.

The equivalences between statements now follow directly from Theorem~1 applied to $\frac{f}{q}$ and $\frac{g}{q}$ as quasiprobability distributions in $L^1(X,\nu)$.
\end{proof}

\subsection{Restatement of results for counting measure}
\label{app:count}
Discrete probability distributions feature in the first formulations of majorization theory and in key results in quantum information theory~\cite{Owari_CV_Nielsen_2008}.
Hence, we explicitly restate our main results for this important special case. 

Consider the $\sigma$-finite measure space $(\mathbb{N}, \mu)$, where $\mu$ is the counting measure, defined as $\mu(A) = |A|$ for $A \in 2^{\mathbb{N}}$, where $|A| = \infty$ if $A$ is not finite.
A basis $\{e_n | n \in \mathbb{N}\}$ in the space $L^1(\mathbb{N}, \mu)$ is given by the Kronecker delta, $(e_n)_m = \delta_{nm}$~\cite{Manjegani2023}, so any quasiprobability distribution $f \in L^1(\mathbb{N}, \mu)$ can be expressed as
\begin{equation}
    f = \sum_{n\in\mathbb{N}} f(n) e_n \,.
\end{equation}
Moreover, we can express the $L^1$-finiteness and normalization conditions as
\begin{align}
    \norm{f}_1 = \sum_{n\in\mathbb{N}} |f(n)| < \infty \,\quad \mathrm{and}\quad 
    \sum_{n\in\mathbb{N}} f(n) = 1 \,,
\end{align}
and so the negative volume is $NV(f) = \frac{1}{2}\left(\sum_{n\in\mathbb{N}} |f(n)| - 1\right)$, which is also known as sum negativity in the resource theory of magic~\cite{veitch2014resource}.

An integral operator $S: L^1(\mathbb{N}, \mu) \rightarrow L^1(\mathbb{N}, \mu)$ is defined via a kernel $S(m,n)$ on $\mathbb{N} \cross \mathbb{N}$,
\begin{equation}
    (Sf)(m) = \sum_{n\in\mathbb{N}} S(m,n) f(n) \,.
\end{equation}
Operator $S$ is positive if $S(m,n) \ge 0$ for all $m,n \in \mathbb{N}$, and stochastic if additionally $\sum_{m\in\mathbb{N}} S(m,n) = 1$ for all $n \in \mathbb{N}$. 
It is semidoubly stochastic (SDS) if it is stochastic and $\sum_{n\in\mathbb{N}} S(m,n) \leq 1$ for all $m\in \mathbb{N}$, and doubly stochastic if the latter inequality holds as equality. 
It is semi-$q$-stochastic (S$q$S) if $\sum_{n\in\mathbb{N}} S(m,n) q(n) \leq q(m)$ for all $m\in \mathbb{N}$.

Let $f,g\in L^1(\mathbb{N},\mu)$ and let $q$ be any strictly positive distribution over $\mathbb{N}$. 
The distribution and codistribution functions of $f/q$ are
\begin{equation}
    D^q_f(t) = \sum_{n: f(n) > tq(n) } q(n) \,\quad \mathrm{and}\quad C^q_f(t) = \sum_{n: f(n) < tq(n) } q(n) \,.
\end{equation}
The rearrangements and Lorenz curves of $f$ relative to $q$ can then be defined as before.
Informally, consider a permutation $\pi$ that sorts $f/q$ into decreasing order. The positive Lorenz curve is piecewise linear, defined by the points $(0,0), (\pi q(0), \pi f(0)), (\pi q(0)+\pi q(1), \pi f(0)+\pi f(1)), \dots$\ . Similarly for the increasing rearrangement and negative Lorenz curve.

We are now in the position to restate our main result in terms of the counting measure.
\begin{corollary}[Restatement of Theorem~2 for the counting measure]\label{relmajor_counting}
    Given two functions $f,g\in L^1(\mathbb{N},\mu)$, where $\mu$ is the counting measure, and a strictly positive, measurable function $q$, we say that $f$ majorizes $g$ relative to $q$, and write $f \succ_q g$, if any of the following equivalent statements holds.
    \begin{enumerate}
    \item $L^{\Dsh}_{f|q}(s) \geq L^{\Dsh}_{g|q}(s)$ and $L^{\Ish}_{f|q}(s) \leq L^{\Ish}_{g|q}(s)$ for all $s \in [0,\nu(\mathbb{N})]$, and $\sum_{n\in\mathbb{N}} f(n) = \sum_{n\in\mathbb{N}} g(n)$;
    \item There exists a sequence of stochastic integral operators $(S_n)_{n \in \mathbb{N}}$ such that $S_n f$ converges to $g$ in $L^1(\mathbb{N},\mu)$ and $S_n q \leq q$ for all $n$;
    \item $\sum_{n\in\mathbb{N}} q(n) \phi\left(\frac{f(n)}{q(n)}\right) \ge \sum_{n\in\mathbb{N}} q(n) \phi\left(\frac{g(n)}{q(n)}\right)$, for all nonnegative convex functions $\phi: \mathbb{R} \rightarrow \mathbb{R}_{\geq 0}$ satisfying $\phi(0)=0$ such that the sums converge, and $\sum_{n\in\mathbb{N}} f(n) = \sum_{n\in\mathbb{N}} g(n)$;
    \item $\sum_{n\in\mathbb{N}} (f(n)-zq(n))^+ \ge \sum_{n\in\mathbb{N}} (g(n)-zq(n))^+$ and $ \sum_{n\in\mathbb{N}} (g(n)+zq(n))^- \ge \sum_{n\in\mathbb{N}} (f(n)+ zq(n))^-$ for all real numbers $z \ge 0$, and $\sum_{n\in\mathbb{N}} f(n) = \sum_{n\in\mathbb{N}} g(n)$.
    \end{enumerate}
\end{corollary}
As in Theorem~2, the inequality in statement 3 extends to linear shifts of such convex functions, and relative majorization reduces to regular majorization for $q(n)=1$.

Considering a finite subset of $\mathbb{N}$, our construction resembles the one commonly used in quantum thermodynamics. 
In quantum thermodynamics, the equivalence between relative and regular majorization is achieved via, firstly, rearranging $f,g$ as per the distribution functions $\smash{D^{q}_{f},D^{q}_{g}}$ (this is known as the $\beta$--ordering of $f,g$, where $q$ is a thermal distribution defined by some temperature $\beta^{-1}$~\cite{lostaglio2019introductory}), and, secondly, embedding $f,g$ from the original probability space to one of larger dimension, where $q$ becomes the uniform distribution while the normalization of $f,g$ is preserved~\cite{brandao2015second,lostaglio2019introductory,koukoulekidis2022constraints}.
Rescaling the measure in the Lebesgue integral reflects the embedding step.

We can write the $\alpha$-R\'{e}nyi entropies and divergences as
\begin{align}
    H_\alpha(f) = \frac{1}{1-\alpha} \log \sum_{n} \abs{f(n)}^\alpha  \,\quad \mathrm{and}\quad 
    D_\alpha(f||q) = \frac{1}{\alpha-1} \log \sum_{n} \abs{f(n)}^\alpha q(n)^{1-\alpha} \,.
\end{align}

\subsection{Extreme points of majorization preorder}\label{app:prop}

We briefly discuss the extreme points of the majorization preorder, i.e. distributions which either majorize or are majorized by all others. To illustrate, suppose $X$ is either a finite or infinite subset of $\mathbb{N}$ equipped with the counting measure.
If $\mu(X)$ is finite, the uniform distribution is the maximally disordered distribution. If infinite, there is no maximally disordered distribution. Probability and quasiprobability distributions alike can become arbitrarily spread out (e.g. consider $p(k)=\frac{1}{n}$ for $ k\in\{1,\dots,n\}$ as $n\rightarrow \infty$).
The maximally ordered probability distributions are precisely the deterministic ones, i.e. those with a single nonzero entry equal to 1. 
By allowing negative values, quasiprobability distributions can concentrate arbitrarily high weight on specific elements, something impossible for proper probability distributions. Hence, it is possible to construct quasiprobability distributions that are more ordered than any deterministic probability distribution (e.g.~consider $p=(1+a,-a,0,\dots)$ for $a>0$). The above discussion is summarized in Supplementary Table~\ref{table:max_ordered_disordered}.

Note that in the continuous case, there is not necessarily a maximally ordered distribution within the class of integrable functions. For example, any Gaussian function is majorized by a Gaussian with smaller variance. In the limit of zero variance, one obtains the Dirac delta, which is the continuous analog of a deterministic distribution, but, strictly speaking, is not in $L^1$.

\begin{table}[h]
\centering
\renewcommand{\arraystretch}{1.2}
\resizebox{0.6\columnwidth}{!}{%
\begin{tabular}{|c|c|c|}
\hline
 \cellcolor[HTML]{EFEFEF} Distributions over subsets of $\mathbb{N}$ & \cellcolor[HTML]{EFEFEF}\textbf{Finite measure} & \cellcolor[HTML]{EFEFEF}\textbf{Infinite measure} \\ \hline
\multicolumn{1}{|c|}{\cellcolor[HTML]{EFEFEF}\textbf{Probabilities}}      & max. ordered                                           & max. ordered                                             \\
\multicolumn{1}{|c|}{\cellcolor[HTML]{EFEFEF}}    & max. disordered                                        & \cancel{max. disordered}                                 \\ \hline
\multicolumn{1}{|c|}{\cellcolor[HTML]{EFEFEF}\textbf{Quasiprobabilities}} & \cancel{max. ordered}                                  & \cancel{max. ordered}                                    \\
\multicolumn{1}{|c|}{\cellcolor[HTML]{EFEFEF}}    & max. disordered                                        & \cancel{max. disordered}                                 \\ \hline
\end{tabular}%
}
\captionof{table}{Structure of different majorization preorders.
When considering discrete probability distributions defined over a finite measure space, both maximally disordered and maximally ordered distributions exist. However, these extremes can break down in two distinct ways: when the underlying space has infinite measure, there is no longer a maximally disordered distribution; and when quasiprobability distributions are allowed, maximally ordered distributions no longer exist.
}
\label{table:max_ordered_disordered}
\end{table}

\section*{Supplementary Note 2: Proof of Theorem~3 }
\label{app:L2-incomparability}
\setcounter{subsection}{0} 
Here, we prove Theorem~3 on the incomparability of bounded integrable functions.
\begin{proof}
Since we assume that $f,g\in L^{1}(X,\mu)\cap L^{\infty}(X,\mu)$, they have well-defined $L^p$-norms for any $p\in[1,\infty]$~\cite{viaclovsky:mit-18.125-lec17}.
The first step of our proof is to study an inner product space formed from pairs of rearrangements.
Let $r_1,r_2\in L^2(\mathbb{R}_{\geq0})$, so that $(r_1,r_2)\in L^2(\mathbb{R}_{\geq 0})\times L^2(\mathbb{R}_{\geq 0})$. 
We define the inner product
\begin{align}
	\big\langle
	(r_1,r_2),(s_1,s_2)
	\big\rangle
	\coloneqq
	\int_{0}^{\infty}
	r_1(u)s_1(u)\mathrm{d}u 
	+
	\int_{0}^{\infty}
	r_2(u)s_2(u) 
	\mathrm{d}u,
    \label{eq:def_inner_product_rearrangements}
\end{align}
which satisfies the three inner-product conditions:
\begin{enumerate}
\item	Symmetry: $\big\langle(r_1,r_2),(s_1,s_2)\big\rangle=\big\langle(s_1,s_2),(r_1,r_2)\big\rangle$
\item Linearity: $\big\langle \alpha(r_1,r_2)+\beta(s_1,s_2),(t_1,t_2)\big\rangle=\alpha\big\langle(r_1,r_2),(t_1,t_2)\big\rangle+\beta\big\langle(s_1,s_2),(t_1,t_2)\big\rangle$
\item Positive-definiteness: $(r_1,r_2)$ is nonzero $\Leftrightarrow\ \big\langle(r_1,r_2),(r_1,r_2)\big\rangle>0$.
\end{enumerate}
It thus obeys the Cauchy--Schwarz inequality:
\begin{align}
	\big\langle
	(r_1,r_2),(s_1,s_2)
	\big\rangle^2
	\leq
	\big\langle
	(r_1,r_2),(r_1,r_2)
	\big\rangle
	\cdot 
	\big\langle
	(s_1,s_2),(s_1,s_2)
	\big\rangle
	\label{eq:cauchy-scharwz}
\end{align}
with equality achieved if and only if $(r_1,r_2)\propto(s_1,s_2)$~\cite{Hassani2013-ue}.

Equipped with this inner product, we now present a method for building Schur-convex functionals.
Given a function $f\in L^{1}(X,\mu)\cap L^{\infty}(X,\mu)$, we denote the decreasing rearrangement of its positive part and the increasing rearrangement of its negative part respectively as $f^{+\downarrow}$ and $f^{-\uparrow}$.
We then define the functional $\Phi_f$ using the inner product previously introduced:
\begin{align}
	\Phi_f(g)
	\vcentcolon=
	\big\langle
	(f^{+\downarrow},f^{-\uparrow}),
	(g^{+\downarrow},g^{-\uparrow})
	\big\rangle.
    \label{eq:schur_convex_functional_phi_f}
\end{align}
By symmetry of the inner product, observe that $\Phi_f(g)=\Phi_g(f)$.
We are now going to show that $\Phi_f$ is Schur-convex for any choice of $f$.
Let $g\in L^{1}(X,\mu)\cap L^{\infty}(X,\mu)$; we can write: 
\begin{align}
		\Phi_f(g)
		&=
		\big\langle
		(f^{+\downarrow},f^{-\uparrow}),
		(g^{+\downarrow},g^{-\uparrow})
		\big\rangle
		\\[1em]
		&=
		\int\limits_{0}^{\infty}
		f^{+\downarrow}(u)
		\underbrace{
		g^{+\downarrow}(u)
		}_{=\frac{\mathrm{d}}{\mathrm{d}u}L_g^{\Dsh}}
		\mathrm{d}u
		+
		\int\limits_{0}^{\infty}
		f^{-\uparrow}(u)
		\underbrace{
		g^{-\uparrow}(u)
		}_{=\frac{\mathrm{d}}{\mathrm{d}u}L_g^{\Ish}}
		\mathrm{d}u
		\\[1em]&=
		\underbrace{
		\Big[
		f^{+\downarrow}(u)
		L^{\Dsh}_{g}(u)
		\Big]_{u=0}^{u=\infty}}_{=0}
		\ -\ 
		\int\limits_{0}^{\infty}
        \left(
            \frac{\mathrm{d}}{\mathrm{d}u}
		f^{+\downarrow}(u)
        \right)
		L^{\Dsh}_{g}(u)
        \mathrm{d}u \label{eq:wigner-to-schur-convex-step1}
        \\&\qquad\qquad\qquad
        +
		\underbrace{\Big[
		f^{-\uparrow}(u)L^{\Ish}_{g}(u)
		\Big]_{u=0}^{u=\infty}}_{=0}
		\ -\ 
		\int\limits_{0}^{\infty}
        \left(
        \frac{\mathrm{d}}{\mathrm{d}u}
		f^{-\uparrow}(u)
        \right)
		L^{\Ish}_{g}(u)
        \mathrm{d}u
        \nonumber
		\\[1em]&=
		\int\limits_{0}^{\infty}
		\underbrace{\left(-\frac{\mathrm{d}}{\mathrm{d}u}
        f^{+\downarrow}(u)\right)}_{\geq 0}
		\ 
        \underbrace{
		L_g^{\Dsh}(u)
        }_{\text{Schur-convex}}
		\mathrm{d}u
		+\int\limits_{0}^{\infty}
		\underbrace{\frac{\mathrm{d}}{\mathrm{d}u}f^{-\uparrow}(u)}_{\geq 0}
		\
        \underbrace{
            \Big(
		-L_g^{\Ish}(u)
        \Big)}_{\text{Schur-convex}}
		\mathrm{d}u.
		\label{eq:wigner-to-schur-convex-step2}
\end{align}
To reach Eq.~\eqref{eq:wigner-to-schur-convex-step1}, we applied integration by parts, under the assumption that the rearrangements $f^{+\downarrow}$ and $f^{-\uparrow}$ are absolutely continuous~\cite{Stromberg2015-au}.
Since $f,g\in L^{1}(X,\mu)\cap L^{\infty}(X,\mu)$, it follows that $\smash{f^{+\downarrow}, f^{-\uparrow}, L^{\Ish}_g, L^{\Dsh}_g}$ are bounded, that $f^{+\downarrow}, f^{-\uparrow}$ go to zero as $u\to\infty$, and that $\smash{L^{\Ish}_g, L^{\Dsh}_g}$ go to zero as $u\to 0$.
We have then that $\smash{f^{+\downarrow}L^{\Dsh}_g,f^{-\uparrow}L^{\Ish}_g}$ go to zero both as $u\rightarrow 0$ and $u\to\infty$, allowing us to simplify two terms in Eq.~\eqref{eq:wigner-to-schur-convex-step1}.
Finally, in Eq.~\eqref{eq:wigner-to-schur-convex-step2}, we used the monotonicity of $\smash{f^{+\downarrow}}$ and $\smash{f^{-\uparrow}}$, together with the fact that $\smash{L^{\Dsh}_g(u)}$ and $\smash{-L^{\Ish}_g(u)}$ are Schur-convex in $g$ for all $u$, to conclude that $\Phi_f(g)$ is itself Schur-convex in $g$, being a nonnegative combination of Schur-convex functionals.

We now come to the last step of the proof.
The $L^2$-norm of $f$, denoted by $\Vert f\Vert_2=\sqrt{\int_{X}f^2\mathrm{d}\mu}$, is related to the norm of the vector $(f^{+\downarrow},f^{-\uparrow})$ as follows:
\begin{align}
	\big\langle
	(f^{+\downarrow}, f^{-\uparrow}),
	(f^{+\downarrow}, f^{-\uparrow})
	\big\rangle
	&=
	\int \big(f^{+\downarrow}(u)\big)^2\mathrm{d}u
	+
	\int \big(f^{-\uparrow}(u)\big)^2\mathrm{d}u
        \\[0.5em]&=
        \int_X \big(f^{+}\big)^2\mathrm{d}\mu
	+
	\int_X \big(f^{-}\big)^2\mathrm{d}\mu
        \\[0.5em]&=
	\int_{X} f^2\mathrm{d}\mu
        \ =\ 
        \Vert f\Vert_2^2.
        \label{eq:equivalence_l2_norm_rearrangement}
\end{align}
As a consequence of the Cauchy--Schwarz inequality (see Eq.~\eqref{eq:cauchy-scharwz}), the Schur-convex functional $\Phi_f$ is upper bounded:
\begin{align}
	\Phi_f(g)
	&=
	\big\langle
	(f^{+\downarrow}, f^{-\uparrow}),
	(g^{+\downarrow}, g^{-\uparrow})	
	\big\rangle
	\\[1em]&\leq
	\sqrt{
	\big\langle
	(f^{+\downarrow}, f^{-\uparrow}),
	(f^{+\downarrow}, f^{-\uparrow})
	\big\rangle}
        \cdot
	\sqrt{ 
	\big\langle
	(g^{+\downarrow}, g^{-\uparrow}),
	(g^{+\downarrow}, g^{-\uparrow})
	\big\rangle
	}
	\\[1em]&=
        \sqrt{
        \int_X f^2\mathrm{d}\mu
        }
        \cdot 
        \sqrt{
        \int_X g^2\mathrm{d}\mu
        }
        =
        \Vert f\Vert_2\cdot\Vert g\Vert_2,
        \label{eq:phi_f_upper_bound_cauchy_schwarz}
\end{align}
and the equality $\Phi_f(g)=\Vert f\Vert_2\cdot\Vert g\Vert_2$ holds if and only if $(f^{+\downarrow}, f^{-\uparrow})\propto(g^{+\downarrow}, g^{-\uparrow})$.

We now assume that $f$ and $g$ have the same $L^2$-norm, i.e. $\Vert f\Vert_2=\Vert g\Vert_2$.
We consider the Schur-convex functionals $\Phi_f$ and $\Phi_g$, associated respectively with $f$ and $g$, and defined from Eq.~\eqref{eq:schur_convex_functional_phi_f}.
From Eq.~\eqref{eq:equivalence_l2_norm_rearrangement} and $\Vert f\Vert_2=\Vert g\Vert_2$, we know that $\Phi_f(f)=\Phi_g(g)$; from the symmetry of the inner product, we know that $\Phi_f(g)=\Phi_g(f)$.
Moreover, from Eq.~\eqref{eq:phi_f_upper_bound_cauchy_schwarz} we have the inequality $\Phi_f(g)\leq\Phi_f(f)$.
At this point, there are two possibilities:
either (i) the inequality is strict or (ii) it is an equality.

\begin{description}[leftmargin=1.8cm, labelsep=0.4cm]
    \item[Case (i)] $\Phi_f(g)<\Phi_f(f)$.

    Since $\Phi_f(g) = \Phi_g(f)$ and $\Phi_f(f) = \Phi_g(g)$, we also have the reversed strict inequality $\Phi_g(g) > \Phi_g(f)$.  
    Since both $\Phi_f$ and $\Phi_g$ are Schur-convex, this implies that $f$ and $g$ are majorization-incomparable, i.e., $f \not\succ g$ and $f \not\prec g$.

    \item[Case (ii)] $\Phi_f(g) = \Phi_f(f)$.

    Equality in the Cauchy--Schwarz inequality holds if and only if the two vectors are linearly dependent, i.e., $(f^{+\downarrow}, f^{-\uparrow}) \propto (g^{+\downarrow}, g^{-\uparrow})$.  
    Since $f$ and $g$ have equal normalization, it must in fact be that $(f^{+\downarrow}, f^{-\uparrow}) = (g^{+\downarrow}, g^{-\uparrow})$.  
    If $f$ and $g$ share the same decreasing and increasing rearrangements, they are majorization-equivalent, i.e., $f \succ g$ and $f \prec g$.
\end{description}
This concludes the proof of Theorem~3.
\end{proof}

\section*{Supplementary Note 3: Quasiprobability representations}\label{app:quasiprob}
\setcounter{subsection}{0} 
We now outline the general structure of quasiprobability representations of states and channels.  Here a choice of quasiprobability representation is an injective linear map $\Upsilon$ from quantum states to real-valued distributions over the \textit{phase space} $X$.  For our purposes we then restrict attention to those distributions that are integrable.
Such a map obeys several important properties, both generic and representation-specific~\cite{brif1999}, which we shall introduce as needed. Note that each choice of representation defines a different subset of functions in $L^1(X, \mu)$, and that in general the number of such choices is vast~\cite{Cohen_2013}.

\subsection{State representations}
The linearity of $\Upsilon$ allows us to express the mapping $\hat\rho \mapsto f_{\hat \rho}$ as
\begin{equation}\label{eq:SW_kernel_abstract}
    f_{\hat \rho}(x) = \Tr[\hat \Delta_{\Upsilon} (x) \hat \rho], \qquad x \in X,
\end{equation}
where $\hat \Delta_\Upsilon(x)$ are \textit{phase-point operators}. These are operators satisfying normalization $\Tr[\hat \Delta_\Upsilon (x)]=1$ and completeness $\int_X \hat \Delta_\Upsilon (x) d\mu(x) = \hat{\id}$. If $\hat \Delta_\Upsilon (x)$ is not trace-class, one can still make sense of the normalization formula by restricting the bases in which the trace is taken, or by considering a sequence of trace-class operators that approximates the given phase-point operator arbitrarily well~\cite{Benedict1995}.
This mapping is invertible through use of the \textit{dual} quasiprobability representation $\Upsilon'$:
\begin{equation}
    \hat \rho = \int_X f_{\hat \rho} (x) \hat \Delta_{\Upsilon'}(x) \, \mathrm{d} \mu(x),
    \label{eq:inverse_map_representation}
\end{equation}
where $\hat \Delta_{\Upsilon'} (x)$ are the phase-point operators associated with $\Upsilon'$. Here duality refers to the following \textit{traciality} relation:
\begin{equation}\label{eq:traciality}
    \Tr\left[\hat A \, \hat \rho\right] = \int_X f'_{\hat A}(x) f_{\hat \rho} (x)  \mathrm{d} \mu(x),
\end{equation}
where $f'_{\hat A}(x) = \Tr\left[\hat \Delta_{\Upsilon'}(x) \, \hat A\right]$ is the dual representation of the operator $\hat A$~\cite{brif1999}.  See also~\cite{Ferrie_Emerson_2009} for a review of these concepts in finite-dimensional Hilbert spaces.

\subsection{Channel representations}
Linear maps acting on states can be represented by integral operators on the states' quasiprobability representations: given $\Upsilon$ and $\Upsilon'$, a quantum channel $\Phi$ has a representation as an integral (kernel) operator
\begin{equation}\label{eq:channel_symbol_abstract}
    K_\Phi(x,y)\coloneqq \Tr[\hat \Delta_{\Upsilon}(x) \, \Phi(\hat \Delta_{\Upsilon'}(y))],
\end{equation}
so that 
\begin{equation}
    f_{\Phi(\hat \rho)}(x) = \int_X  K_{\Phi}(x,y) f_{\hat \rho}(y) \mathrm{d} \mu(y).
\end{equation}

When the kernel is SDS (Definition~5 in the main text) or S$q$S (Definition~8 in the main text) we may apply our results for majorization theory (Theorem~1) or relative majorization theory (Theorem~2), respectively.
This raises the following question: which \textit{quantum channels} $\Phi: \mathcal{D}(\H) \rightarrow \mathcal{D}(\H)$ are represented by an SDS or S$q$S integral operator with respect to $\Upsilon$? 

We now find conditions for a kernel to be SDS. By definition, the kernel is positive (Definition~3) when 
\begin{equation}
\Tr[ \hat \Delta_{\Upsilon}(x) \, \Phi(\hat \Delta_{\Upsilon'}(y))]\geq0.
\end{equation}
If the kernel is positive, stochasticity (Definition~4) follows immediately because the channel is trace-preserving:
\begin{align}
    \int_X \Tr[\hat \Delta_{\Upsilon}(x) \, \Phi(\hat \Delta_{\Upsilon'}(y)) ]\mathrm{d}  \mu(x)&=\Tr[\int_X  \hat \Delta_{\Upsilon}(x) \mathrm{d}  \mu(x) \, \Phi(\hat \Delta_{\Upsilon'}(y)) ]\\
    &=\Tr[\Phi(\hat \Delta_{\Upsilon'}(y)) ]\\
    &=1,
\end{align}
where in the second line we have used the completeness relation and in the third line normalization and trace-preservation.
The kernel is further semidoubly stochastic when
\begin{align}
    \int_Y \Tr[\hat \Delta_{\Upsilon}(x) \, \Phi(\hat \Delta_{\Upsilon'}(y)) ]\mathrm{d}  \mu(y)&=\Tr[ \Phi^*(  \hat \Delta_{\Upsilon}(x))  \,  \int_Y \hat \Delta_{\Upsilon'}(y) \mathrm{d} \mu(y) ]\\
    &=\Tr[\Phi^*(  \hat \Delta_{\Upsilon}(x)) ] \leq 1,
\end{align}
where $(\cdot)^*$ denotes the adjoint.
Since the phase-point operators are normalized and satisfy completeness, $\Tr\left[\Phi^*(\hat \Delta_{\Upsilon}(x)) \right] \leq 1$ if and only if $\Tr\left[\hat \Delta_{\Upsilon}(x) (\hat{\id}-\Phi(\hat{\id})) \right] \geq0$, i.e. $\hat{\id}-\Phi(\hat{\id})$ is nonnegatively represented.

More generally, we find an equivalent condition for the kernel to be S$q$S (Definition~8), where $q(x)=\Tr\left[ \hat{Q} \hat{\Delta}(x)\right]$ for some operator $\hat{Q}$. 
Concretely, $K_\Phi$ is S$q$S if and only if 
\begin{align}
    q(x)&\geq \int_Y \Tr\left[\Phi(\hat \Delta_{\Upsilon'}(y))\, \hat \Delta_{\Upsilon}(x)\right] q(y) \mathrm{d} \mu(y)\\
    &=  \Tr\left[\Phi\left( \int_Y q(y) \hat \Delta_{\Upsilon'}(y) \mathrm{d} \mu(y)\right)\, \hat \Delta_{\Upsilon}(x)\right] \\ 
    &= \Tr\left[\Phi \left[ \hat{Q}\right] \hat \Delta_\Upsilon(x)\right]. 
\end{align}
This condition is satisfied if and only if 
\begin{equation}
    \Tr\left[ (\hat{Q} - \Phi \left[ \hat{Q}\right]) \hat \Delta_\Upsilon(x)\right]\geq 0 \quad \forall x,
\end{equation}
i.e., $\hat{Q} - \Phi [ \hat{Q}]$ is a classical operator in this particular quasiprobabilistic representation. It suffices for $\hat{Q}$ to be a fixed point of the channel. A powerful feature of our majorization framework is that it allows us to study channels with phase-space fixed points \emph{that are not the representation of any density operator}, or indeed even of any bounded operator (see Wigner majorization for an explicit example).

\section*{Supplementary Note 4: The Wigner representation}
\setcounter{subsection}{0} 
An important and common choice of quasiprobability representation $\Upsilon$ is the Wigner-Weyl-Moyal correspondence~\cite{Wigner1932, de_gosson_wigner_2017}.
\subsection{Wigner functions}
\label{sec:wigfunc}
The Wigner function's phase-point operators \eqref{eq:SW_kernel_abstract} can be mapped to the set of displaced parity operators~\cite{Grossmann_1976, royer1977Wigner}, defined as the standard parity operator $\hat{\Pi}\vcentcolon=(-1)^{\hat{n}} = e^{i\pi \hat n}$ evolved under a displacement operator $\hat D(\bm x, \bm p) \vcentcolon= e^{i(\bm p \hat{\bm{x}} - \bm x \hat{\bm{p}})}$ as follows:
\begin{align}
    \hat{\Delta}_{\mathrm{W}}(\bm{x},\bm{p})
    \vcentcolon=
    2^n\,\hat{D}(\bm{x},\bm{p})
    \,\hat{\Pi}\,
    \hat{D}^{\dagger}(\bm{x},\bm{p}).
\end{align}
This yields a map from density matrices $\hat \rho$ on the Hilbert space $L^2(\mathbb R^n)$ to phase space functions $W_{\hat \rho}$ defined as:
\begin{equation}\label{eq:def_wigner_function}
    W_{\hat{\rho}}(\bm{x},\bm{p}) \coloneqq \mathrm{Tr}\big[\hat{\Delta}_{\mathrm{W}}(\bm{x},\bm{p})\;\hat{\rho}\big]  = 
    \int_{\mathbb R^{n}} e^{-i\bm p \cdot \bm y}\ \big\langle \bm x + \tfrac{\bm{y}}{2}\big|\, \hat \rho\, \big| \bm x - \tfrac{\bm y}{2} \big\rangle\ \mathrm{d} \bm y,
\end{equation}
where the second equality is obtained by evaluating the trace in the position basis.

The image $W_{\hat \rho}$ is called the \textit{Wigner function} of $\hat \rho$. 
The Wigner function is real and normalized, but can take negative values. Note that not all quantum states $\hat \rho$ have a finite Wigner negative volume (see for example the Feichtinger algebra~\cite{De_Gosson_De_Gosson_2021_Feichtinger} and its complement), but those that do are dense in the space of all states.

The Wigner representation has the remarkable property of being self-dual (i.e.\ $\Upsilon = \Upsilon'$), meaning that the inverse map is simply obtained as:
\begin{align}
    \hat{\rho}
    &=
    \iint
    W_{\hat{\rho}}(\bm{x},\bm{p})
    \;
    \hat{\Delta}_{\mathrm{W}}(\bm{x},\bm{p})
    \ 
    \frac{\mathrm{d}\bm{x}\mathrm{d}\bm{p}}{(2\pi)^n}.
\end{align}
Comparing with Eq.~\eqref{eq:inverse_map_representation}, this amounts to choosing the measure $\mathrm{d}\mu(\bm{x},\bm{p})=\mathrm{d}\bm{x}\mathrm{d}\bm{p}/(2\pi\hbar)^n$ and setting $\hbar=1$~\cite{brif1999}.
Another consequence of self-duality is that the overlap between quantum operators is directly connected to the overlap of their respective Wigner functions:
\begin{align}
    \mathrm{Tr}\big[\hat{A}\,\hat{\rho}\big]
    =
    \iint
    W_{\hat{A}}(\bm{x},\bm{p})\,
    W_{\hat{\rho}}(\bm{x},\bm{p})\ 
    \frac{\mathrm{d}\bm{x}\mathrm{d}\bm{p}}{(2\pi)^n}.
\end{align}
The above relation, which is the counterpart of Eq.~\eqref{eq:traciality}, is sometimes called the \textit{overlap formula}.
In particular, the overlap formula implies that the purity of a quantum state is the squared $L^2$-norm of its Wigner function:
\begin{equation}
    \Tr[\hat \rho^2] = \int W_{\hat \rho}^2(\bm x, \bm p)\ \frac{\mathrm{d}\bm{x}\mathrm{d}\bm{p}}{(2\pi)^n}.
\end{equation}
Thus an immediate corollary of Theorem~3 is that the Wigner functions of any pair of quantum states with the same purity are either equivalent or incomparable (assuming that the decreasing and increasing rearrangements of the Wigner functions are absolutely continuous, which holds for a large class of states).

Another key property of the Wigner-Weyl-Moyal representation is that of covariance under Gaussian unitaries.  Let $\hat U_\mathrm{G}$ be a Gaussian unitary (i.e.\ an element of the inhomogeneous metaplectic group $\mathrm{IMp}(2n, \mathbb R)$~\cite{de_gosson_wigner_2017}) with associated symplectic matrix $\bm S$ and displacement vector $\bm d$. Then, the Wigner function of any quantum state obeys the following property:
\begin{equation}
    W_{\hat U_{\mathrm G} \hat \rho \hat U^\dagger_{\mathrm G}}(\bm x, \bm p) = W_{\hat \rho}\Big(\bm S^{-1}
    \big(
    \begin{mypsmallmatrix}
        \bm{x}\\\bm{p}
    \end{mypsmallmatrix}
    - \bm d
    \big)
    \Big).
\end{equation}
That is, the action of a Gaussian unitary on Hilbert space $L^2(\mathbb R^n)$ is represented by an affine symplectic coordinate change of the Wigner function; see also~\cite{Dias_Prata_2019_Wigner_coordinate}.  This is relevant to us because such linear area-preserving coordinate transformations are trivial examples of stochastic maps.  In fact, they are doubly stochastic and can intuitively be seen as examples of the continuous-variable analogue of permutation matrices. For example, all coherent states are equivalent in the regular majorization preorder.

The integral kernel \eqref{eq:channel_symbol_abstract} of a quantum channel $\Phi$ in the Wigner representation is
\begin{equation}
    K_{\Phi}(\bm x, \bm p, \bm x', \bm p') = 
     \Tr\big[\Phi\big(\hat \Delta_{\mathrm W}(\bm x', \bm p')\big)\hat \Delta_{\mathrm W}(\bm x, \bm p)\big].
\end{equation}

\subsection{Wigner-positive channels}\label{app:wignerpos}

We now describe several examples of remarkable quantum channels that have Wigner-positive kernels, summarized in Supplementary Figure~\ref{fig:gaussian-dilatable_measure-and-prepare_channels}.

\begin{figure}[h!]
\includegraphics[width=\linewidth]{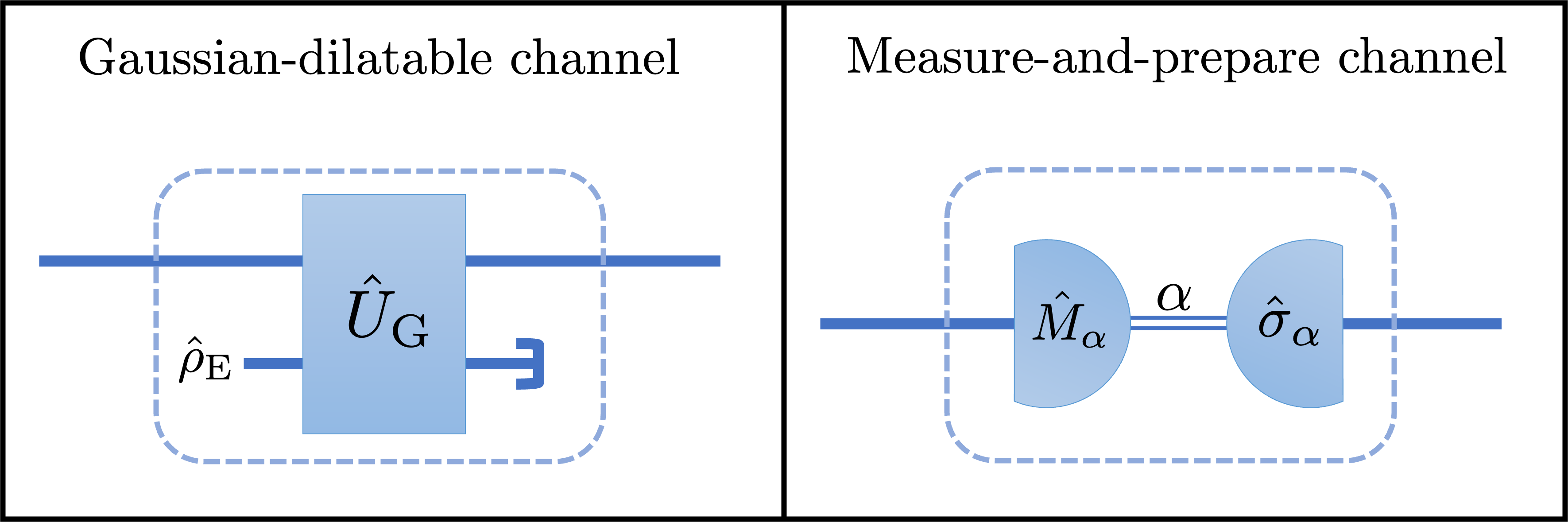}
\centering
\captionof{figure}{
Gaussian-dilatable channel: the input state is coupled to an environment state $\hat{\rho}_{\mathrm{E}}$ and evolves under a Gaussian unitary $\hat{U}_{\mathrm{G}}$, the environment is then traced out.
Measure-and-prepare channel: the input state is measured with a POVM with elements $\lbrace\hat{M}_{\alpha}\rbrace$, then conditionally on the measurement outcome $\alpha$, the output state $\hat{\sigma}_{\alpha}$ is prepared.
}
\label{fig:gaussian-dilatable_measure-and-prepare_channels}
\end{figure}

\paragraph{Gaussian channels.} 
Gaussian channels have been extensively studied~\cite{Cerf2007-chap2, Caruso2008-dt,Weedbrook2012-qu}.
Such channels have a kernel of the form
\begin{align}
K_\Phi(\bm x, \bm p,\bm x', \bm p') = \Tr[\hat \Delta_{\mathrm W}(\bm x, \bm p) \Phi(\hat\Delta_{\mathrm W}(\bm x', \bm p'))] 
\hspace{10em}
\\[0.8em]
= 
\frac{2^n}{\sqrt{\det \bm Y}}\   
\mathrm{exp}\left[
-\Big(
\begin{mypsmallmatrix}
    \bm{x}\\\bm{p}
\end{mypsmallmatrix}
-\bm{X}
\begin{mypsmallmatrix}
\bm{x}'\\\bm{p}'
\end{mypsmallmatrix}
-\bm{\delta}
\right)^\intercal 
\bm{Y}^{-1}
\left(
\begin{mypsmallmatrix}
    \bm{x}\\\bm{p}
\end{mypsmallmatrix}
-
\bm{X}
\begin{mypsmallmatrix}
    \bm{x}'\\\bm{p}'
\end{mypsmallmatrix}
-\bm{\delta}
\Big)
\right],
\label{eq:gaussian_wigner_kernel}
\end{align}
where $\bm X$ is a rescaling of the input quadratures, $\bm Y$ is classical noise added to the covariance matrix, and $\bm{\delta}$ is a phase-space displacement. The kernel is manifestly nonnegative and satisfies $\int  K_\Phi(\bm{x},\bm{p},\bm{x}',\bm{p}') \mathrm{d}\bm{x}\mathrm{d}\bm{p}/(2\pi)^n=1\ \forall\bm{x}',\bm{p}'$. By symmetry and a change of variables, $\int K_\Phi(\bm{x},\bm{p},\bm{x}',\bm{p}')\mathrm{d}\bm{x}'\mathrm{d}\bm{p}'/(2\pi)^n =1/{\abs{\det\bm{X}}}\ \forall \bm{x},\bm{p}$. Thus, the amplifying Gaussian channels, $\abs{\det \bm X}\geq 1$, have semidoubly stochastic kernels. For this subset of channels, we can apply majorization to compare two states and rule out state conversions. Gaussian channels with $\abs{\det \bm X}<1$, the attenuating Gaussian channels, are more `powerful' than their large-determinant counterparts, in the sense that they can take less-ordered Wigner functions to more-ordered ones. A natural and powerful technique to study such channels is to use relative majorization. Rather elegantly, all attenuating Gaussian channels have a density-operator fixed point~\cite{Ivan2011-pm}. Hence, their kernels are S$q$S with $q$ the Wigner function of the fixed point.

Among single-mode Gaussian channels, a common example with $\abs{\det \bm{X}}< 1$ is the pure-loss channel (PLC).
The PLC is ubiquitous in quantum optics as it adequately models the transmission of light in optical fibers and free-space links.
It is described as a beam-splitter mixing the input state with vacuum.
Let the unitary of the beam-splitter be $\hat{U}_{\eta}\vcentcolon=\exp\big(i\,\theta(\hat{x}_1\hat{p}_2-\hat{p}_1\hat{x}_2)\big)$ where $\eta=\cos^2\theta$ is the transmittance. The PLC is then defined as:
\begin{align}
    \mathcal{E}_{\eta}[\hat{\rho}]
    =
    \mathrm{Tr}_2\left[
    \hat{U}_{\eta}
    \big(
    \hat{\rho}
    \otimes\dyad{0}
    \big)
    \hat{U}^{\dagger}_{\eta}
    \right].
\end{align}
It is associated with a kernel with rescaling matrix $\bm{X}=\sqrt{\eta} \;\bm{I}_2$, noise matrix $\bm{Y}=(1-\eta)\;\bm{I}_2$ and displacement vector $\bm{\delta}=0$ ($\bm{I}_2$ is a $2\times 2$ identity matrix).

In fact, the PLC is a particular single-mode instance of a class of Gaussian channels called linear-optical networks (LON). Such channels consist of a global passive Gaussian unitary (i.e. beam splitters and phase shifters) with ancillary environment modes in vacuum~\cite{Rahimi-Keshari2016}. An $M$-mode linear-optical network is associated with a transfer matrix $\bm{L}\in\mathbb{C}^{M\times M}$ and acts on multimode coherent states $\ket{\bm{\gamma}}$, $\bm \gamma \in \mathbb C^M \simeq \mathbb R^{2M}$, as
\begin{equation}
    \dyad{\bm{\gamma}}\rightarrow \dyad{\bm{L}\bm{\gamma}}.
    \label{eq:LON_over_coherent_states}
\end{equation}
The matrix $\bm{L}$ is unitary for a lossless LON while for a lossy LON it is a principal submatrix of a unitary matrix $\bm{U}\in\mathbb{C}^{N\times N}$ (with $N>M$).
As they are particular instances of Gaussian channels, LONs admit a Wigner kernel of the form of Eq.~\eqref{eq:gaussian_wigner_kernel}.
More precisely, their displacement vector $\bm{\delta}$ and matrices $\bm{X}$, $\bm{Y}$ are obtained from $\bm{L}$ as follows:
\begin{align}
    \bm{X}
    =
    \begin{pmatrix}
        \mathrm{Re}\,\bm{L} &-\mathrm{Im}\,\bm{L}\\
        \mathrm{Im}\,\bm{L} &\mathrm{Re}\,\bm{L}
    \end{pmatrix},
    \qquad\qquad
    \bm{Y}
    =
    \bm{I}_{2M}-\bm{X}^{\intercal}\bm{X},
    \qquad\qquad\bm{\delta}=0.
    \label{eq:LON_kernel_matrices}
\end{align}
Ref.~\cite{Rahimi-Keshari2016} shows that in fact LONs have a stochastic kernel for a large family of phase-space quasiprobability distributions, known as $s$-ordered distributions~\cite{cahill1969density}.
The $s$-parameterized kernel is simply obtained with the change $\bm{Y}\to(1-s)\bm{Y}$ in Eq.~\eqref{eq:LON_kernel_matrices} (Wigner corresponds to $s=0$).
As is clear from Eq.~\eqref{eq:LON_over_coherent_states}, the vacuum is a fixed point of any such channel.  

As a final class of Gaussian channels, consider \textit{thermal loss channels}, which can be realized by coupling the system to ancillary modes prepared in thermal states (all at the same temperature), followed by a global passive Gaussian unitary.  Such channels have a thermal state as a fixed point.

\paragraph{Gaussian-dilatable channels with Wigner-positive environment.}
Gaussian-dilatable channels~\cite{Sabapathy_Guassian_dilatable_2017, Lami_Sabapathy_Winter_2018} are a generalization of Gaussian channels whose action can be expressed as a global Gaussian unitary $\hat U_{\mathrm G}$ acting on the input state $\hat \rho_{\mathrm S}$ together with an arbitrary environment state $\hat \rho_{\mathrm E}$:
\begin{align}
    \mathcal{E}[\hat{\rho}_{\mathrm{S}}]
    =
    \mathrm{Tr}_{\mathrm{E}}\Big[
    \hat{U}_{\mathrm{G}}
    \left(
    \hat{\rho}_{\mathrm{S}}
    \otimes
    \hat{\rho}_{\mathrm{E}}
    \right)
    \hat{U}^{\dagger}_{\mathrm{G}}
    \Big].
    \label{eq:def_gaussian_dilatable_channel}
\end{align}
Such channels reduce to Gaussian channels when the environment is Gaussian. However, not all Gaussian-dilatable channels have a Wigner-positive kernel. We derive in Supplementary Note 5 the Wigner kernel of a general Gaussian-dilatable channel and prove that it is nonnegative if the environment $\hat{\rho}_{\mathrm{E}}$ is Wigner positive.  In addition, we observe that the (semi)double stochasticity of the channel only depends on the global Gaussian unitary $\hat{U}_{\mathrm{G}}$. Specifically, consider the symplectic matrix $\bm{S}$ associated to $\hat{U}_{\mathrm{G}}$, and compute the submatrix of $\bm{S}^{-1}$ restricted to the environment modes. If the determinant $\abs{\mathrm{det}\left[(\bm{S}^{-1})_{\mathrm{EE}}\right]}=1$ then the channel is doubly stochastic, while if $\abs{\mathrm{det}\left[(\bm{S}^{-1})_{\mathrm{EE}}\right]}\geq 1$ then it is semidoubly stochastic.

\paragraph{Measure-and-prepare channels.} Consider a general measure-and-prepare channel defined by a positive operator-valued-measure (POVM) $\{\Gamma \ni z \mapsto \hat M_z\}$ for outcomes $z$ in an outcome space $(\Gamma, \mathrm d\mu(z))$ and a set of states $\{\hat \sigma_z\}$ also parametrized by $\Gamma$:
\begin{equation}
    \Phi(\hat\rho)=\int_\Gamma  \Tr[\hat M_z \hat \rho] \, \hat \sigma_z \mathrm d\mu(z).
\end{equation}
This channel averages over the preparations of $\hat \sigma_z$ conditioned on the outcome $z$ upon measuring $\hat \rho$.  The Wigner kernel of such a channel can be quickly derived as follows,
\begin{align}
    W_{\Phi(\hat \rho)}(\bm x, \bm p) &= \Tr[\hat \Delta_{\mathrm W}(\bm x, \bm p) \Phi(\hat \rho)]\\
    &=\int_\Gamma  \Tr[\hat M_z \hat \rho] \underbrace{\Tr[\hat \Delta_{\mathrm W}(\bm x, \bm p) \hat \sigma_z]}_{W_{\hat{\sigma}_z}(\bm x, \bm p)} \mathrm{d} \mu(z)\\ 
    &= \int_\Gamma  \Tr[\hat M_z\ \int_X W_{\hat \rho}(\bm x', \bm p') \hat \Delta_{\mathrm W}(\bm x', \bm p') \frac{\mathrm d\bm x' \mathrm d \bm p'}{(2\pi)^n}]\; W_{\hat\sigma_z}(\bm x, \bm p) \mathrm{d} \mu(z)\\
    &=  \int_X  \underbrace{\int_\Gamma W_{\hat \sigma_z}(\bm x, \bm p) W_{\hat M_z}(\bm x', \bm p') d\mu(z)}_{\coloneqq K_\Phi(\bm x, \bm p, \bm x', \bm p')} W_{\hat\rho}(\bm x', \bm p') \frac{\mathrm{d}\bm x' \mathrm{d}\bm p'}{(2\pi)^n}
\end{align}
Hence it is sufficient for $W_{\hat M_z}$ and $W_{\hat\sigma_z}$ to be nonnegative for all outcomes $z$ to imply that the Wigner kernel is positive.  Stochasticity then follows immediately from trace-preservation. 

Such measure-and-prepare channels also have a satisfying characterization in terms of relative majorization.  For a positive probability distribution $q$ over the phase space $X$, compute
\begin{align}
    \int_X K_\Phi(\bm x, \bm p, \bm x', \bm p') q(\bm x', \bm p') \frac{\mathrm d\bm x' \mathrm d \bm p'}{(2\pi)^n} &= \int_X \int_\Gamma W_{\hat \sigma_z}(\bm x, \bm p) W_{\hat M_z}(\bm x', \bm p') q(\bm x', \bm p') \frac{\mathrm d\bm x' \mathrm d \bm p'}{(2\pi)^n} \mathrm d\mu(z) \\
    &=\int_\Gamma W_{\hat \sigma_z}(\bm x, \bm p) \Tr[\hat M_z \hat{Q}] \mathrm{d} \mu(z). 
\end{align}
Hence the kernel is S$q$S (recall Definition~8) when $\int_\Gamma W_{\hat \sigma_z}(\bm x, \bm p) \Tr[\hat M_z \hat{Q}] \mathrm{d} \mu(z) \leq q(\bm x, \bm p)$ for almost all $(\bm x, \bm p) \in X$.  Intuitively, this is the condition that if $W_{\hat \sigma_z}$ is highly ordered relative to $q$, then $\hat M_z$ must have small Hilbert-Schmidt overlap with the operator $\hat Q$ associated to the lowest-ordered distribution $q$.  That is, the channel cannot replace low-order distributions with high-order ones.

Finally, we note that this discussion of measure-and-prepare channels generalizes immediately beyond the Wigner-Weyl-Moyal model to general quasiprobability representations and their duals \eqref{eq:traciality}.  In particular, the kernel becomes
\begin{equation}
    K_\Upsilon(\bm x, \bm p, \bm x', \bm p') = \int_\Gamma \Tr[\hat \sigma_z \hat\Delta_\Upsilon(\bm x, \bm p)] \Tr[\hat M_z \hat \Delta_{\Upsilon'}(\bm x', \bm p')] d\mu(z),
\end{equation}
and one obtains an analogous condition for the kernel to be S$q$S.

\paragraph{Convex mixtures thereof.}
Finally, observe that semi-$q$-stochasticity is preserved under convex mixing.
For example, consider a mixed-unitary quantum channel,
\begin{equation}\label{eq:mixture_of_gaussians_channel_def}
    \mathcal C_p[\hat \rho] = \int \text{d}p \, p(\hat U_G) \hat U_G \hat \rho \hat U^\dagger_G, \qquad \int p(\hat U_G) \mathrm{d} p = 1,
\end{equation}
where $U_G$ are Gaussian unitaries and $p(\hat U_G)$ is a probability distribution over them.
Since Gaussian unitaries are doubly stochastic, this channel is also doubly stochastic.
The dephasing channel~\cite{Walls2008} is an important mixed-unitary channel with widespread applications. It degrades coherence with respect to the Fock basis,
\begin{align}
    \mathcal{N}_{\gamma}[\hat{\rho}]
    \vcentcolon=
    \sum\limits_{m,n=0}^{\infty}
    \bra{m}\hat{\rho}\ket{n}\ 
    e^{-\frac{1}{2}\gamma(m-n)^2}\ 
    \ket{m}\bra{n}.
\end{align}
It can equivalently be expressed as
\begin{equation}
    \mathcal{N}_{\gamma}[\hat{\rho}]
    =
    \int_{-\infty}^\infty
    e^{-i\varphi\hat{n}}\;
    \hat{\rho}\;
    e^{i\varphi\hat{n}}
    \ 
    p_{\gamma}(\varphi)\ 
    \mathrm{d}\varphi,
    \qquad\text{with}\quad
    p_{\gamma}(\varphi)=\sqrt{\frac{1}{2\pi\gamma}}e^{-\frac{1}{2\gamma}\,\varphi^2}.
\end{equation}
The above expression is of the form \eqref{eq:mixture_of_gaussians_channel_def}, demonstrating that $\mathcal{N}_{\gamma}$ is a convex mixture of Gaussian unitaries.
We emphasize that, like most channels of the form \eqref{eq:mixture_of_gaussians_channel_def}, the dephasing channel is non-Gaussian.

\section*{Supplementary Note 5: Gaussian-dilatable channels}
\label{app:gaussian-dilatable}
\setcounter{subsection}{0} 
We consider here Gaussian-dilatable channels, i.e. the set of all channels $\mathcal{E}$ that admit a decomposition of the form of Eq.~\eqref{eq:def_gaussian_dilatable_channel}, which we recall below for convenience:
\begin{align}
    \mathcal{E}[\hat{\rho}_{\mathrm{S}}]
    =
    \mathrm{Tr}_{\mathrm{E}}\Big[
    \hat{U}_{\mathrm{G}}
    \left(
    \hat{\rho}_{\mathrm{S}}
    \otimes
    \hat{\rho}_{\mathrm{E}}
    \right)
    \hat{U}^{\dagger}_{\mathrm{G}}
    \Big].
\end{align}
In the above equation, $\hat{U}_{\mathrm{G}}$ is any Gaussian unitary, and $\hat{\rho}_{\mathrm{E}}$ any environment state (we will later focus on the case of Wigner-positive environment).
We consider the case where $\hat{\rho}_\mathrm{S}$ and $\hat{\rho}_\mathrm{E}$ are both $n$-mode states.

\subsection{Derivation of the Wigner kernel}

We derive the kernel representation of such channels in the Wigner representation.
More precisely, we find the kernel $K$ such that:
\begin{align}
    W_{\mathcal{E}[\hat{\rho}]}(\bm{x},\bm{p})
    =
    \iint
    W_{\hat{\rho}}(\bm{x}',\bm{p}')
    K(\bm{x},\bm{p},\bm{x'},\bm{p}')
    \frac{\mathrm{d}\bm{x}' \mathrm{d}\bm{p}'}{(2\pi)^n}.
    \label{eq:wigner-kernel}
\end{align}

Recall that Gaussian unitaries act as affine symplectic transformations in phase space.
With $\hat{\rho}'=\hat{U}_{\mathrm{G}}\hat{\rho}\hat{U}^{\dagger}_{\mathrm{G}}$, $W_{\hat{\rho}'}(\bm{x},\bm{p})=W_{\hat{\rho}}(\bm{x}',\bm{p}')$ where $(\bm{x}',\bm{p}')^{\intercal}=\mathbf{S}^{-1}\;\big((\bm{x},\bm{p})^{\intercal}-\bm{d}\big)$ for some symplectic matrix $\bm{S}$ and displacement vector $\bm{d}$.
Let $\bm{S}$ and $\bm{d}$ be the symplectic matrix and displacement vector associated to the Gaussian unitary $\hat{U}_{\mathrm{G}}$.
We define $\bm{T}\vcentcolon=\bm{S}^{-1}$ and $\bm{f}\vcentcolon=\bm{S}^{-1}\bm{d}$ and decompose $\bm{T}$ in four sub-matrices and $\bm{f}$ in two sub-vectors as follows:
\begin{align}
    \begin{pmatrix}
        \bm{x}'\\
        \bm{p}'\\
        \bm{x}'_{\mathrm{E}}\\
        \bm{p}'_{\mathrm{E}}
    \end{pmatrix}
    =
    \underbrace{\begin{pmatrix}
        \bm{T}_{\mathrm{SS}} &\bm{T}_{\mathrm{ES}}
        \\
        \\
        \bm{T}_{\mathrm{SE}} &\bm{T}_{\mathrm{EE}}
    \end{pmatrix}}_{\bm{T}=\bm{S}^{-1}}
    \begin{pmatrix}
        \bm{x}\\
        \bm{p}\\
        \bm{x}_{\mathrm{E}}\\
        \bm{p}_{\mathrm{E}}
    \end{pmatrix}
    -
    \underbrace{
    \begin{pmatrix}
    \bm{f}_{\mathrm{S}}
    \\
    \\
    \bm{f}_{\mathrm{E}}
    \end{pmatrix}}_{\bm{f}=\bm{S}^{-1}\bm{d}}.
    \label{eq:gaussian_dilatable_symplectic_submatrices}
\end{align}

We may then write:
\begin{align}
    W_{\mathcal{E}[\hat{\rho}]}
    (\bm{x},\bm{p})
    &=
    W_{\mathrm{Tr}_{\mathrm{E}}\big[
    \hat{U}_{\mathrm{G}}
    \left(
    \hat{\rho}
    \otimes
    \hat{\rho}_{\mathrm{E}}
    \right)
    \hat{U}^{\dagger}_{\mathrm{G}}
    \big]}
    (\bm{x},\bm{p})
    \\&=
    \iint
    W_{
    \hat{U}_{\mathrm{G}}
    \left(
    \hat{\rho}
    \otimes
    \hat{\rho}_{\mathrm{E}}
    \right)
    \hat{U}^{\dagger}_{\mathrm{G}}}
    (\bm{x},\bm{p},\bm{x}_{\mathrm{E}},\bm{p}_{\mathrm{E}})
    \ 
    \frac{\mathrm{d}\bm{x}_{\mathrm{E}}\mathrm{d}\bm{p}_{\mathrm{E}}}{(2\pi)^n}
    \\&=
    \iint
    W_{\hat{\rho}\otimes\hat{\rho}_{\mathrm{E}}}
    (\bm{x}',\bm{p}',\bm{x}'_{\mathrm{E}},\bm{p}'_{\mathrm{E}})
    \ 
    \frac{\mathrm{d}\bm{x}_{\mathrm{E}}\mathrm{d}\bm{p}_{\mathrm{E}}}{(2\pi)^n}
    \\&=
    \iint
    W_{\hat{\rho}}
    (\bm{x}',\bm{p}')
    \ 
    W_{\hat{\rho}_{\mathrm{E}}}
    (\bm{x}'_{\mathrm{E}},\bm{p}'_{\mathrm{E}})
    \ 
    \frac{\mathrm{d}\bm{x}_{\mathrm{E}}\mathrm{d}\bm{p}_{\mathrm{E}}}{(2\pi)^n}
    \\&=
    \iint
    W_{\hat{\rho}}
    \left(
    \bm{T}_{\mathrm{SS}}(\bm{x},\bm{p})^{\intercal}
    +
    \bm{T}_{\mathrm{ES}}(\bm{x}_\mathrm{E},\bm{p}_{\mathrm{E}})^{\intercal}
    -\bm{f}_{\mathrm{S}}
    \right)\label{eq:gaussian_dilatable_channel_wigpos_env_derivation}
    \\&\qquad\times 
    W_{\hat{\rho}_{\mathrm{E}}}
    \left(
    \bm{T}_{\mathrm{SE}}(\bm{x},\bm{p})^{\intercal}
    +
    \bm{T}_{\mathrm{EE}}(\bm{x}_\mathrm{E},\bm{p}_\mathrm{E})^{\intercal}
    -\bm{f}_{\mathrm{E}}
    \right)
    \ 
    \frac{\mathrm{d}\bm{x}_{\mathrm{E}}\mathrm{d}\bm{p}_{\mathrm{E}}}{(2\pi)^n}.\nonumber
\end{align}
Assuming $\bm{T}_{\mathrm{ES}}$ is invertible, we introduce the following change of variables:
\begin{align}
    \begin{pmatrix}
        \tilde{\bm{x}}
        \\
        \tilde{\bm{p}}
    \end{pmatrix}
    =
    \bm{T}_{\mathrm{SS}}
    \begin{pmatrix}
        \bm{x}\\
        \bm{p}
    \end{pmatrix}
    +
    \bm{T}_{\mathrm{ES}}
    \begin{pmatrix}
        \bm{x}_{\mathrm{E}}\\
        \bm{p}_{\mathrm{E}}
    \end{pmatrix}
    -
    \bm{f}_{\mathrm{S}}
    \quad\Leftrightarrow\quad
    \begin{pmatrix}
        \bm{x}_{\mathrm{E}}\\
        \bm{p}_{\mathrm{E}}
    \end{pmatrix}
    =
    \bm{T}^{-1}_{\mathrm{ES}}
    \left[
    \begin{pmatrix}
        \tilde{\bm{x}}\\
        \tilde{\bm{p}}
    \end{pmatrix}
    -
    \bm{T}_{\mathrm{SS}}
    \begin{pmatrix}
        \bm{x}\\
        \bm{p}
    \end{pmatrix}
    +\bm{f}_{\mathrm{S}}
    \right],
\end{align}
which implies the relation $\mathrm{d}\tilde{\bm{x}}\mathrm{d}\tilde{\bm{p}}=\abs{\det\bm{T}_{\mathrm{ES}}}\mathrm{d}\bm{x}_{\mathrm{E}}\mathrm{d}\bm{p}_{\mathrm{E}}$.
Substituting this into Eq.~\eqref{eq:gaussian_dilatable_channel_wigpos_env_derivation} yields:
\begin{align}
    W_{\mathcal{E}[\hat{\rho}]}(\bm{x},\bm{p})
    =&
    \iint
    \ \frac{\mathrm{d}\tilde{\bm{x}}\mathrm{d}\tilde{\bm{p}}}{(2\pi)^n}
    \ 
    \frac{1}{\abs{\det\bm{T}_{\mathrm{ES}}}}
    \ 
    W_{\hat{\rho}}
    \left(
    \tilde{\bm{x}},\tilde{\bm{p}}
    \right)
    \\[0.5em]&\times
    W_{\hat{\rho}_{\mathrm{E}}}
    \left(
    \bm{T}_{\mathrm{SE}}(\bm{x},\bm{p})^{\intercal}
    +
    \bm{T}_{\mathrm{EE}}
    \bm{T}^{-1}_{\mathrm{ES}}
    \Big[
    (\tilde{\bm{x}},\tilde{\bm{p}})^{\intercal}
    -
    \bm{T}_{\mathrm{SS}}
    (\bm{x},\bm{p})^{\intercal}
    +\bm{f}_{\mathrm{S}}
    \Big]
    -\bm{f}_{\mathrm{E}}
    \right).\nonumber
\end{align}
From this relation, we can identify the Wigner kernel of $\mathcal{E}$, see Eq.~\eqref{eq:wigner-kernel}, as
\begin{align}
    K(\bm{x},\bm{p},\bm{x'},\bm{p}')
    &=
    \frac{1}{\abs{\det\bm{T}_{\mathrm{ES}}}}
    W_{\hat{\rho}_{\mathrm{E}}}
    \left(
    \bm{T}_{\mathrm{SE}}(\bm{x},\bm{p})^{\intercal}
    +
    \bm{T}_{\mathrm{EE}}
    \bm{T}^{-1}_{\mathrm{ES}}
    \Big[
    (\bm{x}',\bm{p}')^{\intercal}
    -
    \bm{T}_{\mathrm{SS}}
    (\bm{x},\bm{p})^{\intercal}
    +\bm{f}_{\mathrm{S}}
    \Big]
    -\bm{f}_{\mathrm{E}}
    \right)
    \\&=
    \frac{1}{\abs{\det\bm{T}_{\mathrm{ES}}}}
    W_{\hat{\rho}_{\mathrm{E}}}
    \left(
    \left[
    \bm{T}_{\mathrm{SE}}
    -
    \bm{T}_{\mathrm{EE}}
    \bm{T}^{-1}_{\mathrm{ES}}
    \bm{T}_{\mathrm{SS}}
    \right]
    (\bm{x},\bm{p})^{\intercal}
    +
    \bm{T}_{\mathrm{EE}}
    \bm{T}^{-1}_{\mathrm{ES}}
    (\bm{x}',\bm{p}')^{\intercal}
    +\tilde{\bm{f}}
    \right),
\end{align}
where $\tilde{\bm{f}}=\bm{T}_{\mathrm{EE}}\bm{T}_{\mathrm{ES}}^{-1}\bm{f}_{\mathrm{S}}-\bm{f}_{\mathrm{E}}$.

\subsection{Properties of the Wigner kernel}
It is clear that $K(\bm{x},\bm{p},\bm{x}',\bm{p}')$ is nonnegative whenever $\hat{\rho}_{\mathrm{E}}$ is Wigner-positive.
We compute:
\begin{align}
    \iint K(\bm{x},\bm{p},\bm{x}',\bm{p}')
    \frac{\mathrm{d}\bm{x}\mathrm{d}\bm{p}}{(2\pi)^n}
    &=
    \iint
    \frac{1}{\abs{\det\bm{T}_{\mathrm{ES}}}}
    W_{\hat{\rho}_{\mathrm{E}}}
    \left(
    \left[
    \bm{T}_{\mathrm{SE}}
    -
    \bm{T}_{\mathrm{EE}}
    \bm{T}^{-1}_{\mathrm{ES}}
    \bm{T}_{\mathrm{SS}}
    \right]
    (\bm{x},\bm{p})^{\intercal}
    \right)
    \,\frac{\mathrm{d}\bm{x}\mathrm{d}\bm{p}}{(2\pi)^n}
    \\[1em]&=
    \frac{1}{\abs{\det\bm{T}_{\mathrm{ES}}}}
    \frac{1}{\abs{\det\bm{T}_{\mathrm{SE}}
    -
    \bm{T}_{\mathrm{EE}}
    \bm{T}^{-1}_{\mathrm{ES}}
    \bm{T}_{\mathrm{SS}}}}
    \iint
    W_{\hat{\rho}_\mathrm{E}}(\bm{x},\bm{p})
    \frac{\mathrm{d}\bm{x}\mathrm{d}\bm{p}}{(2\pi)^n}
    \\[1em]&=
    \abs{
    \det(\bm{T}_{\mathrm{ES}})
    \det(\bm{T}_{\mathrm{SE}}
    -
    \bm{T}_{\mathrm{EE}}
    \bm{T}^{-1}_{\mathrm{ES}}
    \bm{T}_{\mathrm{SS}})
    }^{-1}
    \\[1em]&=
    \abs{\det(\bm{T})}^{-1}
    =
    \abs{\det(\bm{S})}
    =1
\end{align}
where we have used the relation $\det\begin{pmatrix}\bm{A}&\bm{B}\\\bm{C}&\bm{D}\end{pmatrix}=\det(\bm{A}-\bm{B}\bm{D}^{-1}\bm{C})\det(\bm{D})$, that permuting rows and columns of a matrix does not change the absolute value of its determinant, and that a symplectic matrix has a determinant equal to $1$.
Assuming $\bm{T}_{\mathrm{EE}}$ is invertible, we further compute:
\begin{align}
    \iint K(\bm{x},\bm{p},\bm{x}',\bm{p}')
    \frac{\mathrm{d}\bm{x}'\mathrm{d}\bm{p}'}{(2\pi)^n}
    &=
    \iint
    \frac{1}{\abs{\det\bm{T}_{\mathrm{ES}}}}
    W_{\hat{\rho}_{\mathrm{E}}}
    \left(
    \bm{T}_{\mathrm{EE}}
    \bm{T}^{-1}_{\mathrm{ES}}
    (\bm{x}',\bm{p}')^{\intercal}
    \right)
    \,\frac{\mathrm{d}\bm{x}'\mathrm{d}\bm{p}'}{(2\pi)^n}
    \\[1em]&=
    \iint
    \frac{1}{\abs{\det(\bm{T}_{\mathrm{ES}})}}
    \frac{1}{\abs{\det(\bm{T}_{\mathrm{EE}}
    \bm{T}^{-1}_{\mathrm{ES}})}}
    W_{\hat{\rho}_{\mathrm{E}}}
    \left(
    \bm{x}',\bm{p}'
    \right)
    \,\frac{\mathrm{d}\bm{x}'\mathrm{d}\bm{p}'}{(2\pi)^n}
    \\[1em]&=
    \abs{\mathrm{det}(\bm{T}_{\mathrm{ES}})\det(\bm{T}_{\mathrm{EE}})\det(\bm{T}^{-1}_{\mathrm{ES}})}^{-1}
    =
    \abs{\det(\bm{T}_{\mathrm{EE}})}^{-1}
\end{align}
where we have used the relation $\det(\bm{A}\bm{B})=\det(\bm{A})\det(\bm{B})$ and $\det(\bm{A}^{-1})=\det({\bm{A}})^{-1}$.
Recall that $\bm{T}=\bm{S}^{-1}$.
We find that the value of $\abs{\det\left[(\bm{S}^{-1})_{\mathrm{EE}}\right]}$ determines if the channel $\mathcal{E}$ is doubly stochastic or semidoubly stochastic.

\section*{Supplementary Note 6: Majorization on Husimi functions}
\label{app:Husimi}
\setcounter{subsection}{0} 
The \textit{Husimi function} $Q_{\hat \rho}$ is another common quasiprobability representation~\cite{husimi1940some, cahill1969density} over $\mathbb R^{2n}$ related to the Wigner function via a Gaussian convolution:
\begin{equation}
    Q_{\hat \rho}(\bm x, \bm p) = \frac{1}{\pi^n}\int_{X} \mathrm d\bm x' \mathrm d \bm p' \, \text{exp}\left[ -\left| 
    \begin{mypsmallmatrix}
    \bm{x}\\\bm{p}
\end{mypsmallmatrix} - 
\begin{mypsmallmatrix}
    \bm{x}'\\\bm{p}'
\end{mypsmallmatrix} \right|^2  \right] W_{\hat \rho}(\bm x', \bm p').
\end{equation}
This function is directly related to the minimum-uncertainty coherent states $\ket{\bm x, \bm p} \in L^2(\mathbb R^n)$, often denoted by $\ket{\bm \alpha}$ where $\bm \alpha \in \mathbb C^n \simeq \mathbb R^{2n}$.  The phase-point operators \eqref{eq:SW_kernel_abstract} $\hat \Delta_Q(\bm x, \bm p)$ are proportional to the corresponding coherent state projectors:
\begin{equation}
    Q_{\hat \rho}(\bm x, \bm p) = \frac{1}{\pi^n} \Tr[ \ketbra{\bm x, \bm p}{\bm x, \bm p} \, \hat \rho ] = \frac{1}{\pi^n} \langle \bm x, \bm p | \hat \rho | \bm x, \bm p \rangle
\end{equation}
Unlike the displaced parity operators from the Wigner-Weyl-Moyal representation, the coherent state projectors do not form a self-dual representation; see the Sudarshan-Glauber P function~\cite{Glauber_1963, Sudarshan_1963} for more details.

Since the Q function is positive, it can be studied using the usual majorization definitions for probabilities. Wehrl introduced his eponymous entropy as the Shannon entropy of the Q function and conjectured that it is minimized by vacuum~\cite{Wehrl1979}. Lieb and Solovej later proved this conjecture, showing that, in fact, vacuum majorizes every other state~\cite{Lieb1978,Lieb2014}. 

We remark that the Wigner R\'{e}nyi entropies we introduce have the advantage that, unlike the Wehrl entropy, they are invariant under all Gaussian unitaries, including squeezing. Extending existing results, we prove two analytic hierarchies for Q functions of Fock states, visualized in Supplementary Figure~\ref{fig:l_fockrelhusimi} for the first few Fock states. First, we prove that $Q_{\ket{n}}\succ Q_{\ket{n+1}}$ by constructing the appropriate SDS operators. This is consistent with the Lieb-Wehrl theorem. Second, we show that $Q_{\ket{n+1}} \succ_{Q_{\ket{0}}} Q_{\ket{n}}$ by finding an explicit formula for the Lorenz curves. Thus, unlike for regular majorization, for majorization relative to vacuum there is no maximally ordered Husimi function.

\begin{figure}[h!]
\centering
\includegraphics[width=1\linewidth]{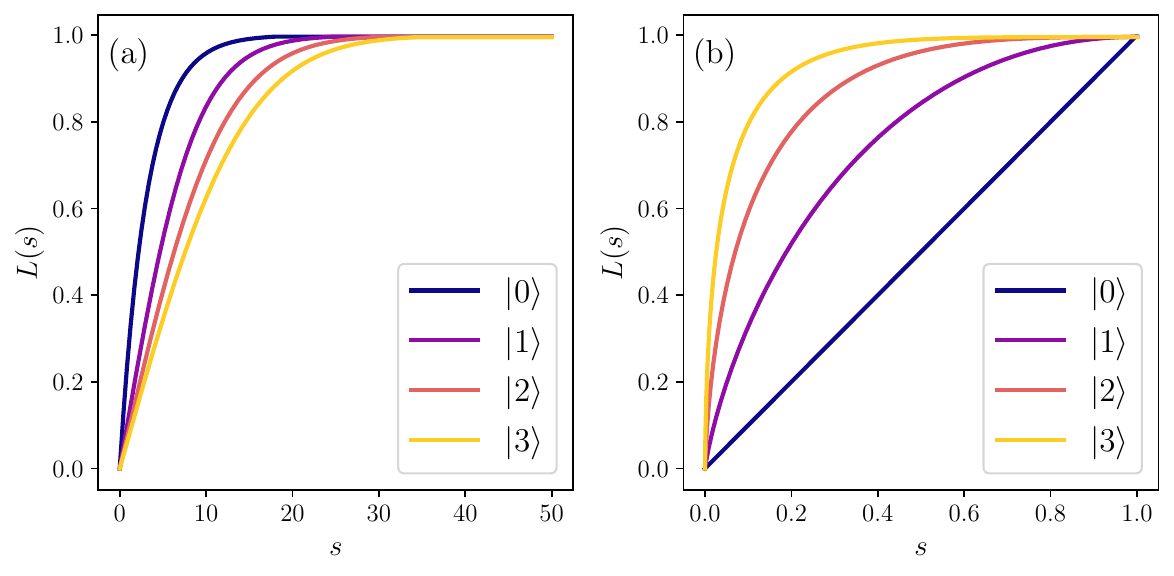}
\captionof{figure}{Regular (a) and relative to vacuum (b) Husimi Lorenz curves of first few Fock states. In both cases they form a majorization hierarchy.}
\label{fig:l_fockrelhusimi}
\end{figure}

\subsection{Majorization hierarchy for Husimi functions of Fock states}

We prove here that the Husimi functions of Fock states obey a majorization hierarchy.  More precisely, we show that $Q_{\ket{n}}\succ Q_{\ket{n+1}}$ where $Q_{\ket{n}}$ is the Husimi function of Fock state $\ket{n}$.
To do so, we employ a similar technique to the one used in Ref.~\cite{VanHerstraeten2023continuous}.

Recall that the Husimi function of the Fock state $\ket{n}$ reads
\begin{align}
    Q_{\ket{n}}(x,p)
    =
    \frac{1}{\pi}
    \frac{1}{n!}
    (x^2 + p^2)^{n}
    \exp(-x^2 - p^2).
\end{align}
This function is rotationally symmetric and depends only on the radial variable $r = \sqrt{x^2 + p^2}$.
We can therefore apply Lemma~1 from Ref.~\cite{VanHerstraeten2023continuous}, which asserts that the majorization relation $Q_{\ket{n}} \succ Q_{\ket{n+1}}$ is equivalent to $f_n \succ f_{n+1}$, where the functions $f_n: \mathbb{R}\to\mathbb{R}$ (extended trivially to the negative real line) are defined by
\begin{align}
    f_n(u)
    =
    \frac{1}{n!}\ 
    u^n\ 
    \exp(-u)\ 
    \Theta(u),
\end{align}
with $\Theta(u)$ denoting the Heaviside step function, i.e., $\Theta(u) = 1$ for $u \geq 0$ and zero otherwise.
Define now the kernel $K(u,v)$ as follows:
\begin{align}
    K(u,v)=\exp(v-u)\Theta(u-v).
\end{align}
A direct calculation shows that $\int K(u,v)\mathrm{d}v=\int K(u,v)\mathrm{d}u=1$.
Since in addition $K(u,v)\geq 0\ \forall u,v$, the kernel is doubly stochastic.
We now compute how $K$ acts on $f_{n}$:
\begin{align}
    \int\limits_{-\infty}^{\infty} K(u,v)f_n(v)\mathrm{d}v
    &=
    \int\limits_{-\infty}^{\infty} 
    \exp(v-u)\ \Theta(u-v)\ 
    \frac{1}{n!}\ 
    v^{n}\ 
    \exp(-v)\ 
    \Theta(v)
    \mathrm{d}v
    \\&=
    \Theta(u)
    \int\limits_{0}^{u} 
    \exp(v-u)
    \frac{1}{n!}\ 
    v^{n}\ 
    \exp(-v)\ 
    \mathrm{d}v
    \\&=
    \Theta(u)\ 
    \frac{1}{n!}\ 
    \exp(-u)\ 
    \underbrace{\int\limits_{0}^{u} 
    v^{n}\ 
    \mathrm{d}v}_{\frac{u^{n+1}}{n+1}}
    \\&=
    \Theta(u)\ 
    \frac{1}{(n+1)!}\ 
    \exp(-u)\ 
    u^{n+1}
    =
    f_{n+1}(u).
\end{align}

Since $f_{n+1}$ can be obtained from $f_{n}$ via the action of a doubly stochastic kernel, it follows that $f_{n}\succ f_{n+1}$.
Applying Lemma~1 from Ref.~\cite{VanHerstraeten2023continuous}, we conclude that $Q_{\ket{n}}\succ Q_{\ket{n+1}}$, which completes the proof.

\subsection{Relative majorization hierarchy for Husimi functions of Fock states}
Here we prove that Fock Husimi functions obey a majorization hierarchy relative to vacuum.
\begin{equation}
    Q_{\ket{n+1}} \succ_{Q_{\ket{0}}} Q_{\ket{n}}.
\end{equation}
The Husimi function is nonnegative so it suffices to compare the positive Lorenz curves in statement 1 of Theorem~2.
Before calculating the distribution function, we note the following change of integration variables for radially symmetric integrands, $\frac{1}{\pi} \int \int \mathrm dx \mathrm dp = 2 \int_0^\infty r \mathrm dr = \int_0^\infty \mathrm dz$.
The distribution function is
\begin{align}
    D^{Q_{\ket{0}}}_{Q_{\ket{n}}}(t)&=\nu\left\{\frac{Q_{\ket{n}}}{Q_{\ket{0}}} > t\right\}\\
    &=\nu\left\{\frac{(x^2+p^2)^n}{n !}>t\right\}\\
    &= \int \int \chi_{\frac{(x^2+p^2)^n}{n !} >t}(x,p) \frac{1}{\pi}e^{-(x^2+p^2)} \mathrm dx \mathrm dp\\
    &=\int_{0}^\infty \chi_{\frac{z^n}{n!}>t}(z)  e^{-z}\mathrm dz\\
    &= \int_{(n ! t)^\frac{1}{n}}^\infty e^{-z}\mathrm dz =e^{-\left[(n ! t)^\frac{1}{n}\right]}\ . 
\end{align}
Then, the decreasing rearrangement is
\begin{align}
    Q_{\ket{n}}^{\downarrow,Q_{\ket{0}}}(u) &=\inf\{ t: D^{Q_{\ket{0}}}_{Q_{\ket{n}}}(t) \leq u \}\\
    &= \inf\{ t: e^{-\left[(n ! t)^\frac{1}{n}\right]}   \leq u \}\\
    &= \frac{(-\ln u)^n}{n!}.
\end{align}
Finally, the Lorenz curve for $n>0$ can be expressed as
\begin{align}
    L^{\Dsh}_{Q_{\ket{n}}|Q_{\ket{0}}}(s) &=\int_0^s Q_{\ket{n}}^{\downarrow,Q_{\ket{0}}}(u) \mathrm du\\
    &=\int_0^s \frac{(-\ln u)^n}{n!} \mathrm du \\
    &= \frac{(-1)^n}{n!} \left[\left. u(\ln u)^n\right\vert_0^s -\int_0^s n (\ln u)^{n-1}   \mathrm du \right] \\ 
    &=s \frac{(-\ln s)^n}{n!} + \int_0^s \frac{(-\ln u)^{n-1}}{(n-1)!}  \mathrm du \\
    &= s \frac{(-\ln s)^n}{n!} + L^{\Dsh}_{Q_{\ket{n-1}}|Q_{\ket{0}}}(s),
\end{align}
where in the third line we integrate by parts. Since $s \frac{(-\ln s)^n}{n!}$ is nonnegative for all $s\in[0,1]$, $L^{\Dsh}_{Q_{\ket{n+1}}|Q_{\ket{0}}}(s)\geq L^{\Dsh}_{Q_{\ket{n}}|Q_{\ket{0}}}(s)$. Thus we prove the desired relative majorization relation by statement 1 of Theorem~2.

Note that $L^{\Dsh}_{Q_{\ket{0}}|Q_{\ket{0}}}(s)=s$, so we get the formula 
\begin{equation}
    L^{\Dsh}_{Q_{\ket{n}}|Q_{\ket{0}}}(s) = s\left(1+ \sum_{k=1}^n \frac{(-\ln s)^k}{k!}\right).
\end{equation}

\section*{Supplementary Note 7: Conventions for Wigner and Husimi functions used in plots}\label{app:numerical}
\setcounter{subsection}{0} 
For reference, we list the Wigner and Husimi functions used in our examples.
Note that in our numerical computations involving Wigner functions, we use a slightly different convention from that of Eq.~\eqref{eq:def_wigner_function}, as the Wigner functions shown below are obtained by setting $\hbar = 1/2$.
This minor difference does not change the majorization preorder.

\begin{align}
W(\ket{n}) &= \frac{2}{\pi} \exp\left(-2(x^2 + p^2)\right) \cdot (-1)^n L_n\left(4(x^2 + p^2)\right)\\
W(\ket{\alpha})&= \frac{2}{\pi}  \exp\left(-2\left([x - \text{Re}(\alpha)]^2 + [p - \text{Im}(\alpha)]^2\right)\right)\\
W(\rho_\textrm{th}(\bar{n})) &=  \frac{2}{\pi(1 + 2 \bar{n})}  \cdot \exp\left(-\frac{2(x^2 + p^2)}{1 + 2 \bar{n}} \right)\\
W(\ket{a,n}_\textrm{ON})&= \frac{1}{1 + |a|^2} \cdot W(\ket{0}) + \frac{|a|^2}{1 + |a|^2} \cdot W(\ket{n}) \\&\quad+ \frac{2^{n+1}}{\pi \sqrt{n!}(1 + |a|^2)}    \exp(-2(x^2 +p^2)) \left[a(x - i p)^n + \overline{a}(x + i p)^n\right] \nonumber\\
W(\ket{\alpha}_\textrm{cat})&= \frac{1}{2\left(1 + \exp(-2 | \alpha |^2)\right)} \left( W(\ket{\alpha}) + W(\ket{-\alpha}) + \frac{4}{\pi} \exp(-2(x^2 + p^2)) \cos(4\alpha p) \right)\\
W(\ket{g,P,s}_\textrm{cubic})&=\frac{4}{\sqrt{8\pi^3 \exp(2s) }}\exp(-\frac{2 x^2}{\exp(2 s)}) \int_0^\infty 2 \cos(2 g y^3+2y(12g x^2-p+P)) \exp(-\frac{y^2}{2\exp(2s)}) dy\\
Q(\ket{n})&=\frac{1}{\pi} \frac{(x^2 + p^2)^n}{n!} \exp{-\left(x^2 + p^2\right)}
\end{align}
where $L_n$ are the Laguerre polynomials and the cat state is the equal superposition of $\ket{\alpha}$ and $\ket{-\alpha}$ with $\alpha$ real.

\bibliography{mainbib.bib}